\documentclass{LMCS}

\def\doi{7 (3:20) 2011}
\lmcsheading%
{\doi}
{1--45}
{}
{}
{Nov.~30, 2009}
{Sep.~28, 2011}
{}   

\usepackage[english]{babel}
\usepackage{mathtools}
\usepackage{xspace}
\usepackage{array}
\usepackage{booktabs}
\usepackage{amssymb}
\usepackage{lmodern}
\usepackage[T1]{fontenc}
\usepackage{tikz}
\usepackage[numbers]{natbib}
\usepackage{enumerate}
\usepackage[breaklinks,pdfpagelabels]{hyperref}
\usepackage{caption}
\usepackage{floatrow}
\usepackage[capitalise]{cleveref}

\newcommand{\dbr}{\discretionary{}{}{}}

\DeclareFloatVCode{toprule}{%
  \hrule height\heavyrulewidth\vskip 1ex%
}
\DeclareFloatVCode{topprule}{%
  \hrule height\heavyrulewidth\vskip 1.5ex%
}
\DeclareFloatVCode{midrule}{%
  \vskip 1ex\hrule height\lightrulewidth
}
\DeclareFloatVCode{endrule}{%
  \vskip 1.5ex\hrule height\heavyrulewidth%
}
\DeclareNewFloatType{algorithm}{placement=tbp,name=Algorithm,fileext=loa,within=section}
\newcommand{\keyw}[1]{\textbf{#1}}

\newcommand\cites[2][]{(see~\citep[][#1]{#2})}
\makeatletter
\renewcommand{\NAT@spacechar}{~}
\makeatother

\crefformat{equation}{#2(#1)#3}
\Crefformat{equation}{#2(#1)#3}
\crefrangeformat{equation}{#3(#1)#4--#5(#2)#6}
\Crefrangeformat{equation}{#3(#1)#4--#5(#2)#6}
\crefmultiformat{equation}{#2(#1)#3}{\crefpairconjunction#2(#1)#3}%
  {, #2(#1)#3}{\creflastconjunction#2(#1)#3}
\Crefmultiformat{equation}{#2(#1)#3}{\crefpairconjunction#2(#1)#3}%
  {, #2(#1)#3}{\creflastconjunction#2(#1)#3}
\crefrangemultiformat{equation}{#3(#1)#4--#5(#2)#6}%
  {\crefpairconjunction#3(#1)#4--#5(#2)#6}{, #3(#1)#4--#5(#2)#6}%
  {\creflastconjunction#3(#1)#4--#5(#2)#6}
\Crefrangemultiformat{equation}{#3(#1)#4--#5(#2)#6}%
  {\crefpairconjunction#3(#1)#4--#5(#2)#6}{, #3(#1)#4--#5(#2)#6}%
  {\creflastconjunction#3(#1)#4--#5(#2)#6}

\crefname{algorithm}{Algorithm}{Algorithms}

\theoremstyle{definition}
\newtheorem{definition}{Definition}[section]
\newtheorem*{definition*}{Definition}
\crefname{definition}{Definition}{Definitions}
\newtheorem{proposition}[definition]{Proposition}
\newtheorem*{proposition*}{Proposition}
\crefname{proposition}{Proposition}{Propositions}
\newtheorem{lemma}[definition]{Lemma}
\newtheorem*{lemma*}{Lemma}
\crefname{lemma}{Lemma}{Lemmas}
\newtheorem{theorem}[definition]{Theorem}
\newtheorem*{theorem*}{Theorem}
\crefname{theorem}{Theorem}{Theorems}
\newtheorem{corollary}[definition]{Corollary}
\newtheorem*{corollary*}{Corollary}
\crefname{corollary}{Corollary}{Corollaries}
\newtheorem{example}[definition]{Example}
\newtheorem*{example*}{Example}
\crefname{example}{Example}{Examples}
\theoremstyle{remark}
\newtheorem{remark}[definition]{Remark}
\newtheorem*{remark*}{Remark}
\crefname{remark}{Remark}{Remarks}
\newtheorem*{claim}{Claim}

\let\phi\varphi

\let\theta\vartheta

\let\epsilon\varepsilon

\newcommand{\restrict}{\upharpoonright}
\newcommand{\overbar}[1]{\mkern 1mu\overline{\mkern-1mu#1\mkern-1mu}\mkern 1mu}
\renewcommand{\vec}[1]{\overbar{#1}}
\renewcommand{\tilde}[1]{\widetilde{#1}}

\newcommand{\coloneq}{\vcentcolon=}
\let\nin\notin
\newcommand{\pow}{\mathcal{P}}

\newcommand{\abs}[1]{\lvert#1\rvert}

\newcommand{\bbN}{\mathbb{N}}
\newcommand{\bbR}{\mathbb{R}}
\newcommand{\calC}{\mathcal{C}}
\newcommand{\calD}{\mathcal{D}}
\newcommand{\calG}{\mathcal{G}}
\newcommand{\calF}{\mathcal{F}}
\newcommand{\calM}{\mathcal{M}}
\newcommand{\frakM}{\mathfrak{M}}
\newcommand{\frakR}{\mathfrak{R}}

\newcommand{\infer}[2]{#1\,$\Rightarrow$\,#2}

\newcommand{\LogSpace}{\textsc{Logspace}\xspace}

\newcommand{\PTime}{\textup{P}\xspace}

\newcommand{\NP}{\textup{NP}\xspace}
\newcommand{\coNP}{\textup{coNP}\xspace}

\newcommand{\UP}{\textup{UP}\xspace}
\newcommand{\coUP}{\textup{coUP}\xspace}
\newcommand{\DP}{\textup{DP}\xspace}

\newcommand{\PNPlog}{\ensuremath{\PTime^{\NP[\log]}}\xspace}

\newcommand{\PSpace}{\textsc{Pspace}\xspace}
\newcommand{\NPSpace}{\textsc{NPspace}\xspace}

\usetikzlibrary{arrows}
\usetikzlibrary{shapes}
\tikzset{every picture/.style={>=stealth,bend angle=20}}
\tikzset{every label/.style={font=\footnotesize}}
\tikzset{every node/.style={font=\footnotesize}}
\tikzset{play/.style={circle,draw,minimum size=#1}}
\tikzset{play/.default=0.85cm}
\tikzset{prob/.style={diamond,draw,minimum size=#1}}
\tikzset{prob/.default=0.9cm}
\tikzset{end/.style={rectangle,draw,minimum size=#1}}
\tikzset{end/.default=0.75cm}

\newcommand{\cf}{cf.\@\xspace}

\newcommand{\eg}{e.g.\@\xspace}

\renewcommand{\iff}{if and only if\xspace}
\newcommand{\ie}{i.e.\@\xspace}
\newcommand{\wrt}{with respect to\xspace}
\newcommand{\wlg}{without loss of generality\xspace}

\DeclareMathOperator{\true}{true}
\DeclareMathOperator{\false}{false}

\DeclareMathOperator{\Prob}{Pr}

\DeclareMathOperator{\Inf}{Inf}
\DeclareMathOperator{\Occ}{Occ}
\DeclareMathOperator{\Reach}{Reach}
\DeclareMathOperator{\Buchi}{B\ddot{u}chi}
\DeclareMathOperator{\CoBuchi}{coB\ddot{u}chi}
\DeclareMathOperator{\Parity}{Parity}
\DeclareMathOperator{\Streett}{Streett}
\DeclareMathOperator{\Rabin}{Rabin}
\DeclareMathOperator{\Muller}{Muller}

\DeclareMathOperator{\init}{init}

\DeclareMathOperator{\val}{val}

\DeclareMathOperator{\FindEC}{FindEC}
\newcommand{\Win}{\mathrm{Win}}

\DeclareMathOperator{\inc}{inc}
\DeclareMathOperator{\dec}{dec}
\DeclareMathOperator{\zero}{zero}
\DeclareMathOperator{\halt}{halt}

\newcommand{\NE}{\text{NE}\xspace}

\newcommand{\EQualNE}{\text{StrQualNE}\xspace}
\newcommand{\PureNE}{\text{PureNE}\xspace}
\newcommand{\FinNE}{\text{FinNE}\xspace}
\newcommand{\PureFinNE}{\text{PureFinNE}\xspace}
\newcommand{\StatNE}{\text{StatNE}\xspace}
\newcommand{\PosNE}{\text{PosNE}\xspace}

\newcommand{\PPAD}{\text{PPAD}\xspace}

\newcommand{\SAT}{\text{SAT}\xspace}

\newcommand{\SqrtSum}{\text{SqrtSum}\xspace}

\newcommand{\TwoSG}{\text{S2G}\xspace}
\newcommand{\TwoSGs}{\text{S2Gs}\xspace}

\newcommand{\TwoSSGs}{\text{SS2Gs}\xspace}

\newcommand{\F}{\mathrm{F}}
\newcommand{\G}{\mathrm{G}}
\renewcommand{\P}{\mathrm{P}}

\newcommand\pl[1]{player~$#1$\xspace}
\newcommand\pli{\pl{i}}
\newcommand\plj{\pl{j}}
\newcommand\Pl[1]{Player~$#1$\xspace}
\newcommand\Pli{\Pl{i}}

\newcommand{\omg}{$\omega$\nobreakdash}

\newcommand{\eps}{$\epsilon$\nobreakdash}

\begin{document}
\title[The Complexity of Nash Equilibria in Stochastic  Multiplayer Games]%
  {The Complexity of Nash Equilibria \\ in Stochastic Multiplayer Games\rsuper*}

\author[M.~Ummels]{Michael Ummels\rsuper a}
\address{{\lsuper a}RWTH Aachen University, Germany}
\email{ummels@logic.rwth-aachen.de}

\author[D.~Wojtczak]{Dominik Wojtczak\rsuper b}
\address{{\lsuper b}CWI Amsterdam, The Netherlands}
\email{d.k.wojtczak@cwi.nl}

\titlecomment{{\lsuper*}Preliminary versions of parts of this paper appeared in the
\emph{Proceedings of the 36th International Colloquium on Automata, Languages
and Programming} (ICALP~2009) and the \emph{Proceedings of the 18th Annual
Conference of the European Association for Computer Science Logic} (CSL~2009).
This work was supported by the DFG Research Training Group~1298
(\textsc{AlgoSyn}) and the ESF Research Networking Programme
``Games for Design and Verification'' (GAMES)}

\begin{abstract}
We analyse the computational complexity of finding Nash equilibria in
turn-based stochastic multiplayer games with \omg-regular objectives. We
show that restricting the search space to
equilibria whose payoffs fall into a certain interval may lead to
undecidability.
In particular, we prove that the following problem is undecidable: Given a
game~$\calG$, does there exist a Nash equilibrium of~$\calG$
where \pl0 wins with probability~$1$? Moreover, this problem remains
undecidable when restricted to pure strategies or (pure) strategies
with finite memory.
One way to obtain a decidable variant of the problem is to restrict
the strategies to be positional or stationary. For the complexity of these two
problems, we obtain a common lower bound of \NP and upper bounds of \NP and
\PSpace respectively. Finally, we single out a special case of the general
problem that, in many cases, admits an efficient solution. In particular, we
prove that deciding the existence of an equilibrium in which each player either
wins or loses with probability~$1$ can be done in polynomial time for games
where the objective of each player is given by a parity
condition with a bounded number of priorities.
\end{abstract}

\maketitle

\section{Introduction}

We study \emph{stochastic games} \citep{NeymanS03} played by multiple
players on a finite, directed graph. Intuitively, a play of such a game evolves
by moving a token along edges of the graph: Each vertex of the graph is either
controlled by one of the players, or it is \emph{stochastic}. Whenever
the token arrives at a non-stochastic vertex, the player who controls this
vertex must move the token to a successor vertex; when the token arrives at a
stochastic vertex, a fixed probability distribution determines the next vertex.
A measurable function maps plays to payoffs.
In the simplest case, which we discuss here, the possible payoffs of a single
play are $0$ and~$1$ (\ie each player either wins or loses a given play).
However,
due to the presence of stochastic vertices, a player's \emph{expected payoff}
(\ie her probability of winning) can be an arbitrary probability.

Stochastic games with \omg-regular objectives have been used as a
formal model for the verification and
synthesis of reactive systems under the influence of random events
\cite{BaierGLBC04}.
Such a system is usually modelled as a game between the
system and its environment, where the environment's objective is the complement
of the system's objective: the environment is considered hostile. Therefore,
the research in this area has traditionally focused on two-player games
where each play is won by precisely one of the two players, so-called
\emph{two-player zero-sum games}. However, the system may consist of
several components with independent objectives, a situation which is naturally
modelled by a multiplayer game.

The most common interpretation of rational behaviour in multiplayer games is
captured by the notion of a \emph{Nash equilibrium} \citep{Nash50}. In a Nash
equilibrium, no player can improve her payoff by unilaterally switching to a
different strategy. \Citet{ChatterjeeJM04} gave an algorithm for
computing a Nash equilibrium in a stochastic multiplayer game with
\omg-regular winning conditions. However, it can be shown that their
algorithm may compute an
equilibrium where all players lose almost surely (\ie receive expected
payoff~0), even when there exist other equilibria where all players win almost
surely (\ie receive expected payoff~$1$).

In applications, one might look for an equilibrium where as many players as
possible win almost surely or where it is guaranteed that the expected payoff
of the equilibrium falls into a certain interval. Formulated as a
decision problem, we want to know, given a $k$-player game~$\calG$ with
initial vertex~$v_0$ and two thresholds $\vec{a},\vec{b}\in [0,1]^k$,
whether $(\calG,v_0)$ has a Nash equilibrium with expected payoff at
least~$\vec{x}$ and at most~$\vec{y}$. This problem, which we call \NE for
short, is a generalisation of the \emph{quantitative decision problem} for
two-player zero-sum games, which asks whether in such a game \pl0 has
a strategy that ensures to win the game with a probability that exceeds
a given threshold.

The problem \NE comes in several variants, depending on the type of strategies
one considers: On the one hand, strategies may be \emph{randomised} (allowing
randomisation over actions) or \emph{pure} (not allowing such randomisation).
On the other hand, one can restrict to strategies that use (unbounded or
bounded) finite memory or even to \emph{stationary} ones (strategies that do not
use any memory at all). For the quantitative decision problem, this distinction
is often not meaningful since in a two-player zero-sum simple stochastic game
with \omg-regular objectives both players have optimal pure strategies
with finite memory. Moreover, in many games even positional (\ie both pure
and stationary) strategies suffice for optimality. However, regarding \NE this
distinction leads to distinct decision problems, which have to be analysed
separately.

Our main result is that \NE is undecidable if we allow either arbitrary
randomised strategies or arbitrary pure strategies. In fact, even
the following, presumably simpler, problem is
undecidable: Given a game~$\calG$, decide whether there exists a Nash
equilibrium (in pure strategies) where \pl0 wins almost surely. Moreover,
the problem remains undecidable if one restricts to randomised or pure
strategies with finite memory.

If we restrict to simpler types of strategies like stationary ones,
\NE becomes decidable. In particular, for
positional strategies the problem is typically \NP-complete, and for
arbitrary stationary strategies it is \NP-hard but typically
contained in \PSpace. To get a better understanding of the latter problem,
we also relate it to the \emph{square root sum problem} (\SqrtSum) by providing
a polynomial-time reduction from \SqrtSum to \NE with the restriction to
stationary strategies. It is a long-standing open problem whether \SqrtSum
falls into the polynomial hierarchy; hence, showing that \NE for stationary
strategies lies inside the polynomial hierarchy would imply a breakthrough
in understanding the complexity of numerical computations.

Finally, we prove decidability for an important fragment of \NE, which we
call the strictly qualitative fragment. This fragment arises from \NE by
restricting the two thresholds to be the same binary payoff. Hence, we are
only interested in equilibria where each player either wins or loses with
probability~$1$. Formally, the task is to decide, given a $k$-player
game~$\calG$ with initial vertex~$v_0$ and a binary payoff
$\vec{x}\in\{0,1\}^k$, whether the game has a Nash equilibrium with expected
payoff~$\vec{x}$. Apart from proving decidability, we show that,
depending on the representation of the objective, this problem is
typically complete for one of the complexity classes \PTime, \NP, \PNPlog and
\PSpace, and that the problem is invariant under restricting the
search space to equilibria in pure finite-state strategies.

\subsection*{Outline}
In \cref{sect:games}, we introduce the model that underlies this work and
survey earlier work on stochastic two-player zero-sum games. In
\cref{sect:equilibria}, we prove that every stochastic multiplayer game has a
Nash equilibrium, thereby addressing an inaccuracy in an earlier proof by
\citet{ChatterjeeJM04}. In \cref{sect:complexity}, we analyse the complexity
of the problem \NE \wrt the six modes of strategies we consider in this work:
positional strategies, stationary strategies, pure finite-state strategies,
randomised finite-state strategies, arbitrary pure strategies, and arbitrary
randomised strategies. Finally, in \cref{sect:qualitative}, we prove that the
strictly qualitative fragment of \NE is decidable and analyse its complexity.

\subsection*{Related Work}
Determining the complexity of Nash equilibria has attracted much
interest in recent years. In particular, a series of papers culminated in the
result that computing a Nash equilibrium of a two-player game in strategic form
is complete for the complexity class \PPAD \citep{DaskalakisGP09,ChenDT09}.
More in the spirit of our work, \citet{ConitzerS03} showed that
deciding whether there exists a Nash equilibrium in a two-player game in
strategic form where player~0 receives payoff at least~$x$ and related
decision problems are all \NP-hard. For non-stochastic infinite games,
a qualitative version of the problem \NE was studied in
\citep{Ummels08}. In particular, it was shown that the problem is \NP-complete
for games with parity winning conditions but in \PTime for games with
B\"uchi winning conditions.

For stochastic games, most results concern the computation of values and
optimal strategies; see \cref{sect:games} for a survey of the most important
results. In the multiplayer case, \citet{ChatterjeeJM04} showed that the problem of
deciding whether a (concurrent) stochastic game with reachability objectives
has a Nash equilibrium in positional strategies with payoff at least~$\vec{x}$
is \NP-complete. We sharpen their hardness result by demonstrating
that the problem remains \NP-hard when it is restricted to games with only
three players (as opposed to an unbounded number of players) where
payoffs are assigned at terminal vertices only (\cf \cref{thm:np-hardness}).

A more restricted model of stochastic games, where questions like ours
have been studied, are \emph{Markov decision processes} (MDPs) with multiple
objectives. These games can be considered
as stochastic games where only one player can influence the outcome of the game.
For MDPs with multiple \omg-regular objectives, \citet{EtessamiKVY08}
showed that questions like the one we ask are decidable. Their~result relies
on the fact that, for MDPs with multiple reachability objectives on terminal
states,
stationary strategies suffice to achieve a payoff that is higher than a given
threshold. Unfortunately, this property does not extend to our model: we give
an example of a stochastic game with the same kind of objectives where every
Nash equilibrium with payoff~$1$ for the fist player requires
infinite memory (see~\cref{prop:inf-mem}).

\section{Stochastic games}
\label{sect:games}

\subsection{Basic definitions}

Let us start by giving a formal definition of the game model that underlies
this paper. The games we are interested in are played by multiple players
taken from a finite set~$\Pi$ of players; we usually refer to them as \pl0,
\pl1, \pl2, and so on.

The \emph{arena} of the game is basically a directed, coloured graph. Intuitively,
the players take turns to form an infinite path through the arena, a \emph{play}.
Additionally, there is an element of chance involved: at some vertices, it is
not a player who decides how to proceed but \emph{nature} who chooses a
successor vertex according to a probability distribution. To model this
scenario, we partition the set~$V$ of vertices into sets~$V_i$ of vertices
\emph{controlled by \pl{i\in\Pi}} and a set of \emph{stochastic vertices}, and
we extend the edge relation to a transition relation that takes probabilities
into account. Formally, an arena for a game with players in~$\Pi$ consists of:
\begin{iteMize}{$-$}
 \item a countable, non-empty set~$V$ of \emph{vertices} or \emph{states},
 \item for each \pli a set $V_i\subseteq V$ of vertices \emph{controlled} by
\pli,
 \item a \emph{transition relation} $\Delta\subseteq V\times
([0,1]\cup\{\bot\})\times V$, and
 \item a \emph{colouring function} $\chi\colon V\to C$ into an
arbitrary set~$C$ of colours.
\end{iteMize}
We make the assumption that every vertex is controlled by at most one player:
$V_i\cap V_j=\emptyset$ if $i\neq j$; vertices that are not controlled by a
player are stochastic. For technical reasons, we also assume that for
each vertex~$v$ the set
\[v\Delta\coloneq\{w\in V:
\text{there exists $p\in(0,1]\cup\{\bot\}$ such that $(v,p,w)\in\Delta$}\}\]
of possible successor vertices is finite and non-empty. Moreover, we require
that probabilities appear only on
transitions originating in stochastic vertices (if $v\in\bigcup_{i\in\Pi}
V_i$ and $(v,p,w)\in\Delta$ then $p=\bot$) and that they are unique:
for every pair of a stochastic vertex~$v$ and an arbitrary vertex~$w$ there
exists precisely one $p\in[0,1]$ such that $(v,p,w)\in\Delta$; we denote
this probability by $\Delta(w\mid v)$. For computational purposes, we assume
that these probabilities are rational numbers. Finally, for each stochastic
vertex~$v$ the probabilities on outgoing transitions must sum up to~$1$:
$\sum_{w\in V}\Delta(w\mid v)=1$. Hence, if $v$~is a stochastic vertex, then
the mapping $V\to [0,1]\colon w\mapsto\Delta(w\mid v)$ is a discrete
probability distribution over~$V$; we~denote the set of all discrete
probability distributions over~$V$ by~$\calD(V)$.

The description of a game is completed by specifying an \emph{objective}
for each player. On~an abstract level, these are just arbitrary sets of
infinite sequences of colours, \ie subsets of $C^\omega$. Since we want to
assign a probability to them, we assume that
objectives are Borel sets over the usual topology on infinite sequences, if not
stated otherwise. Since objectives specify which plays are winning for a
player, they are also called \emph{winning conditions}.

In general, we will identify an objective $\Win\subseteq C^\omega$ over
colours with the corresponding objective $\chi^{-1}(\Win)\coloneq\{\pi\in
V^\omega:\chi(\pi)\in\Win\}\subseteq V^\omega$ over vertices (which is also
Borel since~$\chi$, as a mapping $V^\omega\to C^\omega$, is continuous).
The reason that we allow
objectives to refer to a colouring of the vertices is that the number of
colours can be much smaller than the number of vertices, and it is possible
that an objective can be represented more succinctly as an objective over
colours rather than as an objective over vertices.

If $\Pi$~is a finite set of players, $(V,(V_i)_{i\in\Pi},\Delta,\chi)$ is an
arena and $(\Win_i)_{i\in\Pi}$ is a collection of objectives, we refer to the
tuple $\calG=(\Pi,V,(V_i)_{i\in\Pi},\Delta,\chi,(\Win_i)_{i\in\Pi})$
as a \emph{stochastic multiplayer game (SMG)}. An SMG is \emph{finite}
if the set~$V$ of vertices is finite.

A \emph{play} of~$\calG$ is an infinite path through the arena of~$\calG$,
\ie a sequence $\pi=\pi(0)\pi(1)\ldots$ of vertices such that
for each $k\in\bbN$ there exists $p\in (0,1]\cup\{\bot\}$
with ${(\pi(k),p,\pi(k+1))}\in\Delta$. Finite prefixes of plays are called
\emph{histories}.
We say that a play~$\pi$ of~$\calG$
is \emph{won} by \pli if the corresponding sequence of colours fulfils \pli's
objective, \ie $\chi(\pi)\in\Win_i$; the \emph{payoff} of a play~$\pi$ is the
vector $\vec{x}\in\{0,1\}^\Pi$ defined by $x_i=1$ \iff $\chi(\pi)\in\Win_i$.

Often, it is convenient
to designate an initial vertex $v_0\in V$; we denote the pair
$(\calG,v_0)$ an \emph{initialised SMG}. A play or a history of an
initialised SMG $(\calG,v_0)$ is just a play respectively a history
of~$\calG$ that starts in~$v_0$. In the following, we will refer
to both SMGs and initialised SMGs as SMGs; it should always be clear from the
context whether the game is initialised or not.

SMGs generalise various stochastic models, each of them the subject of
intensive research.
First, there are \emph{Markov chains}, the basic model for stochastic
processes, in which no control is possible. These are just SMGs where the
set~$\Pi$ of players is empty and (consequently) there are only stochastic
vertices.
If we extend Markov chains by a single controller, we arrive at the model
of a \emph{Markov decision process (MDP)}, a model introduced by
\citet{Bellman57} and heavily used in operations research. Formally, an MDP is
an SMG where there is only one player (and only one objective).
Finally, in a \emph{(perfect-information) stochastic two-player zero-sum game
(\TwoSG)}, there are only two players, \pl0 and \pl1, who have opposing
objectives: one player wants to fulfil an objective, while the other one wants
to prevent her from doing so. Hence, one player's objective is the complement
of the other player's objective. Due to their competitive nature, these games
are also known as \emph{competitive Markov decision processes}
\cite{FilarV97}.

The SMG model also incorporates several non-stochastic models. In particular,
we call an SMG \emph{deterministic} if it contains no stochastic vertices.
In the two-player zero-sum setting, the resulting model has found applications in
logic and controller synthesis, to name a few.

\subsection{Objectives}
\label{sect:objectives}

We have introduced objectives as abstract sets of infinite sequences.
In order to be amenable for algorithmic solutions, we need to restrict to
a class of objectives representable by finite objects. The objectives
we consider for this purpose are standard in logic and verification
\cites{GraedelTW02}; for all of them, we require that the set~$C$ of
colours the objective refers to is \emph{finite}.
Moreover, whether an infinite sequence~$\alpha$ fulfils such an objective
only depends on the set $\Occ(\alpha)$ of colours occurring in~$\alpha$ or on
the set $\Inf(\alpha)$ of colours occurring \emph{infinitely often} in~$\alpha$.
In particular, we deal with the following types of objectives:
\begin{iteMize}{$-$}
\item A \emph{reachability objective} is given by a set $F\subseteq C$ of
\emph{good} colours, and the objective requires that a good colour is seen
at least once. The corresponding subset of~$C^\omega$ is
$\Reach(F)\coloneq\{\alpha\in C^\omega:\Occ(\alpha)\cap F\neq\emptyset\}$.

\item A \emph{B\"uchi objective} is again given by a set $F\subseteq C$ of
good colours, but it requires that a good colour is seen infinitely often.
The corresponding subset of~$C^\omega$ is $\Buchi(F)\coloneq\{\alpha\in
C^\omega:\Inf(\alpha)\cap F\neq\emptyset\}$.

\item A \emph{co-B\"uchi objective} is also given by a set $F\subseteq C$ of
good colours; this time, the objective requires that
\emph{from some point onwards} only good colours are seen.
The corresponding subset of~$C^\omega$ is
$\CoBuchi(F)=\{\alpha\in C^\omega:\Inf(\alpha)\subseteq F\}$.

\item A \emph{parity objective} is given by a \emph{priority function}
$\Omega\colon C\to\{0,\dots,d\}$, where ${d\in\bbN}$, which assigns to each
colour a certain \emph{priority}. The objective requires that the 
least priority that occurs infinitely often is even. The corresponding
subset of~$C^\omega$ is $\Parity(\Omega)=
\{\alpha\in C^\omega:\text{$\min(\Inf(\Omega(\alpha)))$ is even}\}$.

\item A \emph{Streett objective} is given by a set~$\Omega$ of
\emph{Streett pairs} $(F,G)$, where ${F,G\subseteq C}$.
The objective requires that, for each of the pairs, if a
colour on the left-hand side is seen infinitely often, then so is
a colour on the right-hand side. The corresponding
subset of~$C^\omega$ is
$\Streett(\Omega)={\{\alpha\in C^\omega:\text{$\Inf(\alpha)\cap F=\emptyset$
or $\Inf(\alpha)\cap G\neq\emptyset$ for all $(F,G)\in\Omega$}\}}$.

\item A \emph{Rabin objective} is given by a set~$\Omega$ of
\emph{Rabin pairs} $(F,G)$, where ${F,G\subseteq C}$; it requires
that for some pair a
colour on the left-hand side is seen infinitely often while all colours on the
right-hand side are seen only finitely often. The corresponding
subset of~$C^\omega$ is
$\Rabin(\Omega)={\{\alpha\in C^\omega:\text{$\Inf(\alpha)\cap F\neq\emptyset$
and $\Inf(\alpha)\cap G=\emptyset$ for some $(F,G)\in\Omega$}\}}$.

\item A \emph{Muller objective} is given by a family~$\calF$ of
\emph{accepting sets} $F\subseteq C$, and it requires that the set of colours
seen infinitely often equals one of these accepting sets. The corresponding
subset of~$C^\omega$ is
$\Muller(\calF)=\{\alpha\in C^\omega:\Inf(\alpha)\in\calF\}$.
\end{iteMize}
Parity, Streett, Rabin and Muller objectives are of particular relevance
because they provide a standard form for arbitrary \omg-regular objectives:
any game with arbitrary \omg-regular objectives can be reduced to one
with parity, Streett, Rabin or Muller objectives (over a larger arena) by
taking the product of its original arena with a suitable deterministic word
automaton for each player's objective \cite{Thomas90}.

In this work, for reasons that will become clear later, we are
particularly attracted to objectives that are invariant
under adding and removing finite prefixes; we call such objectives
\emph{prefix-independent}. More formally, an objective is prefix-independent if
for each $\alpha\in C^\omega$ and $x\in C^*$ the sequence~$\alpha$ satisfies
the objective \iff the sequence~$x\cdot\alpha$ does. From the objectives
listed above, only reachability objectives are, in general,
not prefix-independent. However, many of our results (in particular, many of
our lower bounds) apply to games with a prefix-independent form of
reachability, which we call \emph{terminal reachability}. For these
objectives, we assume that each vertex is coloured by itself, \ie $C=V$, and
$\chi$~is the identity mapping. The terminal reachability objective for a
set~$F\subseteq V$ coincides with the reachability objective for~$F$, but we
require that each $v\in F$ is a \emph{terminal vertex}: $v\Delta=\{v\}$.
For any such set~$F$, we have $\Occ(\pi)\cap F\neq\emptyset$ \iff
$\Inf(\pi)\cap F\neq\emptyset$
for every play~$\pi$. Hence, terminal reachability objectives can be regarded
as prefix-independent objectives.

For \TwoSGs, the distinction between reachability and terminal reachability is
not important: every \TwoSG with a reachability objective can easily be
transformed into an equivalent \TwoSG with a reachability objective on
terminal states.
For SMGs, we believe that any such transformation requires exponential time:
deciding whether in a deterministic game with terminal reachability objectives
there exists a play that fulfils each of the objectives can be done in
polynomial time, whereas the same problem is \NP-complete for deterministic
games with standard reachability objectives \citep{ChatterjeeJM04,Ummels05}.

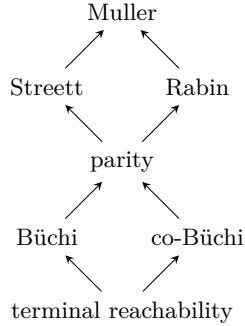
\begin{figure}
\begin{tikzpicture}[x=1.0cm,y=1cm,->]
\node (p0) at (0,0) {terminal reachability};
\node (p1a) at (-1,1) {B\"uchi};
\node (p1b) at (1,1) {co-B\"uchi};
\node (p2) at (0,2) {parity};
\node (p3a) at (-1,3) {Streett};
\node (p3b) at (1,3) {Rabin};
\node (p4) at (0,4) {Muller};

\draw (p0) to (p1a);
\draw (p0) to (p1b);
\draw (p1a) to (p2);
\draw (p1b) to (p2);
\draw (p2) to (p3a);
\draw (p2) to (p3b);
\draw (p3a) to (p4);
\draw (p3b) to (p4);
\end{tikzpicture}
\caption{\label{fig:objectives}A hierarchy of prefix-independent objectives}
\end{figure}

The resulting hierarchy of objectives is depicted in \cref{fig:objectives}.
As explained above, a simple reachability objective can be viewed as
a (co-)\dbr B\"uchi objective.
Any (co-)\dbr B\"uchi objective is equivalent to a parity objective
with only two priorities, and any parity objective is equivalent to both
a Streett and a Rabin objective; in fact, the intersection (union) of two
parity objectives is equivalent to a Streett (Rabin) objective.
Moreover, any Streett or Rabin objective is equivalent to a Muller
objective, although the translation from a set of Streett/Rabin pairs to
an equivalent family of accepting sets is, in~general, exponential.
Finally, the complement of a B\"uchi (Streett) objective is equivalent
to a co-B\"uchi (Rabin) objective, and vice versa, whereas the complement
of a parity (Muller) objective is also a parity (Muller) objective.
In~fact, any objective that is equivalent to both a Streett and a Rabin
objective is equivalent to a parity objective \citep{Zielonka98}.

To denote the class of SMGs (\TwoSGs) with a certain type of objectives,
we prefix the name SMG (\TwoSG) with the name(s) of the  objective; for
instance, we use the term \emph{Streett-Rabin SMG} to denote SMGs where each
player has a Streett or a Rabin objective. For \TwoSGs, we adopt the convention
to name the objective of \pl0 first; hence, in a Streett-Rabin
\TwoSG \pl0 has a Streett objective, while \pl1 has a Rabin objective.
Inspired by \citet{Condon92}, we will refer to SMGs with terminal reachability
objectives and \TwoSGs with a (terminal) reachability objective for \pl0 as
\emph{simple stochastic multiplayer games (SSMGs)} and \emph{simple stochastic
two-player zero-sum games (\TwoSSGs)}, respectively.

\subsubsection*{Drawing an SMG}
When drawing an SMG as a graph, we will use the
following conventions: The initial vertex is marked by a dangling incoming
edge. Vertices that are controlled by a player are depicted as
circles, where the player who controls the vertex is given by the label next
to it. Stochastic vertices are depicted as diamonds, where the
transition probabilities are given by the labels on its outgoing edges (the
default being equal probabilities on all outgoing transitions). Finally,
terminal
vertices are generally represented by their associated payoff vector. In fact,
we allow arbitrary vectors of rational probabilities as payoffs. This does not
increase the power of the model since such a payoff vector can easily be
realised by an SSMG consisting of stochastic and terminal vertices only.

\subsection{Strategies and strategy profiles}

\subsubsection{Randomised and pure strategies}

The notion of a strategy lies at the heart of game theory. Formally, a
\emph{(randomised) strategy of \pli in an SMG~$\calG$} is a mapping
$\sigma\colon V^*V_i\to\calD(V)$ assigning to each sequence
$xv\in V^* V_i$ of vertices ending in a vertex controlled by \pli a
discrete probability distribution over~$V$ such that $\sigma(xv)(w)>0$
only if $(v,\bot,w)\in\Delta$. Instead of $\sigma(xv)(w)$, we usually
write $\sigma(w\mid xv)$.
We say that a play~$\pi$ of~$\calG$ is \emph{compatible} with a
strategy~$\sigma$ of \pli if $\sigma(\pi(k+1)\mid \pi(0)\ldots\pi(k))>0$ for
all $k\in\bbN$ with $\pi(k)\in V_i$. Similarly, a history $v_0\dots v_k$
is compatible with~$\sigma$ if $\sigma(v_{j+1}\mid v_0\ldots v_j)>0$ for
all $0\leq k<n$.

A \emph{(randomised) strategy profile of $\calG$} is a tuple
$\vec{\sigma}=(\sigma_i)_{i\in\Pi}$ where $\sigma_i$ is a strategy of
\pli in~$\calG$. We say that a play or a history of $\calG$ is compatible
with a strategy profile~$\vec{\sigma}$ if it is compatible with
each~$\sigma_i$.
Given a strategy profile $\vec{\sigma}=(\sigma_j)_{j\in\Pi}$ and a
strategy~$\tau$ of \pli, we denote by $(\vec{\sigma}_{-i},\tau)$ the
strategy profile resulting from $\vec{\sigma}$ by replacing
$\sigma_i$ with~$\tau$.

A strategy $\sigma$ of \pli is called \emph{pure} or \emph{deterministic}
if for each $xv\in V^*V_i$ there exists $w\in v\Delta$ with
$\sigma(w\mid xv)=1$; note that a pure strategy of \pli can be identified with
a function $\sigma\colon V^*V_i \to V$. A strategy profile
$\vec{\sigma}=(\sigma_i)_{i\in\Pi}$
is called \emph{pure} (or \emph{deterministic}) if each~$\sigma_i$ is pure.

\subsubsection{The probability space induced by a strategy profile}

Given a game~$\calG$ and a strategy profile $\vec{\sigma}=(\sigma_i)_{i\in\Pi}$
of~$\calG$, the \emph{conditional probability} of $w\in V$ given
$xv\in V^*V$ is the number $\sigma_i(w\mid xv)$ if $v\in V_i$ and the
probability $\Delta(w\mid v)$ if $v$~is a stochastic vertex;
let~us denote this probability by $\vec{\sigma}(w\mid xv)$. Given an
initial vertex $v_0\in V$, the
probabilities $\vec{\sigma}(w\mid xv)$ give rise to a probability measure:
the probability of a basic cylinder set
${v_0\dots v_k\cdot V^\omega}$ equals the product
$\prod_{j=0}^{k-1}\vec{\sigma}(v_{j+1}\mid v_0\dots v_j)$;
basic cylinder sets that start in a vertex different from~$v_0$
have probability~$0$. This definition induces a probability measure on the
algebra of cylinder sets, which---by Carath\'eodory's extension
theorem---can be extended to a probability measure
on the Borel $\sigma$-algebra over~$V^\omega$;
we denote the extended measure by~$\Prob_{v_0}^{\vec{\sigma}}$.
Finally, by viewing the colouring function $\chi\colon V\to C$ as a continuous
function $V^\omega\to C^\omega$, we obtain a probability measure
on the Borel $\sigma$-algebra over $C^\omega$; we abuse notation and denote
this measure also by~$\Prob_{v_0}^{\vec{\sigma}}$.

For a strategy profile~$\vec{\sigma}$, we are mainly interested in the
probabilities $p_i\coloneq\Prob^{\vec{\sigma}}_{v_0}(\Win_i)$ of winning. We
call~$p_i$ the \emph{(expected) payoff of $\vec{\sigma}$ for \pli (from~$v_0$)}
and the vector $(p_i)_{i\in\Pi}$ the \emph{(expected) payoff of $\vec{\sigma}$
(from~$v_0$)}.
Finally, we say that a history~$xv$ of $(\calG,v_0)$ is
\emph{consistent} with~$\vec{\sigma}$ if
$\Prob^{\vec{\sigma}}_{v_0}(xv\cdot V^\omega)>0$, \ie if the basic cylinder
induced by this history has positive probability.

In order to apply known results about Markov chains, we can also view the
stochastic process induced by a strategy profile~$\vec{\sigma}$ as a
countable Markov chain~$\calG^{\vec{\sigma}}$, defined as follows:
The set of states of $\calG^{\vec{\sigma}}$ is the set~$V^+$ of
all non-empty sequences of vertices in~$\calG$.
The only transitions from a state~$xv$, where $x\in V^*$,
$v\in V$, are to states of the form $xvw$, where $w\in V$, and such a
transition occurs with probability~$p>0$ \iff either $v$ is stochastic
and $(v,p,w)\in\Delta$ or $v\in V_i$ and $\sigma_i(w\mid xv)=p$. Finally,
the colouring~$\chi$
of vertices is extended to a colouring of states by setting~$\chi(xv)=\chi(v)$
for all $x\in V^*$ and $v\in V$. With this definition, we could equivalently
define the
payoff of~$\vec{\sigma}$ for \pli as the probability of the event
$\chi^{-1}(\Win_i)$ in $(\calG^{\vec{\sigma}},v_0)$.

For each \pli, the Markov decision process~$\calG^{\vec{\sigma}_{-i}}$ is
defined just as~$\calG^{\vec{\sigma}}$, but states $xv\in V^* V_i$ are
controlled by \pli (the unique player in $\calG^{\vec{\sigma}_{-i}}$), and
there is a transition from such a state to each state of the form~$xvw$, where
$w\in V$, with $(v,\bot,w)\in\Delta$; \pli's objective is the same as
in~$\calG$.

\subsubsection{Strategies with memory}

A \emph{memory structure} for a game~$\calG$ with vertices in~$V$ is a
triple $\frakM=(M,\delta,m_0)$ where $M$ is a set of \emph{memory states},
$\delta\colon M\times V\to M$ is the \emph{update function}, and $m_0\in M$
is the \emph{initial memory}. A \emph{(randomised) strategy with
memory~$\frakM$} of \pli is a function $\sigma\colon M\times V_i\to\calD(V)$
such that $\sigma(m,v)(w)>0$ only if $w\in vE$. The strategy $\sigma$ is
a \emph{pure strategy with memory~$\frakM$} if additionally the following
property holds: for all $m\in M$ and $v\in V$ there exists $w\in V$ such that
$\sigma(m,v)(w)=1$. Hence, a pure strategy with memory~$\frakM$ can be
described by a function $\sigma\colon M\times V_i\to V$.
Finally, a \emph{(pure) strategy profile with
memory~$\frakM$} is a tuple $\vec{\sigma}=(\sigma_i)_{i\in\Pi}$ such that each
$\sigma_i$ is a (pure) strategy with memory~$\frakM$ of \pli.

A (pure) strategy $\sigma$ with memory~$\frakM$ of \pli
defines a (pure) strategy of \pli in the usual sense as follows: Let
$\delta^*(x)$ be the memory state after $x\in V^*$, defined inductively
by $\delta^*(\epsilon)=m_0$ and $\delta^*(xv)=\delta(\delta^*(x),v)$ for
$x\in V^*$ and $v\in V$. If $v\in V_i$, then the distribution (successor
vertex) chosen by the strategy~$\sigma$ for the sequence~$xv$
is~$\sigma(\delta^*(x),v)$. Vice versa, every strategy (profile) of~$\calG$
can be viewed as a strategy (profile) with
memory~$\frakM\coloneq (V^*,\cdot,\epsilon)$.

A \emph{finite-state strategy (profile)} is a strategy (profile) with
memory~$\frakM$ for a finite memory structure~$\frakM$. Note that a strategy
profile is finite-state \iff each of its strategies is finite-state.
If $|M|=1$, we call a strategy (profile) with memory~$\frakM$
\emph{stationary}. Moreover, we call a pure stationary strategy (profile)
a \emph{positional} strategy (profile). A stationary strategy of \pli can be
described by a function $\sigma\colon V_i\to\calD(V)$, and a positional
strategy even by a function $\sigma\colon V_i\to V$.

If $\vec{\sigma}=(\sigma_i)_{i\in\Pi}$ is a strategy profile with
memory~$\frakM$, we coarsen the Markov chain~$\calG^{\vec{\sigma}}$ by
taking $M\times V$ as its domain. The transition relation is
defined as follows:
there is a transition from $(m,v)$ to $(n,w)$ with probability~$p>0$
\iff $\delta(m,v)=n$ and either $v$~is a stochastic vertex of~$\calG$ and
$(v,p,w)\in\Delta$ or $v\in V_i$ and $\sigma_i(m,v)(w)=p$. Finally,
a~state~$(m,v)$ has the same colour as the vertex~$v$ in~$\calG$.
Analogously, we coarsen the Markov decision
process~$\calG^{\vec{\sigma}_{-i}}$ by using $M\times V$ as its domain:
vertices~$(m,v)\in M\times V_i$ are controlled
by \pli, and there is a transition from such a vertex~$(m,v)$ to $(n,w)\in
M\times V$ \iff $n=\delta(m,v)$ and $(v,\bot,w)\in\Delta$.
Note that the arenas of both $\calG^{\vec{\sigma}}$ and
$\calG^{\vec{\sigma}_{-i}}$ are finite if the memory~$\frakM$ and the
original arena of~$\calG$ are finite.

\subsubsection{Residual games and strategies}

Given an SMG~$\calG$ and a sequence $x\in V^*$ (which is usually a
history), the \emph{residual game} $\calG[x]$ has the same arena as~$\calG$ but
different objectives: if $\Win_i\subseteq C^\omega$ is the objective of \pli
in~$\calG$, then her objective in $\calG[x]$ is $\chi(x)^{-1}\Win_i\coloneq
\{\alpha\in C^\omega:\chi(x)\cdot\alpha\in\Win_i\}$. In particular, if all
objectives in~$\calG$ are prefix-independent, then $\calG[x]=\calG$.

If \pli plays according to a strategy~$\sigma$ in~$\calG$, then the
corresponding strategy in~$\calG[x]$ is the \emph{residual
strategy}~$\sigma[x]$, defined by $\sigma[x](yv)=\sigma(xyv)$.
If $\vec{\sigma}=(\sigma_i)_{i\in\Pi}$ is a strategy profile, then the
\emph{residual strategy profile}~$\vec{\sigma}[x]$ is just the profile
of the residual strategies~$\sigma_i[x]$. The following lemma, taken
from \citep{Zielonka04}, shows how to compute probabilities \wrt a
residual strategy profile.

\begin{lemma}\label{lemma:residual-strategies}
Let $\vec{\sigma}$ be a strategy profile of an SMG $(\calG,v_0)$, and let
$xv\in V^*V$. If~$X\subseteq V^\omega$ is a Borel set, then
$\Prob^{\vec{\sigma}}_{v_0}(X\cap xv\cdot V^\omega)=\Prob^{\vec{\sigma}}_{v_0}
(xv\cdot V^\omega)\cdot \Prob^{\vec{\sigma}[x]}_{v}(x^{-1}X)$.
\end{lemma}

\subsection{Subarenas and end components}

Algorithms for stochastic games often employ a \emph{divide-and-conquer}
approach and compute a solution for a complex game from the solution of
several smaller games. These smaller games are usually obtained from
the original game by restricting to a \emph{subarena}. Formally,
given an SMG~$\calG$, a set $U\subseteq V$ is a subarena if
\begin{iteMize}{$-$}
\item $U\neq\emptyset$,
\item $v\Delta\cap U\neq\emptyset$ for each $v\in U$, and
\item $v\Delta\subseteq U$ for each stochastic vertex~$v\in U$.
\end{iteMize}
Clearly, if $U$~is a subarena, then the restriction of~$\calG$
to vertices in~$U$ is again an SMG, which we denote by
$\calG\restrict U$. Formally,
\[\calG\restrict U\coloneq
(\Pi,U,(V_i\cap U)_{i\in\Pi},\Delta\cap(U\times([0,1]\cup\{\bot\})\times U),
\chi_U,(\Win_i)_{i\in\Pi}),\]
where $\chi_U\colon U\to C\colon u\mapsto\chi(u)$ is the restriction of
the colouring function to~$U$.

Of particular interest are the strongly connected subarenas of a game because
they can arise as the sets~$\Inf(\pi)$ of vertices visited
infinitely often in a play; we call these sets \emph{end components}.
Formally, a set $U\subseteq V$ is an end component if $U$~is a
subarena and every vertex $w\in U$ is reachable from every other
vertex~$v\in U$, \ie there exists a sequence
$v=v_1,v_2,\dots,v_n=w$ such that $v_{i+1}\in v_i\Delta$ for
each $0<i<n$. An end component~$U$ is \emph{maximal} in a set $S\subseteq V$
if there is no end component~$U'$ such that $U\subsetneq U'\subseteq S$. For
any finite subset $S\subseteq V$, the set of all end components maximal
in~$S$ can be computed in quadratic time \cite{Alfaro97}.

The theory of end components has been developed by
\citet{Alfaro97,Alfaro98} and \citet{CourcoubetisY95,CourcoubetisY98}.
The  central fact
about end components in finite SMGs is that, under any
strategy profile, the set of vertices visited infinitely often is
almost surely an end component.

\begin{lemma}
\label{lemma:end-components}
Let $\calG$ be a finite SMG. Then
${\Prob^{\vec{\sigma}}_v(\{\pi\in V^\omega:
\text{$\Inf(\pi)$ is an end component}\})=1}$
for each strategy profile~$\vec{\sigma}$ of~$\calG$
and each $v\in V$.
\end{lemma}

Moreover, for any end component~$U$, we can construct a
stationary strategy profile, or alternatively a pure finite-state
strategy profile, that, when started in~$U$, 
guarantees almost surely to visit all and only the vertices in~$U$
infinitely often.
In fact, the stationary profile that chooses for
each vertex in~$U$ a successor in~$U$ uniformly at random
fulfils this property.

\begin{lemma}
\label{lemma:end-component-strategy}
Let $\calG$ be a finite SMG and $U$ one of its end components.
There exists both a stationary and a pure finite-state strategy
profile~$\vec{\sigma}$ such that
$\Prob^{\vec{\sigma}}_v({\{\pi\in V^\omega:\Inf(\alpha)=U\}})=1$ for
every vertex $v\in U$.
\end{lemma}

Given an SMG~$\calG$ with (objectives representable as) Muller objectives
given by a family~$\calF_i$ of accepting sets, we say that an end component~$U$
is \emph{winning} for \pli if $\chi(U)\in\calF_i$; the \emph{payoff} of~$U$ is
the vector $\vec{z}\in\{0,1\}^\Pi$, defined by $z_i=1$ \iff $U$~is winning
for \pli.

\subsection{Values, determinacy and optimal strategies}


Given a strategy~$\tau$ of \pli in~$\calG$ and a vertex $v\in V$, the
\emph{value} of~$\tau$ from~$v$ is the number $\val^\tau(v)\coloneq
\inf_{\vec{\sigma}}\Prob^{\vec{\sigma}_{-i},\tau}_{v}(\Win_i)$, where
$\vec{\sigma}$ ranges over all strategy profiles of~$\calG$. Moreover,
the \emph{value} of~$\calG$ for \pli from~$v$ is the supremum
of these values: $\val_i^\calG(v)\coloneq\sup_\tau \val^\tau(v)$, where
$\tau$~ranges over all strategies of \pli in~$\calG$.
Intuitively, $\val_i^\calG(v)$ is the maximal payoff that \pli can
\emph{ensure} when the game starts from~$v$.

Given an initial vertex $v_0\in V$, a strategy~$\tau$ of \pli in $\calG$
is called \emph{(almost-surely) winning} if $\val^\tau(v_0)=1$. More generally,
$\tau$~is called \emph{optimal} if $\val^\tau(v_0)=\val_i^\calG(v_0)$. For
$\epsilon>0$, it is called \emph{\eps-optimal} if
$\val^\tau(v_0)\geq\val_i^\calG(v_0)-\epsilon$.
A \emph{globally (\eps-)\dbr optimal} strategy is a strategy that
is (\eps-)\dbr optimal for every possible initial vertex $v_0\in
V$. Note that optimal strategies need not exist since the supremum in the
definition of $\val_i^\calG$ is not necessarily attained; in~this case,
only \eps-optimal strategies do exist. Also note that there
exists a globally (\eps-)\dbr optimal strategy whenever there exists
an (\eps-)\dbr optimal strategy for every possible initial vertex.
Finally, we say
that a strategy~$\tau$ of \pli in $(\calG,v_0)$ is \emph{strongly optimal}
if the residual strategy~$\tau[x]$ is optimal in the residual game
$(\calG[x],v)$ for every history~$xv$ of~$(\calG,v_0)$ that is compatible with~$\tau$. Intuitively, a strategy is strongly optimal if it is
also optimal when the other players do not play optimally.
Note that, for games with prefix-independent objectives, any globally
optimal positional strategy profile is also strongly optimal.

Determining values and finding optimal strategies in SMGs actually reduces
to performing the same tasks in \TwoSGs. Formally,
given an SMG~$\calG$, define for each \pli the \emph{coalition game}~$\calG_i$
to be the same game as~$\calG$ but with only two players: \pli acting as \pl0
and the coalition $\Pi\setminus\{i\}$ acting as \pl1. The coalition
controls all vertices that in~$\calG$ are controlled by some \pl{j\neq i},
and its objective is the complement of \pli's objective in~$\calG$.
Clearly, $\calG_i$ is an \TwoSG,
and $\val^{\calG_i}(v)=\val_i^\calG(v)$ for every vertex~$v$. Moreover,
any (strongly, \eps-) optimal strategy for \pli
in~$(\calG,v_0)$ is (strongly, \eps-) optimal
in~$(\calG_i,v_0)$, and vice versa. Hence, when we study values and optimal
strategies, we can restrict to~\TwoSGs.

A celebrated theorem due to \citet{Martin98} and \citet{MaitraS98} states that
\TwoSGs with Borel objectives are \emph{determined}: $\val_0^\calG=
1-\val_1^\calG$. The number $\val^\calG(v)\coloneq\val_0^\calG(v)$ is
consequently called the \emph{value} of~$\calG$ from~$v$. In fact,
an~inspection of the proof shows that for turn-based games both players
not only have randomised \eps-optimal strategies but pure
\eps-optimal strategies.

\begin{theorem}[\citep{Martin98,MaitraS98}]
\label{thm:determinacy}
Every \TwoSG with Borel objectives is determined;
for all $\epsilon>0$, both players have \eps-optimal pure strategies.
\end{theorem}

For finite \TwoSGs with prefix-independent objectives, we can
show a stronger result than \cref{thm:determinacy}: in these games, both
players not only have \eps-optimal pure strategies but optimal ones
\citep{GimbertH10}.
In~fact, the proof reveals the existence of strongly optimal strategies
(see also~\cite{Ummels10}).

\begin{theorem}[\citep{GimbertH10}]
\label{thm:optimal-strategies}
In any finite \TwoSG with prefix-in\-de\-pen\-dent objectives,
both players have strongly optimal pure strategies.
\end{theorem}

For finite \TwoSGs with \omg-regular objectives, more attractive
strategies than arbitrary pure strategies suffice for optimality.
In particular, in any finite Rabin-Streett
\TwoSG there exists a globally optimal positional strategy
for \pl0 \citep{Klarlund94,ChatterjeeAH05}.

\begin{theorem}[\citep{Klarlund94,ChatterjeeAH05}]
\label{thm:positional-optimal}
In any finite Rabin-Streett \TwoSG, \pl0 has a globally optimal
positional strategy.
\end{theorem}

A consequence of \cref{thm:positional-optimal} is that the values of
a finite Rabin-Streett \TwoSG are rational of polynomial bit complexity
in the size of the arena: Given a positional strategy
profile~$\vec{\sigma}$ of~$\calG$, the finite MDP~$\calG^{\vec{\sigma}_{-1}}$
is not larger than the game~$\calG$. Moreover, if $\sigma_0$~is globally
optimal, then for every vertex~$v$ the value of~$\calG$ from~$v$ and the value
of~$\calG^{\vec{\sigma}_{-1}}$ from~$v$ sum up to~$1$. But the values of any
Streett MDP form the optimal solution of a linear programme of polynomial size
\cites{Alfaro97} and are therefore rational of small bit complexity.

Of course, it also follows from \cref{thm:positional-optimal} that
finite parity \TwoSGs are \emph{positionally determined}: both
players have
globally optimal positional strategies. This result was first proven for
deterministic games (even over infinite arenas), independently by
\citet{EmersonJ91} and \citet{Mostowski91a}. For \TwoSSGs, the existence of
optimal positional strategies follows from a result of \citet{BewleyK78}.
Independently, \citet{McIverM02}, \citet{ChatterjeeJH04} and \citet{Zielonka04}
extended these results to parity \TwoSGs.

\begin{corollary}\label{cor:positional-optimal}
In any finite parity \TwoSG, both players have globally optimal
positional strategies.
\end{corollary}

Since every finite \TwoSG with \omg-regular objectives can be
reduced to one with parity objectives, we can conclude
from \cref{cor:positional-optimal} that both players have residually
optimal pure finite-state strategies in finite \TwoSGs with arbitrary
\omg-regular objectives.

\begin{corollary}\label{cor:regular-optimal}
In any finite \TwoSG with \omg-regular objectives, both
players have strongly
optimal pure finite-state strategies.
\end{corollary}

\subsection{Algorithmic problems}
\label{sect:algorithmics}

For the rest of this section, we only consider finite two-player zero-sum
games.
The main computational problems for these games are computing
the value and optimal strategies for one or both players.
Instead of computing the value exactly, we can ask whether the value is
greater than some given rational probability~$p$, a~problem which we call the
\emph{quantitative decision problem}:
\begin{quote}
Given an \TwoSG~$\calG$, a vertex~$v$ and a rational
number $p\in [0,1]$, decide whether $\val^\calG(v)\geq p$.
\end{quote}
In many cases, it suffices to know whether the value is~$1$, \ie whether \pl0
has a strategy to win the game almost surely (asymptotically, at least).
We call the resulting decision problem the \emph{qualitative decision problem}.

Clearly, if we can solve the quantitative decision problem, we can approximate
the values~$\val^\calG(v)$ up to any desired precision by using binary search.
In~fact, for parity \TwoSGs it is well-known that it suffices to solve the
decision problems, since the other problems (computing the values and
optimal strategies) are polynomial-time equivalent to the quantitative decision
problem.

For a Markov decision process whose objective can be represented as a
Muller objective, we can compute the values by an analysis of its
end components: For a given initial vertex~$v$,
the value of the MDP from~$v$ equals the maximal probability of reaching a
winning end component from~$v$; this probability can be computed using linear
programming.

Even though, the number of end components can be exponential, it is
easy to see that the \emph{union}  of all winning end components can be
computed in polynomial time for MDPs with Rabin or Muller
objectives (given by a family of accepting sets).
For MDPs with Streett objectives, \citet{ChatterjeeAH05} gave a polynomial-time
algorithm for computing this set. Hence, for MDPs with any of these
objectives, the quantitative decision problem is solvable in polynomial time.

\begin{theorem}[\citep{Alfaro97,ChatterjeeAH05}]
\label{thm:mdp-ptime}
The quantitative decision problem is in \PTime for Streett, Rabin or
Muller~MDPs.
\end{theorem}

It follows from \cref{thm:positional-optimal,thm:mdp-ptime} that
the quantitative decision problem for Rabin-Streett \TwoSGs is in \NP:
to decide whether $\val^\calG(v)\geq p$, it~suffices to guess a positional
strategy for \pl0 and to check whether in the resulting Streett MDP the
value from~$v$ is not smaller than~$p$.
By determinacy, this result implies that the
quantitative decision problem is in \coNP for Streett-Rabin \TwoSGs and in
$\NP\cap\coNP$ for parity \TwoSGs.

\begin{corollary}
\label{cor:rabin-np}
The quantitative decision problem is
\begin{iteMize}{$-$}
\item in \NP for Rabin-Streett \TwoSGs,
\item in \coNP for Streett-Rabin \TwoSGs, and
\item in $\NP\cap\coNP$ for parity \TwoSGs.
\end{iteMize}
\end{corollary}

A corresponding \NP-hardness result for deterministic Rabin-Streett
\TwoSGs was established by \citet{EmersonJ99}. In particular, this
hardness result also holds for the qualitative decision
problem. Moreover, by determinacy, this result can be turned into a
\coNP-hardness result for (deterministic) Streett-Rabin \TwoSGs.

For \TwoSGs with Muller objectives, \citet{Chatterjee07a} showed that the
quantitative decision problem falls into \PSpace; for deterministic
games, a~polynomial-space algorithm had been given earlier by
\citet{McNaughton93}.
A~matching lower bound for deterministic games with Muller objectives
was provided by \citet{HunterD05}.

\begin{theorem}[\citep{Chatterjee07a,HunterD05}]
\label{thm:muller-pspace}
The quantitative and the qualitative decision problem are
\PSpace-complete for Muller \TwoSGs.
\end{theorem}

\Cref{thm:muller-pspace} does not hold if the Muller objective is given
by a family of subsets of \emph{vertices}: \Citet{Horn08a,Horn08} showed that
the qualitative decision problem for \emph{explicit} Muller \TwoSGs is in
\PTime, and that the quantitative problem is in $\NP\cap\coNP$.

Another class of \TwoSGs for which the qualitative decision problem is
in \PTime is, for each $d\in\bbN$, the class $\Parity[d]$ of all parity
\TwoSGs whose priority function refers to at most $d$~priorities
\citep{AlfaroH00}.
In particular, the qualitative decision problem for \TwoSSGs as well
as (co-)\dbr B\"uchi \TwoSGs is in \PTime.
For general parity \TwoSGs, however, the qualitative decision problem is
only known to lie in $\UP\cap\coUP$ \citep{Jurdzinski98,ChatterjeeJH03}.

\begin{theorem}[\cite{Jurdzinski98,ChatterjeeJH03,AlfaroH00}]
\label{thm:parity-bounded-ptime}
The qualitative decision problem is in ${\UP\cap\coUP}$ for parity \TwoSGs.
For each ${d\in\bbN}$, the qualitative decision problem is in \PTime for
parity \TwoSGs with at most $d$~priorities.
\end{theorem}

\Cref{table:summary-results} summarises the results about the complexity
of the quantitative and the qualitative decision problem for \TwoSGs.
\begin{table}
\begin{tabular}{@{}lll@{}}
\toprule
& Qualitative & Quantitative \\
\midrule
\TwoSSGs &  \PTime-complete & $\NP\cap\coNP$ \\
$\Parity[d]$ & \PTime-complete & $\NP\cap\coNP$ \\
Parity & $\UP\cap\coUP$ & $\NP\cap\coNP$ \\
Rabin-Streett & \NP-complete & \NP-complete \\
Streett-Rabin & \coNP-complete & \coNP-complete \\
Muller &  \PSpace-complete & \PSpace-complete \\
\bottomrule
\end{tabular}
\caption{\label{table:summary-results}The complexity of deciding the value in
\TwoSGs}
\end{table}
\PTime-hardness (via \LogSpace-reductions) for all these problems
follows from the fact that \emph{and-or graph reachability} is \PTime-complete
\citep{Immerman81}.

The results summarised in \cref{table:summary-results} leave open
the possibility that at least one of the following problems is
decidable in polynomial time:
\begin{enumerate}[(1)] 
 \item the qualitative decision problem for parity \TwoSGs,
 \item the quantitative decision problem for \TwoSSGs,
 \item the quantitative decision problem for parity \TwoSGs.
\end{enumerate}
Note that, given that all of them are contained in both \NP and \coNP, it is
unlikely that one of them is \NP-hard or \coNP-hard; such a result would
imply that $\NP=\coNP$, and the polynomial hierarchy would collapse.

For the first problem, \citet{ChatterjeeJH03} gave a polynomial-time
reduction to the qualitative decision problem for \emph{deterministic}
two-player zero-sum parity games. Hence, solving the qualitative decision
problem
for parity \TwoSGs is not harder than deciding which of the two players has
a winning strategy in a deterministic two-player zero-sum parity game. Whether
the latter problem is decidable in polynomial time is a
long-standing open problem. Several years after \citet{EmersonJ91} put the
problem into $\NP\cap\coNP$, \citet{Jurdzinski98} improved
this bound slightly to $\UP\cap\coUP$. Together with Paterson and Zwick
\cite{JurdzinskiPZ08}, he also gave an algorithm that decides the
winner in subexponential time; a~\emph{randomised} subexponential algorithm
had been given earlier by \citet{BjorklundSV03}. On the other hand,
\citet{Friedmann09} recently showed that the most promising candidate
for a polynomial-time algorithm for the general case so far, the
\emph{discrete strategy improvement} algorithm due to \citet{VoegeJ00},
requires exponential time in the worst case.

Regarding the second problem, only some progress towards a polynomial-time
algorithm has been made since \citet{Condon92} proved membership
in $\NP\cap\coNP$; for instance, \citet{BjorklundV05} gave a randomised
subexponential algorithm for solving \TwoSSGs, and \citet{GimbertH09} showed
that the quantitative decision problem for \TwoSSGs is fixed-parameter
tractable \wrt the number of stochastic vertices as the parameter.

For the third problem, \citet{AnderssonM09} recently established a
polynomial-time Turing reduction to the second. Hence, there exists a
polynomial-time algorithm for (2) \iff there exists one for (3). In~particular,
a polynomial-time algorithm for (2) would also give a polynomial-time algorithm
for (1). However, to the best of our knowledge, it is plausible that the
qualitative decision problem for parity \TwoSGs is in \PTime while the
quantitative decision problem for \TwoSSGs is not.

\section{Existence of Nash equilibria}
\label{sect:equilibria}


To capture rational behaviour of selfish players, \citet{Nash50}
introduced the notion of---what is now called---a Nash equilibrium.
Formally, given a strategy profile~$\vec{\sigma}$ of a game $(\calG,v_0)$,
we~call a strategy~$\tau$ of player~$i$ in~$\calG$ a
\emph{best response} to~$\vec{\sigma}$ if $\tau$~maximises the expected
payoff of player~$i$:
$\Prob_{v_0}^{\vec{\sigma}_{-i},\tau'}(\Win_i)\leq
\Prob_{v_0}^{\vec{\sigma}_{-i},\tau}(\Win_i)$ for all strategies~$\tau'$
of player~$i$. A strategy profile $\vec{\sigma}=(\sigma_i)_{i\in\Pi}$ is a
\emph{Nash equilibrium} if each~$\sigma_i$ is a
best response to~$\vec{\sigma}$.

In a Nash equilibrium, no player can improve her payoff
by unilaterally switching to a different strategy. In fact, to
have a Nash equilibrium, it~suffices
that no player can gain from switching to a pure strategy.

\begin{proposition}\label{prop:nash-pure}
A strategy profile~$\vec{\sigma}$ of a game $(\calG,v_0)$ is a Nash equilibrium
\iff, for each \pli and for each \emph{pure} strategy~$\tau$ of \pli in~$\calG$,
$\Prob_{v_0}^{\vec{\sigma}_{-i},\tau}(\Win_i)\leq
\Prob_{v_0}^{\vec{\sigma}}(\Win_i)$.
\end{proposition}

\begin{proof}
Clearly, if $\vec{\sigma}$ is a Nash equilibrium, then
$\Prob_{v_0}^{\vec{\sigma}_{-i},\tau}(\Win_i)\leq\Prob_{v_0}^{\vec{\sigma}}
(\Win_i)$ for each pure strategy~$\tau$ of \pli in~$\calG$.
Now, assume that $\vec{\sigma}$ is not a Nash equilibrium. Hence,
$p\coloneq\sup_\tau \Prob_{v_0}^{\vec{\sigma}_{-i},\tau}(\Win_i)=
\Prob_{v_0}^{\vec{\sigma}}(\Win_i)+\epsilon$ for some \pli and some
$\epsilon>0$. Consider the Markov decision
process~$\calG^{\vec{\sigma}_{-i}}$. Clearly, the value
of~$\calG^{\vec{\sigma}_{-i}}$ from~$v_0$ equals~$p$.
By \cref{thm:determinacy},
there exists an $\epsilon/2$-optimal pure strategy~$\tau$
in $(\calG^{\vec{\sigma}_{-i}},v_0)$. Since the arena of
$\calG^{\vec{\sigma}_{-i}}$ is a forest, we can assume that
$\tau$~is a positional strategy, which can be viewed as a pure strategy
in~$\calG$. We~have
$\Prob_{v_0}^{\vec{\sigma}_{-i},\tau}(\Win_i)
\geq p-\epsilon/2> p-\epsilon
=\Prob_{v_0}^{\vec{\sigma}}(\Win_i)$.
\end{proof}

\noindent
For two-player zero-sum games, a Nash equilibrium is just a pair of
optimal strategies.

\begin{proposition}\label{prop:two-zerosum-nash}
Let $(\calG,v_0)$ be an \TwoSG. A strategy profile
$(\sigma,\tau)$ of $(\calG,v_0)$ is a Nash equilibrium \iff both $\sigma$
and $\tau$ are optimal. In particular, every Nash equilibrium of $(\calG,v_0)$
has payoff~$(\val^\calG(v_0),1-\val^\calG(v_0))$.
\end{proposition}
\begin{proof}
$(\Rightarrow)$ Assume that both $\sigma$ and $\tau$ are optimal, but
that $(\sigma,\tau)$ is not a Nash equilibrium. Hence, one of the players,
say \pl1, can improve her payoff by playing some strategy~$\tau'$. Hence,
$\val^\calG(v_0)=\Prob^{\sigma,\tau}_{v_0}(\Win_0)>
\Prob^{\sigma,\tau'}_{v_0}(\Win_0)$.
However, since $\sigma$ is optimal,
$\val^\calG(v_0)\leq\Prob^{\sigma,\tau'}_{v_0}(\Win_0)$, a contradiction.
The reasoning in the case that \pl0 can improve is analogous.

$(\Leftarrow)$ Let $(\sigma,\tau)$ be a Nash equilibrium of $(\calG,v_0)$,
and let us first assume that $\sigma$~is not optimal, \ie
$\val^\sigma(v_0)<\val^\calG(v_0)$. By the definition of $\val^\calG$,
there exists another strategy~$\sigma'$ of \pl0
such that $\val^\sigma(v_0)<\val^{\sigma'}(v_0)\leq\val^\calG(v_0)$.
We have
\[
\Prob^{\sigma,\tau}_{v_0}(\Win_0)
\leq\val^\sigma(v_0)
<\val^{\sigma'}(v_0)=
\inf\nolimits_{\tau'}\Prob^{\sigma',\tau'}_{v_0}(\Win_0)
\leq \Prob^{\sigma',\tau}_{v_0}(\Win_0),
\]
where the first inequality follows from the fact that $(\sigma,\tau)$
is a Nash equilibrium.
Thus, \pl0 can improve her payoff by playing~$\sigma'$ instead of~$\sigma$,
a contradiction to $(\sigma,\tau)$ being a Nash equilibrium.
The argumentation in the case that $\tau$~is not optimal is
analogous.
\end{proof}

In general, a Nash equilibrium can give a player a higher payoff than her
value. However, the payoff a player receives in a Nash equilibrium can never
be lower than her value, and this is true for every history that is
consistent with the equilibrium.
Formally, we say that a strategy profile~$\vec{\sigma}$ of a game $(\calG,v_0)$
is \emph{favourable} if
$\Prob^{\vec{\sigma}}_{v_0}(\Win_i\mid xv\cdot V^\omega)\geq
\val_i^{\calG[x]}(v)$
for each \pli and every history~$xv$ that is consistent with~$\vec{\sigma}$.

\begin{lemma}\label{prop:nash-value}
Let $(\calG,v_0)$ be an SMG. Every Nash equilibrium of $(\calG,v_0)$
is favourable.
\end{lemma}

\begin{proof}
Assume there exists a history~$xv$ of $(\calG,v_0)$ that is
consistent with~$\vec{\sigma}$,
but $p\coloneq\Prob^{\vec{\sigma}}_{v_0}(\Win_i\mid xv\cdot V^\omega)<
\val_i^{\calG[x]}(v)$. By the definition of $\val_i^{\calG[x]}$,
there exists a strategy~$\tau$ of \pli in~$\calG[x]$ such that
$\val^{\tau}(v)>p$. We define a new strategy~$\sigma'$ for \pli
in~$\calG$ as follows: $\sigma'$~is defined as $\sigma_i$ for histories
that do not begin with~$xv$. For histories of the form~$xvy$, however,
we set $\sigma'(xvy)=\tau(vy)$.
Clearly, $\Prob_{v_0}^{\vec{\sigma}_{-i},\sigma'}(xv\cdot V^\omega)
=\Prob_{v_0}^{\vec{\sigma}}(xv\cdot V^\omega)$.
Moreover, it is easy to see that
$\Prob^{\vec{\sigma}_{-i},\sigma'}_{v_0}(X\setminus xv\cdot V^\omega)
=\Prob^{\vec{\sigma}}_{v_0}(X\setminus xv\cdot V^\omega)$
for every Borel set $X\subseteq V^\omega$.
Using \cref{lemma:residual-strategies}, we can conclude that
\begin{align*}
& \Prob^{\vec{\sigma}_{-i},\sigma'}_{v_0}(\Win_i) \\
=\; & \Prob^{\vec{\sigma}_{-i},\sigma'}_{v_0}(\Win_i\setminus xv\cdot V^\omega)
+ \Prob^{\vec{\sigma}_{-i},\sigma'}_{v_0}(\Win_i\cap xv\cdot V^\omega) \\
=\; & \Prob^{\vec{\sigma}}_{v_0}(\Win_i\setminus xv\cdot V^\omega)
+ \Prob^{\vec{\sigma}[x]_{-i},\sigma'[x]}_v(x^{-1}\Win_i)\cdot
\Prob^{\vec{\sigma}_{-i},\sigma'}_{v_0}(xv\cdot V^\omega) \\
=\; & \Prob^{\vec{\sigma}}_{v_0}(\Win_i\setminus xv\cdot V^\omega)
+ \Prob^{\vec{\sigma}[x]_{-i},\tau}_v(x^{-1}\Win_i)\cdot
\Prob^{\vec{\sigma}}_{v_0}(xv\cdot V^\omega) \\
\geq\; & \Prob^{\vec{\sigma}}_{v_0}(\Win_i\setminus xv\cdot V^\omega)
+ \val^{\tau}(v)\cdot\Prob^{\vec{\sigma}}_{v_0}(xv\cdot V^\omega) \\
>\; & \Prob^{\vec{\sigma}}_{v_0}(\Win_i\setminus xv\cdot V^\omega)
+ p\cdot\Prob^{\vec{\sigma}}_{v_0}(xv\cdot V^\omega) \\
=\; & \Prob^{\vec{\sigma}}_{v_0}(\Win_i\setminus xv\cdot V^\omega)
+ \Prob^{\vec{\sigma}}_{v_0}(\Win_i\mid xv\cdot V^\omega)
  \cdot\Prob^{\vec{\sigma}}_{v_0}(xv\cdot V^\omega) \\
=\; & \Prob^{\vec{\sigma}}_{v_0}(\Win_i\setminus xv\cdot V^\omega)
+ \Prob^{\vec{\sigma}}_{v_0}(\Win_i\cap xv\cdot V^\omega) \\
=\; & \Prob^{\vec{\sigma}}_{v_0}(\Win_i)\,.
\end{align*}
Hence, \pli can improve her payoff by switching to~$\sigma'$,
a contradiction to $\vec{\sigma}$~being a Nash equilibrium.
\end{proof}

It follows from \cref{thm:optimal-strategies,prop:two-zerosum-nash} that
every finite two-player zero-sum stochastic game with prefix-independent
objectives has a Nash equilibrium in pure strategies.
Is this still true if the two-player zero-sum assumption is relaxed?

By \cref{prop:nash-value}, a pure strategy profile can only be a Nash
equilibrium if it is favourable. The next lemma shows that, conversely, we can
turn every favourable pure strategy profile into a Nash equilibrium.
The proof uses so-called \emph{threat} strategies (or \emph{trigger}
strategies), which are added on top of the given strategy profile: each player
threatens to change her behaviour when one of the other players deviates from
the prescribed strategy profile.
Before being applied to stochastic games, this concept proved
fruitful in the related area of \emph{repeated games}
(see \citep[Chapter~8]{OsborneR94} and \citep{Aumann81}).

\begin{lemma}\label{lemma:nash-threat}
Let $(\calG,v_0)$ be a finite SMG with prefix-in\-de\-pen\-dent objectives.
If $\vec{\sigma}$~is a favourable pure strategy profile of $(\calG,v_0)$,
then $(\calG,v_0)$~has a pure Nash equilibrium~$\vec{\sigma}^*$ with $\Prob_{v_0}^{\vec{\sigma}}=\Prob_{v_0}^{\vec{\sigma}^*}$.
\end{lemma}

\begin{proof}
By \cref{thm:optimal-strategies}, for each \pli we can fix a globally optimal
pure strategy~$\tau_i$ of the coalition $\Pi\setminus\{i\}$ in the coalition
game~$\calG_i$; denote by~$\tau_{j,i}$ the corresponding pure strategy of
\pl{j\neq i} in~$\calG$.
To simplify notation, we also define $\tau_{i,i}$ to be an arbitrary
pure strategy of \pli in~$\calG$. \Pli's equilibrium
strategy~$\sigma^*_i$ is defined as follows: For~histories~$xv$ that
are compatible with~$\vec{\sigma}$, we set $\sigma^*_i(xv)=\sigma_i(xv)$.
If $xv$~is not compatible with~$\vec{\sigma}$, then
decompose~$x$ into $x=x_1\cdot x_2$, where $x_1$~is the longest prefix
of~$x$ that is compatible with~$\vec{\sigma}$, and let~$j$ be the player
who has deviated, \ie $x_1$~ends in~$V_j$;
we set $\sigma_i^*(xv)=\tau_{i,j}(x_2v)$.
Intuitively, $\sigma_i^*$~behaves like~$\sigma_i$ as long as no other
\plj deviates from playing~$\sigma_j$, in~which case $\sigma_i^*$~starts
to behave like~$\tau_{i,j}$.

Note that $\Prob^{\vec{\sigma}^*}_{v_0}=\Prob^{\vec{\sigma}}_{v_0}$. We
claim that $\vec{\sigma}^*$~is additionally a Nash
equilibrium of $(\calG,v_0)$. Let $i\in\Pi$, and let $\rho$ be a
pure strategy of \pli in~$\calG$; by~\cref{prop:nash-pure}, it suffices
to show that
$\Prob^{\vec{\sigma}^*_{-i},\rho}_{v_0}(\Win_i)\leq
\Prob^{\vec{\sigma}^*}_{v_0}(\Win_i)$.

Let us call a history $xv\in V^* V_i$ a
\emph{deviation history} if $xv$~is compatible with both $\vec{\sigma}$
and $(\vec{\sigma}_{-i},\rho)$, but $\sigma_i(xv)\neq\rho(xv)$; we denote
the set of all deviation histories consistent with~$\vec{\sigma}$ by~$D$.
Clearly, $\Prob_{v_0}^{\vec{\sigma}}(xv\cdot V^\omega)=
\Prob_{v_0}^{\vec{\sigma}^*}(xv\cdot V^\omega)=
\Prob_{v_0}^{\vec{\sigma}^*_{-i},\rho}(xv\cdot V^\omega)$ for all
$xv\in D$.

\begin{claim}
$\Prob^{\vec{\sigma}^*_{-i},\rho}_{v_0}(X\setminus D\cdot V^\omega)=
\Prob^{\vec{\sigma}}_{v_0}(X\setminus D\cdot V^\omega)$ for every Borel
set~$X\subseteq V^\omega$.
\end{claim}

\proof
This claim can be proved by an induction over the structure of
Borel set.

\begin{claim}
$\Prob^{\vec{\sigma}^*_{-i},\rho}_{v_0}(\Win_i\mid xv\cdot V^\omega)
\leq\val_i^\calG(v)$ for every $xv\in D$.
\end{claim}

\proof
By the definition of the strategies $\tau_{j,i}$, we have that
$\Prob^{(\tau_{j,i})_{j\neq i},\rho}_{v}(\Win_i)\leq\val_i^\calG(v)$ for
\emph{every} vertex $v\in V$ and \emph{every} strategy~$\rho$ of \pli.
Moreover, if $xv$~is a deviation
history, then for each \pl{j\neq i} the residual strategy
$\sigma^*_j[xv]$ is equal to $\tau_{j,i}$ on histories that start
in $w\coloneq\rho(xv)$. Hence, by \cref{lemma:residual-strategies} and since
$\Win_i$~is prefix-independent,
\begin{align*}
& \Prob^{\vec{\sigma}^*_{-i},\rho}_{v_0}(\Win_i\mid xv\cdot V^\omega) \\
=\; & \Prob^{\vec{\sigma}^*_{-i},\rho}_{v_0}
(\Win_i\mid xvw\cdot V^\omega) \\
%
%
=\; & \Prob^{\vec{\sigma}^*_{-i}[xv],\rho[xv]}_w(\Win_i) \\
%
%
\leq\; & \val_i^\calG(w) \\
\leq\; & \val_i^\calG(v)\,.
\end{align*}

\medskip
\noindent
Using the previous two claims, we prove that
$\Prob^{\vec{\sigma}^*_{-i},\rho}_{v_0}(\Win_i)\leq
\Prob^{\vec{\sigma}^*}_{v_0}(\Win_i)$ as follows:
\begin{align*}
& \Prob^{\vec{\sigma}^*_{-i},\rho}_{v_0}(\Win_i) \\
=\; & \Prob^{\vec{\sigma}^*_{-i},\rho}_{v_0}(\Win_i\setminus D\cdot V^\omega)
+\sum_{\mathmakebox[0.5cm][c]{xv\in D}}
\Prob^{\vec{\sigma}^*_{-i},\rho}_{v_0}(\Win_i\cap xv\cdot V^\omega) \\
=\; & \Prob^{\vec{\sigma}}_{v_0}(\Win_i\setminus D\cdot V^\omega)
+\sum_{\mathmakebox[0.5cm][c]{xv\in D}}
\Prob^{\vec{\sigma}^*_{-i},\rho}_{v_0}(\Win_i\cap xv\cdot V^\omega) \\
=\; & \Prob^{\vec{\sigma}}_{v_0}(\Win_i\setminus D\cdot V^\omega)
+\sum_{\mathmakebox[0.5cm][c]{xv\in D}}
\Prob^{\vec{\sigma}^*_{-i},\rho}_{v_0}(\Win_i\mid xv\cdot V^\omega)
\cdot\Prob^{\vec{\sigma}^*_{-i},\rho}_{v_0}(xv\cdot V^\omega) \\
=\; & \Prob^{\vec{\sigma}}_{v_0}(\Win_i\setminus D\cdot V^\omega)
+\sum_{\mathmakebox[0.5cm][c]{xv\in D}}
\Prob^{\vec{\sigma}^*_{-i},\rho}_{v_0}(\Win_i\mid xv\cdot V^\omega)
\cdot\Prob^{\vec{\sigma}}_{v_0}(xv\cdot V^\omega) \\
\leq\; & \Prob^{\vec{\sigma}}_{v_0}(\Win_i\setminus D\cdot V^\omega)
+\sum_{\mathmakebox[0.5cm][c]{xv\in D}} \val_i^\calG(v)
\cdot\Prob^{\vec{\sigma}}_{v_0}(xv\cdot V^\omega) \\
\leq\; & \Prob^{\vec{\sigma}}_{v_0}(\Win_i\setminus D\cdot V^\omega)
+\sum_{\mathmakebox[0.5cm][c]{xv\in D}}
\Prob^{\vec{\sigma}}_{v_0}(\Win_i\mid xv\cdot V^\omega)
\cdot\Prob^{\vec{\sigma}}_{v_0}(xv\cdot V^\omega) \\
=\; & \Prob^{\vec{\sigma}}_{v_0}(\Win_i\setminus D\cdot V^\omega)
+\sum_{\mathmakebox[0.5cm][c]{xv\in D}}
\Prob^{\vec{\sigma}}_{v_0}(\Win_i\cap xv\cdot V^\omega) \\
=\; & \Prob^{\vec{\sigma}}_{v_0}(\Win_i) \\
=\; & \Prob^{\vec{\sigma}^*}_{v_0}(\Win_i),
\end{align*}
where the second inequality follows from the assumption that $\vec{\sigma}$~is
favourable.
\end{proof}

A variant of \cref{lemma:nash-threat} handles games with
prefix-independent \omg-regular objectives and finite-state strategies.

\begin{lemma}\label{lemma:nash-threat-regular}
Let $(\calG,v_0)$ be a finite SMG with prefix-in\-de\-pen\-dent
\omg-regular objectives.
If $\vec{\sigma}$~is a favourable pure finite-state strategy profile of
$(\calG,v_0)$, then $(\calG,v_0)$ has a pure finite-state Nash
equilibrium~$\vec{\sigma}^*$ with
$\Prob_{v_0}^{\vec{\sigma}}=\Prob_{v_0}^{\vec{\sigma}^*}$.
\end{lemma}

\begin{proof}
The proof is analogous to the proof of \cref{lemma:nash-threat}. Since,
by \cref{cor:regular-optimal}, there exist optimal pure finite-state
strategies in every finite SMG with \omg-regular objectives, the
strategies~$\tau_{j,i}$ defined there can be assumed to be pure
finite-state strategies.
Consequently, the equilibrium profile~$\vec{\sigma}^*$ can be implemented
using finite-state strategies as well.
\end{proof}

Using \cref{lemma:nash-threat,thm:optimal-strategies}, we can easily prove the
existence of pure Nash equilibria in finite SMGs with prefix-independent
objectives.

\begin{theorem}\label{thm:nash-existence}
There exists a pure Nash equilibrium in any finite SMG with
prefix-independent objectives.
\end{theorem}

\begin{proof}
Let $\calG$ be a finite SMG with prefix-independent objectives
and initial vertex~$v_0$.
By \cref{thm:optimal-strategies} and the correspondence between
$\calG$ and the coalition game~$\calG_i$, each \pli has a
strongly optimal strategy~$\sigma_i$ in~$\calG$.
Let~$\vec{\sigma}=(\sigma_i)_{i\in\Pi}$. For every history~$xv$
that is consistent with~$\vec{\sigma}$ and each \pli,
we~have $\Prob_{v_0}^{\vec{\sigma}}(\Win_i\mid xv\cdot V^\omega)=
\Prob_v^{\vec{\sigma}[x]}(\Win_i)\geq\val^\calG_i(v)$.
Hence, $\vec{\sigma}$~is favourable, and \cref{lemma:nash-threat}
implies the existence of a pure Nash equilibrium.
\end{proof}

For finite SMGs with \omg-regular objectives,
we can even show the existence of a pure finite-state equilibrium.

\begin{theorem}\label{thm:nash-existence-regular}
There exists a pure finite-state Nash equilibrium in any finite SMG with
\omg-regular objectives.
\end{theorem}

\begin{proof}
Since any SMG with \omg-regular objectives can be reduced to one
with parity objectives using finite memory, it suffices to consider parity
SMGs.
For these games, the~claim follows from \cref{cor:positional-optimal} and
\cref{lemma:nash-threat-regular} using the same argumentation as in the
proof of \cref{thm:nash-existence}.
\end{proof}

\Cref{thm:nash-existence-regular} and a variant of \cref{thm:nash-existence}
appeared originally in \citep{ChatterjeeJM04}. However, their proof contains an
inaccuracy:
Essentially, they claim that any profile of optimal strategies can be extended
to a Nash equilibrium with the same payoff (by adding threat strategies
on~top). This is, in general, not true, as the following example demonstrates.

\begin{example}\label{expl:optimal-no-nash}
Consider the deterministic two-player game~$(\calG,v_0)$ depicted in
\cref{fig:optimal-no-nash} and played by players $0$~and~$1$ (with payoffs
given in this order).
\begin{figure}
\begin{tikzpicture}[x=1.6cm,y=1.3cm,->]
\node[play,label={above:$0$}] (1) at (0,0) {$v_0$};
\node (p1) at (0,-1) {$(0,0)$};
\node[play,label={above:$1$}] (2) at (1,0) {$v_1$};
\node (p2) at (1,-1) {$(1,0)$};
\node (p3) at (2,0) {$(1,1)$};

\draw (-0.7,0) to (1);
\draw (1) to (p1); \draw (1) to (2);
\draw (2) to (p2); \draw (2) to (p3);
\end{tikzpicture}
\caption{\label{fig:optimal-no-nash}A two-player game with a pair of
optimal strategies that cannot be extended to a Nash equilibrium}
\end{figure}
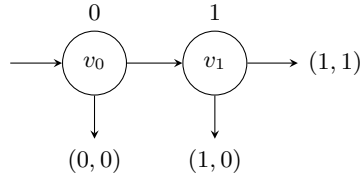
Clearly, the value $\val_0^\calG(v_0)$ for \pl0 from~$v_0$ equals~$1$, and
\pl0's optimal strategy~$\sigma$ is to play from~$v_0$ to~$v_1$.
For \pl1, the value from~$v_0$ is~$0$, and both of her positional
strategies are optimal.
In particular, her strategy~$\tau$ of
playing from~$v_1$ to the terminal vertex with payoff~$(1,0)$ is optimal
(albeit not globally optimal). The
payoff of the strategy profile~$(\sigma,\tau)$ is~$(1,0)$. However, there is no
Nash equilibrium of $(\calG,v_0)$ with payoff~$(1,0)$: In any Nash
equilibrium of $(\calG,v_0)$, \pl0 will move from~$v_0$ to~$v_1$ with
probability~$1$. \Pl1's best response is to play from~$v_1$ to
the terminal vertex with payoff~$(1,1)$ with probability~$1$. Hence, every
Nash equilibrium of this game has payoff~$(1,1)$.
\end{example}

\section{Complexity of Nash equilibria}
\label{sect:complexity}

For the rest of this paper, we consider only \emph{finite} SMGs.
Previous research on algorithms for finding Nash equilibria in such games
has focused on computing \emph{some} Nash equilibrium~\cite{ChatterjeeJM04}.
However, a game may have several Nash equilibria with different payoffs, and
one might not be interested in \emph{any} Nash equilibrium but in one whose
payoff fulfils certain requirements. For example, one might look for a
Nash equilibrium where certain players win almost surely while certain others
lose almost surely. This idea leads us to the following decision problem, which
we call \NE:
\begin{quote}
Given an SMG $(\calG,v_0)$ and thresholds
$\vec{x},\vec{y}\in[0,1]^\Pi$, decide whether there exists a
Nash equilibrium of $(\calG,v_0)$ with payoff $\geq\vec{x}$ and
$\leq\vec{y}$.
\end{quote}
To obtain meaningful results, we assume that all
transition probabilities in~$\calG$ as well as the thresholds $\vec{x}$
and~$\vec{y}$ are rational numbers (with numerator and denominator given in
binary) and that all objectives are \omg-regular.
A \emph{qualitative} variant of the problem, which omits the thresholds,
just asks about a Nash
equilibrium where some distinguished player, say \pl0, wins with
probability~$1$:
\begin{quote}
Given an SMG $(\calG,v_0)$, decide whether there
exists a Nash equilibrium of $(\calG,v_0)$ where \pl0 wins almost surely.
\end{quote}
Clearly, every instance of the qualitative variant can easily be turned into
an instance of~\NE (by adding the thresholds $\vec{x}=(1,0,\ldots,0)$ and
$\vec{y}=(1,\ldots,1)$). Hence, \NE is, a priori, more general than its
qualitative variant.

Note that we have so far not put any restriction on the type of strategies
that realise the equilibrium.
It is natural to restrict the search space to profiles of pure, finite-state,
pure finite-state, stationary or positional strategies.
We denote the corresponding decision problems by \PureNE, \FinNE, \PureFinNE,
\StatNE and \PosNE, respectively.
In the rest of this paper, we are going to prove upper and
lower bounds on the complexity of these problems, where all lower
bounds even hold for the qualitative variants of these problems.

Our first observation is that neither stationary nor pure strategies are
sufficient to implement any Nash equilibrium, even in SSMGs and even if we are
only interested in whether a player wins or loses almost surely in the
equilibrium. Together with another result from this section
(\cref{prop:inf-mem}), this demonstrates that the problems
\NE, \PureNE, \FinNE, \PureFinNE, \StatNE and \PosNE are distinct problems,
which have to be analysed separately. This is in sharp contrast to the
situation for \TwoSSGs where all these problems coincide because \TwoSSGs
admit globally optimal positional strategies.

\begin{proposition}\label{prop:pure-not-enough}
There exists an SSMG with a stationary Nash equilibrium where
\pl0 wins almost surely, but with no pure Nash equilibrium where \pl0 wins
with positive probability.
\end{proposition}

\begin{proof}
Consider the SSMG depicted in \cref{fig:no-pure-nash}, played by
three players $0$, $1$ and~$2$ (with payoffs given in this order).
\begin{figure}
\begin{floatrow}
\ffigbox[.47\textwidth]{%
\begin{tikzpicture}[x=1.6cm,y=1.4cm,->]
\node (init) at (-0.7,0) {};
\node (0) at (0,0) [play,label={above:$1$}] {$v_0$};
\node (1) at (1,0) [play,label={above:$2$}] {$v_1$};
\node (2) at (2,0) [play,label={above:$0$}] {$v_2$};
\node (3) at (2,-1) {$(1,1,0)$};
\node (4) at (3,0) {$(1,0,1)$};
\node (5) at (0,-1) {$(0,\frac{1}{2},0)$};
\node (6) at (1,-1) {$(0,0,\frac{1}{2})$};

\draw (init) -- (0);
\draw (0) -- (1);
\draw (0) -- (5);
\draw (1) -- (2);
\draw (1) -- (6);
\draw (2) -- (3);
\draw (2) -- (4);
\end{tikzpicture}}%
{\caption{\label{fig:no-pure-nash}An SSMG with no pure Nash equilibrium
where \pl0 wins with positive probability}}
\ffigbox[.47\textwidth]{%
\begin{tikzpicture}[x=1.6cm,y=1.4cm,->]
\node (init) at (-0.7,0) {};
\node (0) at (0,0) [play,label={above:1}] {$v_0$};
\node (1) at (1,0) [play,label={above:2}] {$v_1$};
\node (2) at (2.1,0) {$(1,0,0)$};
\node (3) at (-0.1,-1) {$(0,1,0)$};
\node (4) at (1,-1) [play,label={below:$0$}] {$v_2$};
\node (5) at (2.1,-1) {$(0,0,1)$};

\draw (init) -- (0);
\draw (0) -- (1);
\draw (0) -- (4);
\draw (1) -- (2);
\draw (1) -- (4);
\draw (4) -- (3);
\draw (4) -- (5);
\end{tikzpicture}}%
{\caption{\label{fig:no-stat-nash}An SSMG with no stationary Nash
equilibrium where \pl0 wins with positive probability}}
\end{floatrow}
\end{figure}
Clearly, the stationary strategy profile where from vertex~$v_2$ \pl0
selects both outgoing transitions with probability~$1/2$ each,
\pl1 plays from~$v_0$ to~$v_1$ and \pl2 plays from~$v_1$ to~$v_2$ is
a Nash equilibrium where \pl0 wins almost surely. However, for
any pure strategy profile where \pl0 wins with positive probability,
either \pl1 or \pl2 receives payoff~$0$ and could improve her payoff by
switching her strategy at $v_0$ or~$v_1$, respectively.
\end{proof}

\begin{proposition}\label{prop:stationary-not-enough}
There exists an SSMG with a pure finite-state Nash equilibrium where \pl0
wins almost surely, but with no stationary Nash equilibrium where \pl0
wins with positive probability.
\end{proposition}

\begin{proof}
Consider the (deterministic) SSMG~$\calG$ depicted in
\cref{fig:no-stat-nash}, also played by three players 0, 1 and 2.
Clearly, the pure finite-state strategy profile that leads to the
terminal vertex with payoff $(1,0,0)$
and where at~$v_2$ \pl0 plays ``right'' if \pl1 has
played to~$v_0$ and ``left'' if \pl2 has played to~$v_0$ is a Nash equilibrium
of $(\calG,v_0)$. Now consider any stationary
equilibrium of $(\calG,v_0)$ where \pl0 wins with positive
probability.
If at~$v_2$ the stationary strategy of \pl0 prescribes to play
``right''
with positive probability, then \pl2 can improve her payoff by
playing to~$v_2$ with probability~$1$, and otherwise \pl1 can
improve her payoff by playing to~$v_2$ with probability~$1$,
a contradiction.
\end{proof}

\subsection{Positional equilibria}
\label{section:positional}

In this subsection, we analyse the complexity of the (presumably) simplest of
the decision problems introduced so far, namely \PosNE.
Not surprisingly, this problem is decidable; in fact, it is
\NP-complete for all types of objectives we consider in this paper. Let
us start by proving membership to \NP. Since terminal reachability,
(co-)B\"uchi and parity objectives can easily be translated to Rabin or
Streett objectives, it~suffices to consider Streett-Rabin and Muller SMGs.

\begin{theorem}\label{thm:posne-np}
\PosNE is in \NP for Streett-Rabin~SMGs and Muller~SMGs.
\end{theorem}

\begin{proof}
To decide \PosNE, on input $\calG,v_0,\vec{x},\vec{y}$, we can guess a
positional strategy profile~$\vec{\sigma}$, \ie a mapping
$\bigcup_{i\in\Pi} V_i \to V$; then, we verify whether $\vec{\sigma}$~is a
Nash equilibrium with the desired payoff. To~do this, we first
compute the payoff~$z_i$ of~$\vec{\sigma}$ for each \pli by
computing the probability of the event~$\Win_i$ in the (finite) Markov chain
$(\calG^{\vec{\sigma}},v_0)$. Once each~$z_i$ is computed, we can
easily check whether $x_i\leq z_i\leq y_i$.
To~verify that $\vec{\sigma}$~is a Nash equilibrium, we additionally
compute, for each \pli, the value~$r_i$ of the (finite)
MDP~$\calG^{\vec{\sigma}_{-i}}$ from~$v_0$. Clearly,
$\vec{\sigma}$~is a Nash equilibrium \iff $r_i\leq z_i$ for each \pli.
Since we can compute the value of an MDP (or a Markov chain) with a
Streett, Rabin or Muller objective in polynomial time
(\cref{thm:mdp-ptime}), all
these checks can be carried out in polynomial time.
\end{proof}

To establish \NP-completeness, we still need to show \NP-hardness.
In fact, the reduction we are going to present does not only work for \PosNE,
but also for \StatNE, where we allow arbitrary stationary equilibria.

\begin{theorem}\label{thm:np-hardness}
\PosNE and \StatNE are \NP-hard, even for SSMGs with only
two players (three players for the qualitative variants).
\end{theorem}

\proof
The proof is by reduction from \SAT. Given a Boolean formula
$\phi=C_1\wedge\ldots\wedge C_m$ in conjunctive normal form
over propositional variables $X_1,\ldots,X_n$, where
\wlg $m\geq 1$ and each clause is non-empty, we show how to construct
a two-player SSMG $(\calG,v_0)$ such that the following statements
are equivalent:
\begin{enumerate}[(1)]
 \item\label{item:reduction-1} $\phi$ is satisfiable.
 \item\label{item:reduction-2} $(\calG,v_0)$ has a positional Nash
  equilibrium with payoff $(1,\frac{1}{2})$.
\vspace*{1pt}
 \item\label{item:reduction-3} $(\calG,v_0)$ has a stationary Nash
  equilibrium with payoff $(1,\frac{1}{2})$.
\end{enumerate}
Provided that the game can be constructed in polynomial time, these
equivalences establish both reductions. The game~$\calG$ is depicted in
\cref{fig:reduction-np}.
\begin{figure}
\begin{tikzpicture}[x=1.6cm,y=1.5cm,->,bend angle=15]
\tikzset{action/.style={}}

\node (init) at (-0.7,0) {};
\node (start) at (0,0) [prob] {$v_0$};
\node (phi) at (1,0) [prob] {$\phi$};
\node (end) at (0,3.5) {$(1,0)$};

\draw (init) to (start);
\draw (start) to node[above,action] {$\frac{1}{2^{n+1}}$} (phi);
\draw (start) to node[left,action] {$\frac{1}{2^{n+1}}$} (end);

\node (c1) at (1.7,2) [play,label={above:$1$}] {$C_1$};
\node (clauses1) at (1.7,1) {$\vdots$};
\node (clauses2) at (1.7,-1) {$\vdots$};
\node (cm) at (1.7,-2) [play,label={below:$1$}] {$C_m$};
\node (pay) at (5.5,0) {$(1,1)$};

\draw (phi) to node[right,near start,action] {$\frac{1}{m+1}$} (c1);
\draw (phi) to node[right,near start,action] {\ $\frac{1}{m+1}$} (cm);
\draw (phi) to node[above,very near start,action] {$\frac{1}{m+1}$} (pay);

\begin{scope}[yshift=1.5cm]
\node (x1) at (3,2.75) [play,label={above:$0$}] {$X_1$};
\node (x1z) at (3.6,3.55) [play,label={above:$1$}] {$\top$};
\node (x1b) at (4.5,3) [prob] {};
\node (x1bf) at (4.5,2.25) [prob] {$\bot$};
\node (pay1z) at (5.5,3.75) {$(0,1)$};
\node (pay1) at (5.5,3) {$(1,1)$};
\node (pay1f) at (5.5,2.25) {$(1,0)$};

\node (nx1) at (3,1) [play,label={above:$0$}] {\tiny$\neg X_1$};
\node (nx1b) at (4.5,0.75) [prob] {};
\node (nx1bf) at (4.5,1.5) [prob] {$\bot$};
\node (nx1z) at (3.6,0.2) [play,label={below:$1$}] {$\top$};
\node (npay1) at (5.5,0.75) {$(1,1)$};
\node (npay1f) at (5.5,1.5) {$(1,0)$};
\node (npay1z) at (5.5,0) {$(0,1)$};

\draw (x1) to [bend left=10] (x1z);
\draw (x1z) to (pay1z);
\draw (x1z) [bend left=10] to (x1b);
\draw (x1) to [bend left] (x1bf);
\draw (x1b) to [bend right=30] (x1);
\draw (x1bf) to [bend left] (x1);
\draw (x1b) to (pay1);
\draw (x1bf) to (pay1f);

\draw (nx1) to [bend right=10] (nx1z);
\draw (nx1z) to [bend right=10] (nx1b);
\draw (nx1z) to (npay1z);
\draw (nx1) to [bend right] (nx1bf);
\draw (nx1b) to [bend left=30] (nx1);
\draw (nx1bf) to [bend right] (nx1);
\draw (nx1b) to (npay1);
\draw (nx1bf) to (npay1f);
\end{scope}

\draw[dashed] (c1) to (x1); \draw[dashed] (c1) to (nx1);
\draw[dashed] (cm) to (x1); \draw[dashed] (cm) to (nx1);

\node (vars1) at (3,0.6) {$\vdots$};
\node (vars2) at (3,-0.6) {$\vdots$};
\node (nodes3) at (4.5,0.6) {$\vdots$};
\node (nodes4) at (4.5,-0.6) {$\vdots$};

\begin{scope}[yshift=-7.125cm]
\node (x2) at (3,2.75) [play,label={below:$0$}] {$X_n$};
\node (x2z) at (3.6,3.55) [play,label={above:$1$}] {$\top$};
\node (x2b) at (4.5,3) [prob] {};
\node (x2bf) at (4.5,2.25) [prob] {$\bot$};
\node (pay2z) at (5.5,3.75) {$(0,1)$};
\node (pay2) at (5.5,3) {$(1,1)$};
\node (pay2f) at (5.5,2.25) {$(1,0)$};

\node (nx2) at (3,1) [play,label={below:$0$}] {\tiny$\neg X_n$};
\node (nx2b) at (4.5,0.75) [prob] {};
\node (nx2bf) at (4.5,1.5) [prob] {$\bot$};
\node (nx2z) at (3.6,0.2) [play,label={below:$1$}] {$\top$};
\node (npay2) at (5.5,0.75) {$(1,1)$};
\node (npay2f) at (5.5,1.5) {$(1,0)$};
\node (npay2z) at (5.5,0) {$(0,1)$};

\draw (x2) to [bend left=10] (x2z);
\draw (x2z) to (pay2z);
\draw (x2z) [bend left=10] to (x2b);
\draw (x2) to [bend left] (x2bf);
\draw (x2b) to [bend right=30] (x2);
\draw (x2bf) to [bend left] (x2);
\draw (x2b) to (pay2);
\draw (x2bf) to (pay2f);

\draw (nx2) to [bend right=10] (nx2z);
\draw (nx2z) to [bend right=10] (nx2b);
\draw (nx2z) to (npay2z);
\draw (nx2) to [bend right] (nx2bf);
\draw (nx2b) to [bend left=30] (nx2);
\draw (nx2bf) to [bend right] (nx2);
\draw (nx2b) to (npay2);
\draw (nx2bf) to (npay2f);
\end{scope}

\draw[dashed] (c1) to (x2); \draw[dashed] (c1) to(nx2);
\draw[dashed] (cm) to (x2); \draw[dashed] (cm) to (nx2);

\draw (start) to [bend left=30] node[above left,near start,action]%
 {$\frac{1}{4}$} (x1);
\draw (start) to [bend left=15] node[above left,near start,action]%
 {$\frac{1}{4}$} (nx1);
\draw (start) to [bend right=15] node[below left,pos=0.4,action]%
 {$\frac{1}{2^{n+1}}$} (x2);
\draw (start) to [bend right=30] node[below left,near start,action]%
 {$\frac{1}{2^{n+1}}$} (nx2);
\end{tikzpicture}
\caption{\label{fig:reduction-np}Reducing \SAT to \PosNE and \StatNE}
\end{figure}
The game proceeds from the initial vertex~$v_0$ to $X_i$ or~$\neg X_i$ with
probability~$1/2^{i+1}$ each, and to vertex~$\phi$ with
probability~$1/2^{n+1}$; with the remaining probability
of~$1/2^{n+1}$ the game proceeds to a terminal vertex with
payoff~$(1,0)$. From~$\phi$, the game proceeds to each vertex~$C_j$
with probability $1/(m+1)$; with the remaining probability of $1/(m+1)$, the
game proceeds to a terminal vertex with payoff~$(1,1)$. From vertex~$C_j$
(controlled by \pl1), there is a transition to a literal~$L$, \ie
$L=X_i$ or $L=\neg X_i$, \iff $L$~occurs inside the clause~$C_j$.
Obviously, the game~$\calG$ can be constructed
from~$\phi$ in polynomial time. We~conclude the proof by showing that
(\ref{item:reduction-1})--(\ref{item:reduction-3}) are equivalent.

\noindent(\infer{1}{2}) Assume that $\alpha\colon\{X_1,\ldots,X_n\}\to
\{\true,\false\}$ is a satisfying assignment of~$\phi$.
Consider the positional strategy profile where \pl0 moves from a
literal~$L$ to the neighbouring $\top$-labelled vertex \iff $L$~is
mapped to true by~$\alpha$, and \pl1 moves from vertex~$C_j$ to a
literal~$L$ that is contained in~$C_j$ and mapped to true by~$\alpha$
(which is possible since $\alpha$~is a satisfying assignment); at
$\top$-labelled vertices, \pl1 never plays to a terminal vertex. Obviously,
\pl0 wins almost surely in this strategy profile. In~order to compute \pl1's
payoff, note that for each variable~$X$ \pl1 either receives payoff~$1$
from~$X$ and payoff~$0$ from~$\neg X$, or she receives payoff~$1$
from~$\neg X$ and payoff~$0$ from~$X$ (because \pl0 plays
according to a well-defined assignment). Moreover, \pl1 wins almost
surely from~$\phi$ since that assignment satisfies~$\phi$. Hence, \pl1's
payoff equals
\[\frac{1}{2^{n+1}}+\sum_{i=1}^n \frac{1}{2^{i+1}}
=\frac{1}{2^{n+1}}+\frac{1}{2}\bigg(\sum_{i=1}^n \frac{1}{2^i}\bigg)
=\frac{1}{2^{n+1}}+\frac{1}{2}\bigg(1-\frac{1}{2^n}\bigg)
=\frac{1}{2},\]
Obviously, changing her strategy cannot give \pl1 a better payoff. Therefore,
we have identified a Nash equilibrium.\medskip

\noindent(\infer{2}{3}) Trivial.\medskip

\noindent(\infer{3}{1}) Let $\vec{\sigma}=(\sigma_0,\sigma_1)$ be a stationary
Nash equilibrium of $(\calG,v_0)$ with payoff $(1,\frac{1}{2})$. Our
first aim is to show that $\sigma_0$~is actually a positional strategy.
Consider any literal~$L$ such that $\sigma_0(L)$ assigns probability
$q>0$ to the neighbouring $\top$-labelled vertex.  Since
\pl0 wins almost surely, we know that \pl1 never plays to a
terminal vertex with payoff $(0,1)$. Hence, the expected payoff for \pl1
from~$L$ equals~$q$. However, by playing to a terminal vertex with
payoff $(0,1)$, \pl1 can get payoff $2q/(1+q)$ from~$L$. Since
$\vec{\sigma}$~is a Nash equilibrium, we have
$2q/(1+q)\leq q$, which implies that $q=1$.

Now we define a \emph{pseudo
assignment} $\alpha\colon\{X_1,\neg X_1,\ldots,X_n,\neg X_n\}\to
\{\true,\false\}$ by setting $\alpha(L)=\true$ \iff $\sigma_0$~prescribes
to go from vertex~$L$ to the neighbouring $\top$-labelled vertex. Our next
aim is to show that $\alpha$~is actually an assignment: $\alpha(X_i)=\true$
\iff $\alpha(\neg X_i)=\false$. To see this, note that we can compute
\pl1's expected payoff from~$v_0$ as follows:
\[\frac{1}{2}=\frac{p}{2^{n+1}}+\sum_{i=1}^n \frac{a_i}{2^{i+1}},\quad a_i=
\begin{cases}
 0 & \text{if $\alpha(X_i)=\alpha(\neg X_i)=\false$,} \\
 1 & \text{if $\alpha(X_i)\neq\alpha(\neg X_i)$,} \\
 2 & \text{if $\alpha(X_i)=\alpha(\neg X_i)=\true$,}
\end{cases}\]
where $p$~is the expected payoff for \pl1 from vertex~$\phi$. By the
construction of~$\calG$, we~have $p>0$, and the equality only holds
if $p=1$ and $a_i=1$ for all $i=1,\ldots,n$, which proves that $\alpha$~is
an assignment.

Finally, we claim that $\alpha$ satisfies~$\phi$. If this were not
the case, there would exist a clause~$C$ such that \pl1's expected payoff
from vertex~$C$ equals~$0$, and therefore $p<1$. This
is a contradiction to $p=1$, as we have shown above.

To show that the qualitative variants of \PosNE and \StatNE are also \NP-hard,
it suffices to modify the game~$\calG$ as follows: First, we add one new
player, \pl2,  who wins at precisely those terminal vertices where \pl1 loses.
Second, we add two new vertices $v_1$ and~$v_2$. At~$v_1$, \pl1 has the choice
to leave the game; if she decides to stay inside the game,
the play proceeds to~$v_2$, where \pl2 has the choice to leave the game; if
she also decides to stay inside the game, the play proceeds to vertex~$v_0$
from where the game continues normally; if \pl1 or \pl2 decides to
leave the game, then each of them receives payoff~$\frac{1}{2}$, but
\pl0 receives payoff~$0$. Let us denote the modified game by~$\calG'$.
It~is straightforward to see that the following statements are equivalent:
\begin{enumerate}[(1)]
 \item $(\calG',v_1)$ has a stationary Nash equilibrium where \pl0 wins almost
surely.
 \item $(\calG,v_0)$ has a stationary Nash equilibrium with
payoff~$(1,\frac{1}{2})$.
 \item $\phi$~is satisfiable.
 \item $(\calG,v_0)$ has a positional Nash equilibrium with
payoff~$(1,\frac{1}{2})$.
 \item $(\calG',v_1)$ has a positional Nash equilibrium where \pl0
wins almost surely.\qed
\end{enumerate}

\subsection{Stationary equilibria}
\label{section:stationary}

To prove the decidability of \StatNE, we appeal to results
established for the \emph{existential theory of the reals}, the set of
all existential first-order sentences (over the appropriate signature) that
hold in the ordered field $\frakR\coloneq(\bbR,+,\cdot,0,1,\leq)$. The~best
known upper bound for the complexity of the associated decision problem is
\PSpace \citep{Canny88}, which leads to the following theorem.

\begin{theorem}\label{thm:statne-pspace}
\StatNE is in \PSpace for Streett-Rabin~SMGs and Muller~SMGs.
\end{theorem}

\begin{proof}
Since $\PSpace=\NPSpace$, it suffices to provide a nondeterministic algorithm
with polynomial space requirements for deciding \StatNE. On input $\calG,v_0,
\vec{x},\vec{y}$, where \wlg $\calG$~is an SMG with Muller
objectives given by $\calF_i\subseteq\pow(C)$, the algorithm starts by
guessing the \emph{support} $S\subseteq V\times V$ of a stationary strategy
profile~$\vec{\sigma}$ of~$\calG$, \ie
$S=\{(v,w)\in V\times V:\vec{\sigma}(w\mid v)>0\}$.
From the set~$S$ alone, by standard graph algorithms, one can
compute for each \pli the following sets in polynomial time
(see \cite[Chapter~10]{BaierK08}):
\begin{enumerate}[(1)]
 \item the union~$F_i$ of all bottom SCCs~$U$ of the
Markov chain~$\calG^{\vec{\sigma}}$ with $\chi(U)\in\calF_i$,

 \item the set~$R_i$ of vertices~$v$ such that $\Prob_v^{\vec{\sigma}}
(\Reach(F_i))>0$,

 \item the union~$T_i$ of all end components~$U$ of the
MDP~$\calG^{\vec{\sigma}_{-i}}$ with $\chi(U)\in\calF_i$.
\end{enumerate}

After computing all these sets, the algorithm evaluates an existential
first-order sentence~$\psi$, which can be computed in polynomial time from
$\calG$, $v_0$, $\vec{x}$, $\vec{y}$, $S$, $(R_i)_{i\in\Pi}$,
$(F_i)_{i\in\Pi}$ and $(T_i)_{i\in\Pi}$, over $\frakR$ and returns the
answer to this~query.

How does~$\psi$ look like? Let
$\vec{\alpha}=(\alpha_{vw})_{v,w\in V}$, $\vec{r}=(r^i_v)_{i\in\Pi,v\in V}$
and $\vec{z}=(z^i_v)_{i\in\Pi, v\in V}$ be three sets of variables, and
let $V_*=\bigcup_{i\in\Pi} V_i$. The~formula
\begin{align*}
\phi(\vec{\alpha})&\coloneq
\bigwedge_{v\in V_*}
\bigg( \bigwedge_{w\in v\Delta} \alpha_{vw} \geq 0 \wedge
\bigwedge_{\mathmakebox[0.7cm][c]{w\in V\setminus v\Delta}}
\alpha_{vw}=0\wedge\sum_{\mathmakebox[0.6cm][c]{w \in v\Delta}}\alpha_{vw} = 1\bigg)\:\wedge \\
& \qquad
\bigwedge_{\mathmakebox[0.7cm][c]{\substack{v\in V\setminus V_*\\ w\in V}}}
\alpha_{vw} = \Delta(w\mid v) \wedge
\bigwedge_{\mathmakebox[0.7cm][c]{(v,w)\in S}} \alpha_{vw} > 0 \wedge
\bigwedge_{\mathmakebox[0.7cm][c]{(v,w)\nin S}} \alpha_{vw}=0
\end{align*}
states that the mapping~$\vec{\sigma}\colon V\to[0,1]^V$, defined by
$\vec{\sigma}(v)(w)=\alpha_{vw}$,
constitutes a valid stationary strategy profile of~$\calG$ whose support
is~$S$. Provided that $\phi(\vec{\alpha})$ holds in~$\frakR$, the~formula
\[\eta_i(\vec{\alpha},\vec{z})\coloneq
\bigwedge_{v \in F_i} z^i_v= 1 \wedge
\bigwedge_{\mathmakebox[0.7cm][c]{v \in V\setminus R_i}} z^i_v= 0 \wedge
\bigwedge_{\mathmakebox[0.7cm][c]{v \in V\setminus F_i}} z^i_v =
\sum_{\mathmakebox[0.6cm][c]{w \in v\Delta}} \alpha_{vw}\cdot z^i_w\]
states that $z^i_v=\Prob_v^{\vec{\sigma}}(\Win_i)$ for each $v\in V$,
where $\vec{\sigma}$ is defined as above. This follows from a well-known
result about Markov chains, namely that the vector of the aforementioned
probabilities is the unique solution of the given system of equations
(see \cite[Chapter~10]{BaierK08}). Finally, the formula
\[\theta_i(\vec{\alpha},\vec{r})\coloneq
\bigwedge_{v\in V} r^i_v\geq 0 \wedge
\bigwedge_{v\in T_i} r^i_v=1 \wedge
\bigwedge_{\mathmakebox[0.6cm][c]{\substack{v\in V_i \\ w\in v\Delta}}}
r^i_v\geq r^i_w\wedge
\bigwedge_{\mathmakebox[0.7cm][c]{v\in V\setminus V_i}}
r^i_v=\sum_{\mathmakebox[0.6cm][c]{w\in v\Delta}}\alpha_{vw}\cdot r^i_w\]
states that $r^i_v\geq\sup_{\tau}\Prob^{\vec{\sigma}_{-i},\tau}_v(\Win_i)$
for all $v\in V$ (see \cite[Chapter~10]{BaierK08}).

The desired sentence~$\psi$ is the existential closure of the conjunction
of~$\phi$ and, for each \pli, the formulae $\eta_i$ and~$\theta_i$ combined
with formulae stating that \pli cannot improve her payoff and that the
expected payoff for \pli lies in between the given thresholds:
\[\psi\coloneq\exists\vec{\alpha}\,\exists\vec{r}\,\exists\vec{z}\,
\Big(\phi(\vec{\alpha})\wedge\bigwedge_{i\in\Pi}(\eta_i(\vec{\alpha},\vec{z})
\wedge\theta_i(\vec{\alpha},\vec{r})\wedge r^i_{v_0}\leq z^i_{v_0}\wedge
x_i\leq z^i_{v_0}\leq y_i)\Big)\,.\]
Clearly, $\psi$~holds in~$\frakR$ \iff $(\calG,v_0)$ has a
stationary Nash equilibrium with payoff at least~$\vec{x}$ and
at most~$\vec{y}$ whose support is~$S$. Consequently, the algorithm is
correct.
\end{proof}

In \cref{section:positional}, we showed that \StatNE is
\NP-hard, leaving a considerable gap to our upper bound of \PSpace.
Towards gaining a better understanding, we relate \StatNE
to the \emph{square root sum problem} (\SqrtSum) of deciding, given
numbers $d_1,\ldots,d_n,k\in\bbN$, whether $\sum_{i=1}^n \sqrt{d_i}\geq k$.

Recently, \citet{AllenderBKM09} showed that \SqrtSum belongs to the fourth
level of the \emph{counting hierarchy}, a slight improvement
over the previously known \PSpace upper bound.
However, it has been an open question since the 1970s as to whether \SqrtSum
falls into the polynomial hierarchy \citep{GareyGJ76,EtessamiY10}. We identify
a polynomial-time reduction from \SqrtSum to \StatNE for SSMGs.\footnote{Some
authors define \SqrtSum using $\leq$ instead of~$\geq$. With this definition, we
would reduce from the complement of \SqrtSum instead.}
Hence, \StatNE is at least as hard as \SqrtSum, and showing that
\StatNE resides inside the polynomial hierarchy would imply a major
breakthrough in understanding the complexity of numerical computation.

\begin{theorem}\label{thm:sqrt-sum}
$\SqrtSum$ is polynomial-time reducible to \StatNE,
even for 4-player SSMGs.
\end{theorem}

Before we start with the proof of the theorem, let us first examine the
game $\calG(p)$, where $0\leq p\leq 1$, played by players
$0$, $1$, $2$ and~$3$ and depicted in~\cref{fig:sqrt-reduction-1}.
\begin{figure}
\begin{floatrow}
\ffigbox[.34\textwidth]{%
\begin{tikzpicture}[x=1.5cm,y=1.4cm,->]
\node (dummy) at (0.3,2) {\phantom{$(1,\frac{1}{2},0,1)$}};
\node (init) at (-1.9,0) {};
\node (0) at (-1.2,0) [play,label={above:$3$}] {$v_0$};
\node (t) at (-1.2,-1) {$(0,0,0,\frac{k}{dn})$};
\node (1) at (0,0) [prob] {$v_1$};
\node (3) at (50:1.2) {$\calG(p_1)$};
\node (4) at (-50:1.2) {$\calG(p_n)$};
\node (dots) at (0.75,0.09) {$\vdots$};

\draw (init) to (0);
\draw (0) to (1); \draw (0) to (t);
\draw (1) to node[above left] {$\frac{1}{n}$} (3);
\draw (1) to node[below left] {$\frac{1}{n}$} (4);
\end{tikzpicture}}%
{\caption{\label{fig:sqrt-reduction-2}Reducing \SqrtSum to \StatNE}}
\ffigbox[.48\textwidth]{%
\begin{tikzpicture}[x=1.5cm,y=1.4cm,->,bend angle=15]
\node (init) at (-0.8,0) {};
\node (1) at (0,0) [play,label={above left:$1$}] {$s_1$};
\node (1a) at (1.2,0) {$(0,\frac{1}{2},0,0)$};
\node (2) at (0.5,1) [prob] {$r_1$};
\node (2a) at (0.3,2) {$(1,\frac{1}{2},0,1)$};
\node (3) at (1.7,1) [play,label={right:$0$}] {$t_1$};
\node (3a) at (1.9,2) {$(1,1,0,0)$};
\node (4) at (2.2,0) [play,label={above right:$2$}] {$s_2$};
\node (4a) at (3.4,0) {$(0,0,\frac{1}{2},0)$};
\node (5) at (1.7,-1) [prob] {$r_2$};
\node (5a) at (1.9,-2) {$(1,0,\frac{1}{2},1)$};
\node (6) at (0.5,-1) [play,label={left:$0$}] {$t_2$};
\node (6a) at (0.3,-2) {$(1,0,1,0)$};

\draw (init) to (1);
\draw (1) to [bend left] (2); \draw (1) to (1a);
\draw (2) to [bend left=20] node[above] {$1-p$} (3);
\draw (2) to node[left] {$p$} (2a);
\draw (3) to [bend left] (4); \draw (3) to (3a);
\draw (4) to [bend left] (5); \draw (4) to (4a);
\draw (5) to [bend left=20] node[below] {$1-p$} (6);
\draw (5) to node[right] {$p$} (5a);
\draw (6) to [bend left] (1); \draw (6) to (6a);
\end{tikzpicture}}%
{\caption{\label{fig:sqrt-reduction-1}The game~$\calG(p)$}}
\end{floatrow}
\end{figure}

\begin{lemma}\label{lemma:statne-reduction}
The maximal payoff \pl3 receives in a stationary Nash
equilibrium of $(\calG(p),s_1)$ where \pl0 wins almost surely
equals~$\sqrt{p}$.
\end{lemma}

\proof
In the following, assume \wlg that $0<p<1$ (otherwise the statement is
trivial), and define $q\coloneq 1-p$.
For any stationary strategy profile~$\vec{\sigma}$ of~$\calG(p)$
where \pl0 wins almost surely, let $x_1=\sigma_0(s_2\mid t_1)$
and $x_2=\sigma_0(s_1\mid t_2)$ be
the probabilities that \pl0 ``stays inside the game'' at~$t_1$,
respectively~$t_2$. Given $x_1$ and~$x_2$, for $i=1,2$ we can compute
the payoff $f_i(x_1,x_2)\coloneq\Prob^{\vec{\sigma}}_{s_i}(\Win_i)$
for \pli from~$s_i$ by
\[f_i(x_1,x_2)=\frac{p/2+q(1-x_i)}{1-q^2x_1x_2}\,.\]
To have a Nash equilibrium, it must be the case that
$f_1(x_1,x_2),f_2(x_1,x_2)\geq\frac{1}{2}$ since otherwise \pl1 or \pl2
would prefer to leave the game at $s_1$ or~$s_2$, respectively, which would
give the respective player payoff~$\frac{1}{2}$ immediately.
Vice versa, if $f_1(x_1,x_2),f_2(x_1,x_2)\geq \frac{1}{2}$ then
$\vec{\sigma}$ is a Nash equilibrium with expected payoff
\[f(x_1,x_2)\coloneq\frac{p+q x_1p}{1-q^2x_1x_2}\]
for \pl3. Hence,
to determine the maximum payoff for \pl3 in a stationary Nash equilibrium
where \pl0 wins almost surely, we have to maximise
$f(x_1,x_2)$ under the constraints $f_1(x_1,x_2),f_2(x_1,x_2)\geq\frac{1}{2}$
and $0\leq x_1,x_2\leq 1$. We claim that the maximum is reached only if
$x_1=x_2$. If \eg $x_1>x_2$, then we can achieve a higher payoff for \pl3
by setting $x_2'\coloneq x_1$, and the constraints are still satisfied:
\[\frac{p/2+q(1-x_2')}{1-q^2x_1x_2'}=\frac{p/2+q(1-x_1)}{1-q^2 x_1^2}
 \geq\frac{p/2+q(1-x_1)}{1-q^2 x_1 x_2}\geq\frac{1}{2}\,.\]
Hence, it suffices to maximise $f(x,x)$ subject to
$f_1(x,x)\geq\frac{1}{2}$ and $0\leq x\leq 1$, which is
equivalent to maximising $f(x,x)$ subject to $(1-p)x^2-2x+1\geq 0$
and $0\leq x\leq 1$. The~roots of the polynomial are
$(1\pm\sqrt{p})/(1-p)$, but $(1+\sqrt{p})/(1-p)>1$ for $p>0$.
Therefore, any solution~$x$
must satisfy $x\leq x_0\coloneq(1-\sqrt{p})/(1-p)$. Since
$0\leq x_0\leq 1$ for $0<p<1$ and $f(x,x)$ is strictly increasing on $[0,1]$,
the optimal solution is~$x_0$, and the maximal payoff for \pl3
in a stationary Nash equilibrium of $(\calG(p),s_1)$ where \pl0
wins almost surely equals indeed
\[f(x_0,x_0)=\frac{p+qx_0 p}{1-q^2x_0^2}
=\frac{p}{1-q x_0}
=\frac{p}{1-(1-p)x_0}
=\frac{p}{1-(1-\sqrt{p})}=\sqrt{p}\,.\tag*{\qedsymbol}\]

\medskip
\begin{proof}[Proof of \cref{thm:sqrt-sum}]
Given an instance $(d_1,\dots,d_n,k)$ of \SqrtSum,
where \wlg $n>0$, $d_i>0$ for each $i=1,\dots,n$, and
$d\coloneq\sum_{i=1}^n d_i$,
we construct a 4-player SSMG $(\calG,v_0)$
such that $(\calG,v_0)$ has a stationary Nash equilibrium where
\pl0 wins almost surely \iff $\sum_{i=1}^n\sqrt{d_i}\geq k$.
Define $p_i\coloneq d_i/d^2$ for $i=1,\dots,n$.
For the reduction, we use $n$~copies of the game~$\calG(p)$, where in the
$i$th copy we set $p$ to~$p_i$. The complete
game~$\calG$ is depicted in \cref{fig:sqrt-reduction-2}.
By \cref{lemma:statne-reduction}, the maximal payoff \pl3 receives
in a stationary Nash equilibrium of $(\calG(p_i),s_1)$ where \pl0
wins almost surely equals $\sqrt{p_i}=\sqrt{d_i}/d$. Hence, the
maximal payoff \pl3 receives in a stationary Nash equilibrium of
$(\calG,v_1)$ where \pl0 wins almost surely equals
\[\sum_{i=1}^n\frac{1}{n}\cdot\frac{\sqrt{d_i}}{d}
=\frac{1}{dn}\cdot\sum_{i=1}^n\sqrt{d_i}\,.\]
If $\sum_{i=1}^n\sqrt{d_i}\geq k$, then we can extend such an
equilibrium to a
stationary Nash equilibrium of $(\calG,v_0)$ where \pl0 wins
almost surely by letting \pl3 play from $v_0$ to~$v_1$ with
probability~$1$. On the other hand, if $\sum_{i=1}^n\sqrt{d_i}<k$,
then in any stationary Nash equilibrium of $(\calG,v_0)$ \pl3 plays
to~$v_1$ with probability~$0$, and \pl0 loses almost surely.
\end{proof}

\begin{remark}
The positive results of \cref{section:positional,section:stationary} can easily
be extended to equilibria in pure or randomised strategies with a memory of a
fixed size $k\in\bbN$: a
nondeterministic algorithm can guess a memory~$\frakM$ of size~$k$
and then look for a positional, respectively stationary, equilibrium
in the product of the original game~$\calG$ with the memory~$\frakM$.
Hence, for any fixed
$k\in\bbN$, we~can decide in \PSpace (\NP) the existence of a randomised (pure)
equilibrium of size~$k$ with payoff $\geq\vec{x}$ and $\leq\vec{y}$.
\end{remark}

\subsection{Pure and randomised equilibria}
\label{section:undecidability}

In this section, we show that the problems \NE and \PureNE are undecidable,
by exhibiting a reduction from an undecidable problem about
\emph{two-counter machines}. Our construction is inspired by a
construction used by \citet{BrazdilBFK06} to prove the
undecidability of stochastic games with \emph{branching-time objectives}
(see \cref{remark:pctl} below).

Let $\Gamma\coloneq\{\inc(j),\dec(j),\zero(j):j=1,2\}$ (the set
of \emph{instructions}).
A two-counter machine is of the form $\calM=(Q,q_0,\delta)$, where
\begin{iteMize}{$-$}
\item $Q$ is a finite set of \emph{states},
\item $q_0\in Q$ is the \emph{initial state}, and
\item $\delta\subseteq Q\times\Gamma\times Q$ is the
\emph{transition relation}.
\end{iteMize}
For $q\in Q$ let
$\delta(q)\coloneq\{(\gamma,q')\in\Gamma\times Q: (q,\gamma,q')\in\delta\}$.
We call~$\calM$ \emph{deterministic} if for each $q\in Q$ either
$\delta(q)=\emptyset$, or $\delta(q)=\{(\inc(j),q')\}$ for some
$j\in\{1,2\}$ and $q'\in Q$, or
$\delta(q)=\{(\zero(j),q_1),(\dec(j),q_2)\}$ for some $j\in\{1,2\}$ and
$q_1,q_2\in Q$.

A configuration of~$\calM$ is a triple $C=(q,i_1,i_2)\in
Q\times\bbN\times\bbN$, where~$q$ denotes the current state and $i_j$~denotes
the current value of counter~$j$.
A~configuration~$C'=(q',i_1',i_2')$ is a \emph{successor} of
configuration~$C=(q,i_1,i_2)$, denoted by
$C\vdash C'$, if there exists a ``matching'' transition
$(q,\gamma,q')\in\delta$. For example, $(q,i_1,i_2)\vdash(q',i_1+1,i_2)$ \iff
$(q,\inc(1),q')\in\delta$. The instruction $\zero(j)$ performs a
\emph{zero test}: $(q,i_1,i_2)\vdash (q',i_1,i_2)$ \iff
$i_1=0$ and $(q,\zero(1),q')\in\delta$, or $i_2=0$ and
$(q,\zero(2),q')\in\delta$.

A \emph{partial computation} of~$\calM$ is a finite or infinite sequence
$\rho=\rho(0)\rho(1)\ldots$ of configurations such that
$\rho(0)\vdash\rho(1)\vdash\cdots$ and $\rho(0)=(q_0,0,0)$ (the
\emph{initial configuration}). A partial computation of~$\calM$ is a
\emph{computation} of~$\calM$ if it is infinite
or ends in a configuration~$C$ for which there  for which there exists
no successor configuration.
Note that each deterministic two-counter machine has a unique computation.

The \emph{halting problem} is to decide, given a machine~$\calM$,
whether $\calM$~has a finite computation. It is well-known that
deterministic two-counter machines are Turing powerful, which makes the
halting problem and its dual, the \emph{non-halting problem}, undecidable,
even when restricted to deterministic two-counter machines. In fact,
the non-halting problem for deterministic two-counter machines is not
recursively enumerable.

\begin{theorem}\label{thm:undecidability}
\NE and \PureNE are not recursively enumerable, even for 10-player SSMGs.
\end{theorem}

To prove \cref{thm:undecidability}, we give a reduction from the
non-halting problem for deterministic two-counter machines.
Our aim is thus to compute from a machine~$\calM$ a 10-player SSMG
$(\calG,v_0)$ such that the computation of~$\calM$ is infinite \iff
$(\calG,v_0)$ has a (pure) Nash equilibrium in which \pl0 wins almost surely.
Without loss of generality, we assume that in~$\calM$ there is no zero test
that is followed by another zero test: if $(\zero(j),q')\in\delta(q)$, then
$\abs{\delta(q')}\leq 1$.

The game~$\calG$ is played by players $0$, $1$ and eight other players
$A_j^t$ and~$B_j^t$, indexed by $j\in\{1,2\}$ and $t\in\{0,1\}$.
Intuitively, \pl0 and \pl1 build up the computation of~$\calM$: \pl0 updates
the counters, and \pl1 chooses transitions. The other players
make sure that \pl0 updates the counters correctly:
players $A_j^0$ and~$A_j^1$ ensure that, in each step, the value of
counter~$j$ is not too high, and players
$B_j^0$ and~$B_j^1$ ensure that, in each step, the value of counter~$j$ is
not too low. More precisely, $A_j^0$ and $B_j^0$ monitor the odd steps of the computation, while $A_j^1$ and $B_j^1$ monitor the even steps.

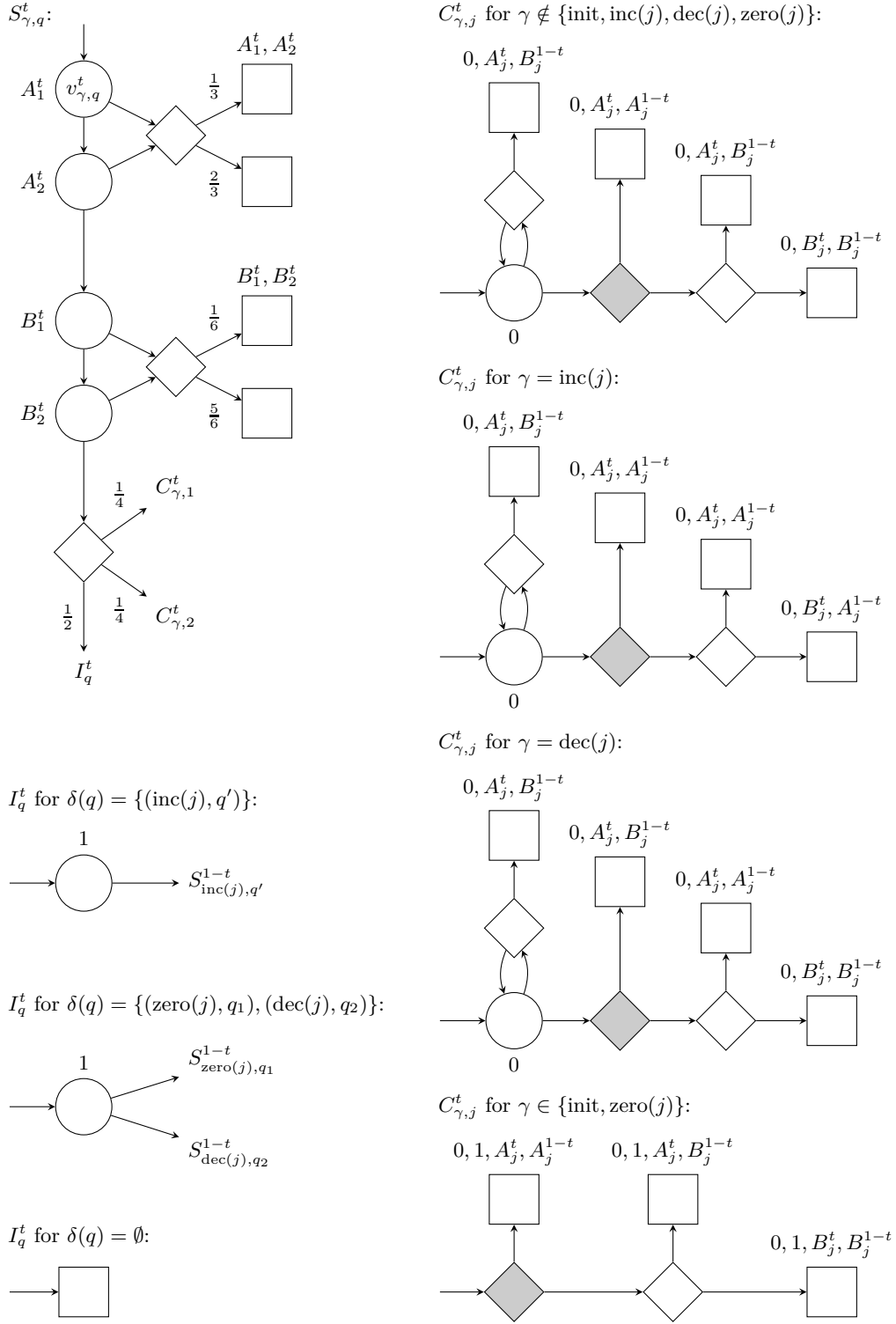
\begin{figure}
\begin{tikzpicture}[x=1.6cm,y=1.4cm,->]

\begin{scope} 
\node (caption) at (-0.8,0.8) [anchor=west] {$S_{\gamma,q}^t$:};

\node (1) at (0,0) [play,label=left:$A_1^t$,inner sep=0cm] {$v_{\gamma,q}^t$};
\node (1a) at (0.87,-.5) [prob] {};
\node (1b) at (1.73,0) [end,label=above:{$A_1^t,A_2^t$}] {};
\node (2) at (0,-1) [play,label=left:$A_2^t$] {};
\node (2b) at (1.73,-1) [end] {};
\node (3) at (0,-2.5) [play,label=left:$B_1^t$] {};
\node (3a) at (0.87,-3) [prob] {};
\node (3b) at (1.73,-2.5) [end,label=above:{$B_1^t,B_2^t$}] {};
\node (4) at (0,-3.5) [play,label=left:$B_2^t$] {};
\node (4b) at (1.73,-3.5) [end] {};
\node (10) at (0,-5) [prob] {};
\node (11) at (0.87,-4.3) {$C_{\gamma,1}^t$};
\node (12) at (0.87,-5.7) {$C_{\gamma,2}^t$};
\node (13) at (0,-6.3) {$I_q^t$};

\draw (0,0.7) to (1);
\draw (1) to (2); \draw (1) to (1a);
\draw (1a) to node[above left,near end] {$\frac{1}{3}$} (1b);
\draw (1a) to node[below left,near end] {$\frac{2}{3}$} (2b);
\draw (2) to (3); \draw (2) to (1a);
\draw (3) to (4); \draw (3) to (3a);
\draw (3a) to node[above left,near end] {$\frac{1}{6}$} (3b);
\draw (3a) to node[below left,near end] {$\frac{5}{6}$} (4b);
\draw (4) to (10); \draw (4) to (3a);
\draw (10) to node[left] {$\frac{1}{2}$} (13);
\draw (10) to node[above left,near end] {$\frac{1}{4}$} (11);
\draw (10) to node[below left,near end] {$\frac{1}{4}$} (12);
\end{scope}

\begin{scope}[yshift=-12cm] 
\node  (caption) at (-0.8,0.9) [anchor=west]
 {$I_q^t$ for $\delta(q)=\{(\inc(j),q')\}$:};

\node (1) at (0,0) [play,label={above:$1$}] {};
\node (2) at (0.9,0) [anchor=west] {$S_{\inc(j),q'}^{1-t}$};

\draw (-0.7,0) to (1);
\draw (1) to (2);
\end{scope}

\begin{scope}[yshift=-15.1cm] 
\node  (caption) at (-0.8,0.9) [anchor=west]
 {$I_q^t$ for $\delta(q)=\{(\zero(j),q_1),(\dec(j),q_2)\}$:};

\node (1) at (0,-0.2) [play,label={above:$1$}] {};
\node (2) at (0.9,0.3) [anchor=west] {$S_{\zero(j),q_1}^{1-t}$};
\node (3) at (0.9,-0.7) [anchor=west] {$S_{\dec(j),q_2}^{1-t}$};

\draw (-0.7,-0.2) to (1);
\draw (1) to (2);
\draw (1) to (3);
\end{scope}

\begin{scope}[yshift=-18.6cm] 
\node  (caption) at (-0.8,0.9) [anchor=west]
 {$I_q^t$ for $\delta(q)=\emptyset$:};

\node (1) at (0,0.3) [end] {};

\draw (-0.7,0.3) to (1);
\end{scope}

\begin{scope}[xshift=6.5cm,yshift=0cm] 
\node (caption) at (-0.8,0.8) [anchor=west]
 {$C_{\gamma,j}^t$ for $\gamma\nin\{\init,\inc(j),\dec(j),\zero(j)\}$:};

\node (0) at (0,-2.2) [play,label=below:$0$] {};
\node (1) at (1,-2.2) [prob,fill=black!20] {};
\node (1a) at (1,-0.7) [end,label=above:{$0,A_j^t,A_j^{1-t}$}] {};
\node (2) at (2,-2.2) [prob] {};
\node (2a) at (2,-1.2) [end,label=above:{$0,A_j^t,B_j^{1-t}$}] {};
\node (2b) at (3,-2.2) [end,label=above:{$0,B_j^t,B_j^{1-t}$}] {};
\node (3) at (0,-1.2) [prob] {};
\node (3a) at (0,-0.2) [end,label=above:{$0,A_j^t,B_j^{1-t}$}] {};

\draw (-0.7,-2.2) to (0);
\draw (0) to (1);
\draw (1) to (2);
\draw (1) to (1a);
\draw (2) to (2a);
\draw (2) to (2b);
\draw (0) to [bend right] (3);
\draw (3) to [bend right] (0);
\draw (3) to (3a);
\end{scope}

\begin{scope}[xshift=6.5cm,yshift=-5.5cm] 
\node (caption) at (-0.8,0.8) [anchor=west]
 {$C_{\gamma,j}^t$ for $\gamma=\inc(j)$:};

\node (0) at (0,-2.2) [play,label=below:$0$] {};
\node (1) at (1,-2.2) [prob,fill=black!20] {};
\node (1a) at (1,-0.7) [end,label=above:{$0,A_j^t,A_j^{1-t}$}] {};
\node (2) at (2,-2.2) [prob] {};
\node (2a) at (2,-1.2) [end,label=above:{$0,A_j^t,A_j^{1-t}$}] {};
\node (2b) at (3,-2.2) [end,label=above:{$0,B_j^t,A_j^{1-t}$}] {};
\node (3) at (0,-1.2) [prob] {};
\node (3a) at (0,-0.2) [end,label=above:{$0,A_j^t,B_j^{1-t}$}] {};

\draw (-0.7,-2.2) to (0);
\draw (0) to (1);
\draw (1) to (2);
\draw (1) to (1a);
\draw (2) to (2a);
\draw (2) to (2b);
\draw (0) to [bend right] (3);
\draw (3) to [bend right] (0);
\draw (3) to (3a);
\end{scope}

\begin{scope}[xshift=6.5cm,yshift=-11cm] 
\node (caption) at (-0.8,0.8) [anchor=west]
 {$C_{\gamma,j}^t$ for $\gamma=\dec(j)$:};

\node (0) at (0,-2.2) [play,label=below:$0$] {};
\node (1) at (1,-2.2) [prob,fill=black!20] {};
\node (1a) at (1,-0.7) [end,label=above:{$0,A_j^t,B_j^{1-t}$}] {};
\node (2) at (2,-2.2) [prob] {};
\node (2a) at (2,-1.2) [end,label=above:{$0,A_j^t,A_j^{1-t}$}] {};
\node (2b) at (3,-2.2) [end,label=above:{$0,B_j^t,B_j^{1-t}$}] {};
\node (3) at (0,-1.2) [prob] {};
\node (3a) at (0,-0.2) [end,label=above:{$0,A_j^t,B_j^{1-t}$}] {};

\draw (-0.7,-2.2) to (0);
\draw (0) to (1);
\draw (1) to (2);
\draw (1) to (1a);
\draw (2) to (2a);
\draw (2) to (2b);
\draw (0) to [bend right] (3);
\draw (3) to [bend right] (0);
\draw (3) to (3a);
\end{scope}

\begin{scope}[xshift=6.5cm,yshift=-16.5cm] 
\node (caption) at (-0.8,0.8) [anchor=west]
 {$C_{\gamma,j}^t$ for $\gamma\in\{\init,\zero(j)\}$:};

\node (1) at (0,-1.2) [prob,fill=black!20] {};
\node (1a) at (0,-0.2) [end,label=above:{$0,1,A_j^t,A_j^{1-t}$}] {};
\node (2) at (1.5,-1.2) [prob] {};
\node (2a) at (1.5,-0.2) [end,label=above:{$0,1,A_j^t,B_j^{1-t}$}] {};
\node (2b) at (3,-1.2) [end,label=above:{$0,1,B_j^t,B_j^{1-t}$}] {};

\draw (-0.7,-1.2) to (1);
\draw (1) to (2);
\draw (1) to (1a);
\draw (2) to (2a);
\draw (2) to (2b);
\end{scope}

\end{tikzpicture}
\caption{\label{fig:simulation}Simulating a two-counter machine}
\end{figure}

Let $\Gamma'\coloneq\Gamma\cup\{\init\}$. For each
$q\in Q$, each $\gamma\in\Gamma'$, each $j\in\{1,2\}$ and each
$t\in\{0,1\}$, the game~$\calG$ contains the gadgets $S_{\gamma,q}^t$,
$I_q^t$ and~$C_{\gamma,j}^t$, which are depicted in
\cref{fig:simulation}.
For better readability, terminal vertices
are depicted as squares; the label indicates which players win.
The initial vertex of~$\calG$ is~$v_0\coloneq v_{\init,q_0}^0$.
Note that in the gadget~$S_{\gamma,q}^t$, each of the players $A_j^t$
and~$B_j^t$ may \emph{quit} the game, which
gives her a payoff of $\frac{1}{3}$ or~$\frac{1}{6}$,
respectively, but payoff~$0$ to players $0$ and~$1$.

It will turn out that \pl1 will play a pure strategy in any Nash equilibrium
of $(\calG,v_0)$ where \pl0 wins almost surely, except possibly for histories
that are not consistent with the equilibrium. Formally, we say that a strategy
profile~$\vec{\sigma}$ of~$(\calG,v_0)$ is
 \index{strategy profile!safe}%
\emph{safe} if for all histories~$xv$
consistent with~$\vec{\sigma}$ and ending in a vertex $v\in I_q^t$ there exists
$w\in V$ with $\sigma_1(w\mid xv)=1$.

For any safe strategy profile~$\vec{\sigma}$ of~$\calG$ where \pl0 wins
almost surely, let $x_0v_0\prec x_1v_1\prec x_2v_2\prec\cdots$
(where $x_i\in V^*$, $v_i\in V$ and $x_0=\epsilon$) be the unique sequence
containing all histories~$xv$ of~$(\calG,v_0)$ that are consistent
with~$\vec{\sigma}$ and end in a vertex~$v$ of the form $v=v_{\gamma,q}^t$.
This sequence is infinite because \pl0 wins almost surely.
Additionally, let $q_0,q_1,\dots$ and
$\gamma_0,\gamma_1,\dots$ be the corresponding sequences of states and
instructions, respectively,
\ie $v_n=v_{\gamma_n,q_n}^0$ or $v_n=v_{\gamma_n,q_n}^1$ for all $n\in\bbN$.
For each $j\in\{1,2\}$ and $n\in\bbN$, we~set:
\begin{align*}
a_j^n &\coloneq\Prob_{v_0}^{\vec{\sigma}}(\text{\pl{A_j^{n\bmod 2}} wins}\mid
 x_n v_n\cdot V^\omega)\,; \\
b_j^n &\coloneq\Prob_{v_0}^{\vec{\sigma}}(\text{\pl{B_j^{n\bmod 2}} wins}\mid
 x_n v_n\cdot V^\omega)\,.
\end{align*}
Note that at every terminal vertex of the counter gadgets $C_{\gamma,j}^t$
and~$C_{\gamma,j}^{1-t}$ either \pl{A_j^t} or \pl{B_j^t} wins. For each~$j$,
the conditional probability that, given the history~$x_n v_n$, we reach
such a vertex is $\sum_{k\in\bbN} 1/2^k\cdot\frac{1}{4}=\frac{1}{2}$.
Hence, $a_j^n=\frac{1}{2}-b_j^n$ for all $n\in\bbN$.
We say that $\vec{\sigma}$~is \emph{stable} if $a_j^n=\frac{1}{3}$ or,
equivalently, $b_j^n=\frac{1}{6}$ for each $j\in\{1,2\}$ and for all
$n\in\bbN$.

Finally, for each $j\in\{1,2\}$ and $n\in\bbN$, we define a
number~$c_j^n\in [0,1]$ as follows: After the history $x_n v_n$, with
probability~$\frac{1}{4}$ the play enters the counter
gadget~$C_{\gamma_n,j}^{n\bmod 2}$. The number~$c_j^n$ is defined as the
probability of subsequently reaching a grey-coloured vertex. Note
that, by the construction of~$\calG$, it holds that $c_j^n=1$
if $\gamma_n=\zero(j)$ or $\gamma_n=\init$;
in~particular, $c_1^0=c_2^0=1$.

\begin{lemma}\label{lemma:reduction}
Let $\vec{\sigma}$ be a safe strategy profile of $(\calG,v_0)$ in
which \pl0 wins almost surely. Then $\vec{\sigma}$~is stable \iff
\begin{equation}\label{eq:counter-update}
c_j^{n+1}=\begin{cases}
  \frac{1}{2}\cdot c_j^n & \text{if $\gamma_{n+1}=\inc(j)$,} \\
  2\cdot c_j^n & \text{if $\gamma_{n+1}=\dec(j)$,} \\
  c_j^n=1 & \text{if $\gamma_{n+1}=\zero(j)$,} \\
  c_j^n & \text{otherwise,}
  \end{cases}
\end{equation}
for each $j\in\{1,2\}$ and $n\in\bbN$.
\end{lemma}

To prove the lemma, consider a safe strategy
profile~$\vec{\sigma}$ of~$(\calG,v_0)$ in which \pl0 wins almost surely.
For each $j\in\{1,2\}$ and $n\in\bbN$, set
\[
p_j^n\coloneq\Prob_{v_0}^{\vec{\sigma}}(\text{\pl{A_j^{n\bmod 2}} wins}\mid
 x_n v_n\cdot V^\omega\setminus x_{n+2} v_{n+2}\cdot V^\omega)\,.
\]
The following claim relates the numbers $a_j^n$ and~$p_j^n$.

\begin{claim}
Let $j\in\{1,2\}$. Then $a_j^n=\frac{1}{3}$ for all $n\in\bbN$ \iff
$p_j^n=\frac{1}{4}$ for all $n\in\bbN$.
\end{claim}

\begin{proof}
($\Rightarrow)$ Assume that $a_j^n=\frac{1}{3}$ for all $n\in\bbN$.
We have $a_j^n=p_j^n+\frac{1}{4}\cdot a_j^{n+2}$ and therefore
$\frac{1}{3}=p_j^n+\frac{1}{12}$ for all $n\in\bbN$. Hence,
$p_j^n=\frac{1}{4}$ for all $n\in\bbN$.

($\Leftarrow$) Assume that $p_j^n=\frac{1}{4}$ for all $n\in\bbN$.
Since $a_j^n=p_j^n+\frac{1}{4}\cdot a_j^{n+2}$ for all $n\in\bbN$,
the numbers~$a_j^n$ must satisfy the following recurrence: $a_j^{n+2}=
4a_j^n-1$. Since all the numbers~$a_j^n$ are
probabilities, $0\leq a_j^n\leq 1$ for all $n\in\bbN$. It is easy
to see that the only values for $a_j^0$ and~$a_j^1$
such that $0\leq a_j^n\leq 1$ for all $n\in\bbN$ are $a_j^0=a_j^1=
\frac{1}{3}$. But this implies that $a_j^n=\frac{1}{3}$ for all
$n\in\bbN$.
\end{proof}

\begin{proof}[Proof of \cref{lemma:reduction}]
By the previous claim, we only need to show that $p_j^n=\frac{1}{4}$
\iff \cref{eq:counter-update} holds.
Let $j\in\{1,2\}$, $n\in\bbN$ and $t=n\bmod 2$. The probability~$p_j^n$ can be
expressed as the sum of the probability that the play reaches a terminal vertex
that is winning for \pl{A_j^t} inside~$C_{\gamma_n,j}^t$ and the
probability that the play reaches such a vertex
inside~$C_{\gamma_{n+1},j}^{1-t}$.
The first probability does not depend on~$\gamma_n$, but the second depends
on~$\gamma_{n+1}$. Let us consider the case that $\gamma_{n+1}=\inc(j)$.
In~this case,
\[p_j^n=\tfrac{1}{4}\cdot\big(1-\tfrac{1}{4}\cdot c_j^n\big)
+\tfrac{1}{8}\cdot c_j^{n+1}
=\tfrac{1}{4}-\tfrac{1}{16}\cdot c_j^n+\tfrac{1}{8}\cdot c_j^{n+1}\,.\]
Obviously, this sum is equal to~$\frac{1}{4}$ \iff
$c_j^{n+1}=\frac{1}{2}\cdot c_j^n$. For any other value
of~$\gamma_{n+1}$, the argumentation is similar.
\end{proof}

\noindent
To establish the reduction, we need to show that the following statements
are equivalent:
\begin{enumerate}[(1)]
 \item the computation of~$\calM$ is infinite;
 \item $(\calG,v_0)$ has a pure Nash equilibrium in which \pl0 wins
almost surely;
 \item $(\calG,v_0)$ has a Nash equilibrium in which \pl0 wins almost surely.
\end{enumerate}

(\infer{1}{2}) Assume that the computation $\rho=\rho(0)\rho(1)\dots$
of~$\calM$ is infinite. We define a pure strategy profile~$\vec{\sigma}$
as follows: (1)~For a history that ends at the unique vertex
$v\in C_{\gamma,j}^t$ controlled by \pl0 after visiting a vertex of
the form $v_{\gamma',q}^t$ or~$v_{\gamma',q}^{1-t}$ exactly $n>0$ times and
$v$ exactly $k\geq 0$ times, \pl0 plays to the grey-coloured successor vertex if
$k$~is greater than or equal to the value of counter~$j$ in configuration
$\rho(n-1)$; otherwise, \pl0 plays to the other successor vertex.
(2)~For a history that ends in one of the instruction
gadgets~$I_q^t$ for $\delta(q)=\{(\zero(j),q_1),(\dec(j),q_2)\}$
after visiting a vertex of the form $v_{\gamma,q'}^t$
or~$v_{\gamma,q'}^{1-t}$ exactly $n>0$ times, \pl1 plays
to~$S_{\zero(j),q_1}^{1-t}$ if the value
of counter~$j$ in configuration~$\rho(n-1)$ is zero and
to~$S_{\dec(j),q_2}^{1-t}$ if the value of counter~$j$ in
configuration~$\rho(n-1)$ is not zero.
(3)~Any other player's pure strategy is defined as follows: after a history
ending in~$S_{\gamma,q}^t$, the strategy prescribes to quit the game \iff
the history is not compatible with~$\rho$ (\ie if the corresponding sequence
of instructions does not match~$\rho$).

Note that the resulting strategy profile~$\vec{\sigma}$ is safe.
Moreover, since the players follows the computation of~$\calM$, a terminal
vertex inside one of the counter gadgets~$C_{\gamma,j}^t$ is reached with
probability~$1$ in~$\vec{\sigma}$. Hence, \pl0 wins almost surely.
Moreover, by the definition of $\vec{\sigma}$, \cref{eq:counter-update} holds,
and we can conclude from \cref{lemma:reduction} that $\vec{\sigma}$~is stable.
We claim that $\vec{\sigma}$~is, in fact, a Nash equilibrium of
$(\calG,v_0)$: It is obvious that \pl0 cannot improve her payoff.
If \pl1 deviates, we reach a
history that is not compatible with~$\rho$. Hence, player $A_1^0$ or~$A_2^0$
will quit the game, which ensures that \pl1 will not receive a
higher payoff. Finally, since $\vec{\sigma}$~is stable, none of the players
$A_j^t$ or~$B_j^t$ can improve her payoff.

(\infer{2}{3}) Trivial.

(\infer{3}{1}) Assume that $\vec{\sigma}$~is a Nash equilibrium of
$(\calG,v_0)$ in which \pl0 wins almost surely. In order to apply
\cref{lemma:reduction}, we first prove that $\vec{\sigma}$~is safe.
By contradiction, assume that there exists a history~$xv$ consistent
with~$\vec{\sigma}$ and ending in a vertex $v\in I_q^t$ such that
$\sigma_1(xv)$~assigns positive probability to two distinct successor
vertices. Hence, $\delta(q)=\{(\zero(j),q_1),(\dec(j),q_2)\}$ for some
$j\in\{1,2\}$ and $q_1,q_2\in Q$. By our assumption that there are no
consecutive zero tests and since \pl0 wins almost surely,
\begin{alignat*}{2}
& \Prob_{v_0}^{\vec{\sigma}}(\text{\pl1 wins}\mid
 xv\cdot v_{\zero(j),q_1}^{1-t}\cdot V^\omega)
&&\geq \tfrac{1}{4}\,, \\
\shortintertext{but}
& \Prob_{v_0}^{\vec{\sigma}}(\text{\pl1 wins}\mid
 xv\cdot v_{\dec(j),q_2}^{1-t}\cdot V^\omega)
&&\leq \tfrac{1}{6}\,.
\end{alignat*}
Hence, \pl1 could improve her payoff by playing to~$v_{\zero(j),q_1}^{1-t}$
with probability~$1$, a~contradiction to $\vec{\sigma}$~being a Nash
equilibrium.

To apply \cref{lemma:reduction} and obtain \cref{eq:counter-update}, it
remains to be shown that $\vec{\sigma}$~is stable. In~order to derive a
contradiction, assume that there exists $j\in\{1,2\}$ and $n\in\bbN$ such that
either $a_j^n<\frac{1}{3}$ or $a_j^n>\frac{1}{3}$ (\ie
$b_j^n<\frac{1}{6}$). In the former case, \pl{A_j^{n\bmod 2}} could
improve her payoff by quitting the game after history~$x_n v_n$, while in the
latter case, \pl{B_j^{n\bmod 2}} could improve her payoff by quitting the game,
again a contradiction to $\vec{\sigma}$~being a Nash equilibrium.

From $c_j^0=1$ and \cref{eq:counter-update}, it follows that
each~$c_j^n$ is of the form $c_j^n=1/2^i$ where $i\in\bbN$. We denote
by~$i_j^n$ the unique number~$i$ such that $c_j^n=1/2^i$ and set
$\rho(n)=(q_n,i_1^n,i_2^n)$ for each $n\in\bbN$. We claim that
$\rho\coloneq\rho(0)\rho(1)\dots$ is in fact the computation of~$\calM$. In
particular, this computation is infinite. It suffices to verify the following
two properties:
\begin{iteMize}{$-$}
 \item $\rho(0)=(q_0,0,0)$;
 \item $\rho(n)\vdash\rho(n+1)$ for all $n\in\bbN$.
\end{iteMize}
The first property is immediate. To prove the second property, let
$\rho(n)=(q,i_1,i_2)$ and $\rho(n+1)=(q',i_1',i_2')$. Hence, $v_n$~lies
inside~$S_{\gamma,q}^t$, and $v_{n+1}$~lies inside~$S_{\gamma',q'}^{1-t}$ for
suitable $\gamma,\gamma'$ and $t=n\bmod 2$.
We only prove the claim for $\delta(q)=\{(\zero(1),q_1),(\dec(1),q_2)\}$; the
other cases are similar. Note that, by the construction of the
gadget~$I_q^t$, it must be the case that either $q'=q_1$ and
$\gamma'=\zero(1)$, or $q'=q_2$ and $\gamma'=\dec(1)$. By
\cref{eq:counter-update}, if $\gamma'=\zero(1)$, then $i_1'=i_1=0$ and
$i_2'=i_2$, and if $\gamma'=\dec(1)$, then $i_1'=i_1-1$ and $i_2'=i_2$. This
implies $\rho(n)\vdash\rho(n+1)$: on the one hand, if $i_1=0$, then
$i_1'\neq i_1-1$, which implies $\gamma'\neq\dec(1)$ and thus
$\gamma'=\zero(1)$, $q'=q_1$ and $i_1'=i_1=0$; on the other hand, if $i_1>0$,
then $\gamma'\neq\zero(1)$ and thus $\gamma'=\dec(1)$, $q'=q_2$ and
$i_1'=i_1-1$.\qed

\begin{remark}
For the problem \PureNE, we can strengthen
\cref{thm:undecidability} slightly by showing undecidability already for
9-player~SSMGs. This~can be achieved by merging \pl0 and \pl1
in the game described in the proof of \cref{thm:undecidability}.
\end{remark}

\begin{remark}\label{remark:pctl}
The proof of \cref{thm:undecidability} can also be viewed as a proof for
the undecidability of a problem about the logic
PCTL (\emph{probabilistic computation tree logic}), introduced by
\citet{HanssonJ94}. PCTL is evaluated over labelled Markov chains and replaces
the universal and existential path quantifiers of CTL by a family of
probabilistic quantifiers~$\P^{\sim x}$, where $\sim$~is a comparison
operator and ${x\in [0,1]}$ is a rational probability. For example, the
formula $\P^{=1/2}\F\,Q$ holds in state~$v$ if (and only if) the probability
of reaching a state labelled with~$Q$ from~$v$ equals~$\frac{1}{2}$.

\Citet{BrazdilBFK06} proved the
undecidability of the following problem: given a labelled Markov decision
process~$(\calG,v_0)$ and a PCTL formula~$\phi$, decide whether the controller
has a strategy~$\sigma$ such that the Markov chain~$(\calG^{\sigma},v_0)$ is a
model of~$\phi$. We can prove a stronger result, namely that there
exists a \emph{fixed} PCTL formula~$\phi$, which only contains the
quantifiers $\P^{=x}\F$ and $\P^{=x}\G$, for which the problem is
undecidable. It suffices to add propositions $A_1^0$, $A_1^1$, $A_2^0$,
$A_2^1$, $Q$, $Q_1$, $Q_2$, $T$, $Z_0$ and~$Z_1$ according to the following
rules:
\begin{enumerate}[(1)]
\item if $v$~is a terminal vertex that is winning for
player~$A\in\{A_1^0,A_1^1,A_2^0,A_2^1\}$, then label~$v$ with~$A$;

\item If $v$~is controlled by \pl0 and $\abs{v\Delta}=2$, then label~$v$
with~$Q$ and label one of its successors with~$Q_1$ and the other
with~$Q_2$.

\item if $v$~is a terminal vertex that is winning for \pl0, then label~$v$
with~$T$;

\item if $v=v_{\gamma,q}^0$, then label~$v$ with~$Z_0$; if
$v=v_{\gamma,q}^1$, then label~$v$ with~$Z_1$.
\end{enumerate}
To obtain an MDP, we make all non-stochastic vertices controlled by \pl0.
Finally, the PCTL formula for which we prove undecidability is
\[\P^{=1}\F\,T\wedge\bigwedge_{\mathmakebox[0.5cm][c]{t=0,1}}
\P^{=1}\G\,\Big(Z_t\to\bigwedge_{\mathmakebox[0.5cm][c]{j=1,2}}
\P^{=1/3}\F\,A_j^t\Big)\wedge \P^{=1}\G\,\Big(Q\to
\bigvee_{\mathmakebox[0.5cm][c]{i=0,1}}\P^{=1}\F\,Q_i\Big)\,.\]
The first part of the formula states that \pl0 wins almost surely, the
second part requires the strategy to be stable, and the last part of the
formula requires the strategy to be safe.
\end{remark}

\subsection{Finite-state equilibria}
\label{section:finite-state}

We can use the construction in the proof of \cref{thm:undecidability} to show
that Nash equilibria may require infinite memory, even if
we are only interested in whether a player wins with probability~$0$ or~$1$.

\begin{proposition}\label{prop:inf-mem}
There exists a finite SSMG that has a pure Nash equilibrium where \pl0
wins almost surely, but that has no finite-state Nash equilibrium where
\pl0 wins with positive probability.
\end{proposition}
\begin{proof}
Consider the game~$(\calG,v_0)$ constructed in the proof of
\cref{thm:undecidability} for the machine~$\calM$ with the single
transition~$(q_0,\inc(1),q_0)$. We modify this game by adding a new initial
vertex~$v_1$ which is controlled by a new player, \pl2, and from where she
can either move to~$v_0$ or to a new terminal vertex where she receives
payoff~$1$ and every other player receives payoff~$0$. Additionally, \pl2 wins
at every terminal vertex of the game~$\calG$ that is winning for \pl0. Let
us denote the modified game by $\calG'$.

Since the computation of~$\calM$ is infinite, the game~$(\calG,v_0)$ has a pure
Nash equilibrium where \pl0 wins almost surely. This equilibrium
induces a pure Nash equilibrium of~$(\calG',v_1)$ where both \pl0 and \pl2
win almost surely.

Now assume that there exists a finite-state Nash equilibrium of~$(\calG',v_1)$
where \pl0 wins with positive probability. Such an equilibrium induces a
finite-state Nash equilibrium~$\vec{\sigma}$ of~$(\calG,v_0)$ where \pl2, and
thus also \pl0, wins almost surely; otherwise, \pl2 would prefer to play
from~$v_1$ to the new terminal vertex. Using the same notation as in the proof
of \cref{thm:undecidability}, it follows from \cref{lemma:reduction} that
$c_1^n=1/2^n$ for each $n\in\bbN$. But this is impossible if
$\vec{\sigma}$~is a finite-state strategy profile.
\end{proof}

\Cref{prop:inf-mem,prop:pure-not-enough} imply that the decision problems \NE,
\FinNE, \PureNE and \PureFinNE are pairwise distinct.
Another way to see that \PureNE and \PureFinNE are distinct is to observe
that \PureFinNE is recursively enumerable: to decide whether an
SMG $(\calG,v_0)$ has a pure finite-state Nash equilibrium with payoff
$\geq\vec{x}$ and $\leq\vec{y}$, one can just enumerate all possible pure
finite-state profiles~$\vec{\sigma}$ and check for each of them whether
it constitutes a Nash equilibrium with the desired properties by analysing the
finite Markov chain~$\calG^{\vec{\sigma}}$ and the finite
MDPs~$\calG^{\vec{\sigma}_{-i}}$. Hence, to prove that \PureFinNE is
undecidable, we cannot reduce from the non-halting problem. Instead, we
reduce from the halting problem (which is recursively enumerable
itself). The same reduction proves that \FinNE is undecidable.

\begin{theorem}\label{thm:undecidability-fin-ne}
\FinNE and \PureFinNE are undecidable, even for 14-player SSMGs.
\end{theorem}

\begin{proof}[Proof sketch]
The construction is similar to the one for proving the undecidability of \NE.
Given a two-counter machine~$\calM$, we modify the SSMG~$\calG$ constructed in
the proof of \cref{thm:undecidability} by adding another counter (together
with four more players for checking whether the counter is updated correctly)
that has to be incremented in each step. Moreover, the gadget~$I_{\gamma,q}$
for $\delta(q)=\emptyset$ is replaced by the gadget shown in
\cref{fig:simulation-2}, and a new instruction~$\halt$ is added, together with
a suitable gadget~$C_{\halt,j}^t$, also depicted in \cref{fig:simulation-2}.
\begin{figure}
\begin{tikzpicture}[x=1.6cm,y=1.4cm,->]

\begin{scope} 
\node (caption) at (-0.8,0.9) [anchor=west]
 {$I_q^t$ for $\delta(q)=\emptyset$:};

\node (10) at (0,-1.2) [prob] {};
\node (11) at (0,-0.1) {$C_{\halt,1}^{1-t}$};
\node (12) at (0,-2.3) {$C_{\halt,2}^{1-t}$};
\node (13) at (1.1,-1.2) [prob] {};
\node (13a) at (1.1,-0.2) [end,label=above:{$0,A_1^{1-t},A_2^{1-t}$}] {};
\node (14) at (2.1,-1.2) [prob] {};
\node (14a) at (2.9,-0.4) [end,label={[text width=1.7cm,text centered]above:%
 {$0,A_1^t,A_2^t,$ $B_1^{1-t},B_2^{1-t}$}}] {};
\node (14b) at (2.9,-2) [end,label={[text width=1.7cm,text centered]below:%
 {$0,B_1^t,B_2^t,$ $A_1^{1-t},A_2^{1-t}$}}] {};

\draw (-0.7,-1.2) to (10);
\draw (10) to node[above] {$\frac{1}{2}$} (13);
\draw (10) to node[right] {$\frac{1}{4}$} (11);
\draw (10) to node[right] {$\frac{1}{4}$} (12);
\draw (13) to (14); \draw (13) to (13a);
\draw (14) to node[above left] {$\frac{2}{3}$} (14a);
\draw (14) to node[below left] {$\frac{1}{3}$} (14b);
\end{scope}

\begin{scope}[xshift=7.5cm,yshift=0cm] 
\node (caption) at (-0.8,0.9) [anchor=west]
 {$C_{\gamma,j}^t$ for $\gamma=\halt$:};

\node (0) at (0,-2.2) [play,label=below:$0$] {};
\node (1) at (1,-2.2) [prob,fill=black!20] {};
\node (1a) at (1,-1.2) [end,label=above:{$0,A_j^{1-t}$}] {};
\node (2) at (2,-2.2) [end,label=above:{$0,B_j^{1-t}$}] {};
\node (3) at (0,-1.2) [prob] {};
\node (3a) at (0,-0.2) [end,label=above:{$0,B_j^{1-t}$}] {};

\draw (-0.7,-2.2) to (0);
\draw (0) to (1);
\draw (1) to (2);
\draw (1) to (1a);
\draw (0) to [bend right] (3);
\draw (3) to [bend right] (0);
\draw (3) to (3a);
\end{scope}
\end{tikzpicture}
\caption{\label{fig:simulation-2}Reducing from the halting problem}
\end{figure}
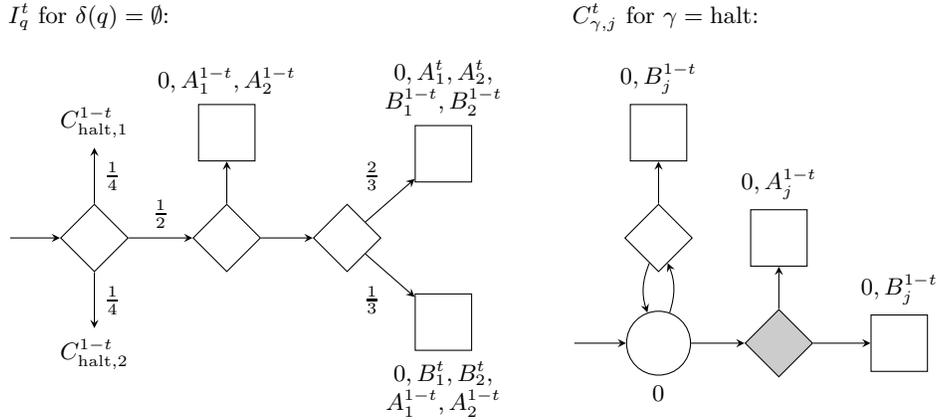
Let us denote the new game by~$\calG'$. If $\calM$~does not halt, any Nash
equilibrium of $(\calG',v_0)$ where \pl0 wins with probability~$1$ needs
infinite memory: to win almost surely, \pl0 must follow the computation
of~$\calM$ and increment the new counter at each step, which requires infinite
memory. On the other hand, if $\calM$~halts, there exists a pure finite-state
Nash equilibrium of $(\calG',v_0)$ in which \pl0 wins almost surely.
(The arguments for the existence of such an equilibrium are the same as in the
proof of \cref{thm:undecidability}; since $\calM$~halts, the equilibrium can be
implemented with finite memory).
\end{proof}

\begin{remark}
With the same reasoning as for \PureNE, we can eliminate one
player in the reduction for \PureFinNE. Hence, this problem
is already undecidable for SSMGs with 13~players.
\end{remark}

\section{The strictly qualitative fragment}
\label{sect:qualitative}

In this section, we prove that the fragment of \NE that arises from restricting
the thresholds to be the same binary payoff (\ie each entry is either $0$
or~$1$) is decidable for games with \omg-regular objectives; we denote this
problem by \EQualNE. Formally, \EQualNE is defined as follows:
\begin{quote}
Given an SMG $(\calG,v_0)$ and $\vec{x}\in\{0,1\}^\Pi$, decide whether
$(\calG,v_0)$~has a Nash equilibrium with payoff~$\vec{x}$.
\end{quote}
To prove decidability, we first characterise the existence of a
Nash equilibrium with a binary payoff in games with prefix-independent
objectives.

\subsection{Characterisation of existence}
\label{sect:nash-qual-characterisation}

Given an SMG~$\calG$ and a \pli, we denote by~$W_i$ the set of all
vertices $v\in V$ such that $\val_i^\calG(v)>0$.

\begin{proposition}\label{prop:qual-nash}
Let $(\calG,v_0)$ be any SMG with prefix-independent objectives,
and let $\vec{x}\in\{0,1\}^\Pi$. Then the following statements
are equivalent:
\begin{enumerate}[(1)]
 \item $(\calG,v_0)$ has a Nash equilibrium with payoff~$\vec{x}$;
 \item there exists a strategy profile~$\vec{\sigma}$ of $(\calG,v_0)$ with
payoff~$\vec{x}$ such that
$\Prob^{\vec{\sigma}}_{v_0}(\Reach(W_i))=0$ for each \pli with $x_i=0$;
 \item there exists a pure strategy profile~$\vec{\sigma}$ of $(\calG,v_0)$
with payoff~$\vec{x}$ such that $\Prob^{\vec{\sigma}}_{v_0}
(\Reach(W_i))=0$ for each \pli with $x_i=0$;
 \item $(\calG,v_0)$ has a pure Nash equilibrium with payoff~$\vec{x}$.
\end{enumerate}
If additionally all objectives are \omg-regular, then each of the
above statements is equivalent to each of the following statements:
\begin{enumerate}[(1)]\setcounter{enumi}{4}\raggedright
 \item There exists a pure finite-state strategy profile~$\vec{\sigma}$ with
payoff~$\vec{x}$ such that $\Prob^{\vec{\sigma}}_{v_0}
(\Reach(W_i))=0$ for each \pli with $x_i=0$.
 \item $(\calG,v_0)$ has a pure finite-state Nash equilibrium with
payoff~$\vec{x}$.
\end{enumerate}
\end{proposition}

\begin{proof}
(\infer{1}{2})  Let $\vec{\sigma}$ be a Nash equilibrium of $(\calG,v_0)$ with
payoff~$\vec{x}$. We claim that $\vec{\sigma}$~is already the strategy profile
we are looking for: $\Prob^{\vec{\sigma}}_{v_0}(\Reach(W_i))=0$ for each
\pli with $x_i=0$. Let $i\in\Pi$ be a player with $x_i=0$. By
\cref{prop:nash-value} and since $\Win_i$~is prefix-independent, we have
$0=\Prob^{\vec{\sigma}}_{v_0}(\Win_i\mid xv\cdot V^\omega)\geq
\val_i^{\calG}(v)$
for all histories~$xv$ that are consistent with~$\vec{\sigma}$. Hence,
$v\in V\setminus W_i$ for all such histories~$xv$, and
$\Prob^{\vec{\sigma}}_{v_0}(\Reach(W_i))=0$.

(\infer{2}{3}) Let $\vec{\sigma}$ be a strategy profile
of~$(\calG,v_0)$ with payoff~$\vec{x}$ such that
$\Prob^{\vec{\sigma}}_{v_0}(\Reach(W_i))=0$ for each \pli with $x_i=0$.
Consider the MDP~$\calM$ that is obtained from~$\calG$ by removing all
vertices~$v\in V$ such that $v\in W_i$ for some \pli with $x_i=0$, merging
all players into one, and imposing the objective
\[\Win=\bigcap_{\substack{i\in\Pi \\ x_i=1}} \Win_i\cap
  \bigcap_{\substack{i\in\Pi \\ x_i=0}} C^\omega\setminus\Win_i\,.\]
The MDP~$\calM$ is well-defined since its domain is a subarena of~$\calG$.
Moreover, the~value $\val^{\calM}(v_0)$ of $\calM$ from~$v_0$ equals~$1$
because the strategy profile~$\vec{\sigma}$ induces a
strategy~$\sigma$ in~$\calM$ satisfying $\Prob_{v_0}^\sigma(\Win)=1$.
Since each of the objectives~$\Win_i$ is prefix-independent, so is the
objective~$\Win$.
Hence, by \cref{thm:optimal-strategies}, $(\calM,v_0)$ admits an optimal pure
strategy~$\tau$. Since $\val^{\calM}(v_0)=1$, we have
$\Prob_{v_0}^\tau(\Win)=1$, and $\tau$~induces
a pure strategy profile of $(\calG,v_0)$ with the desired properties.

(\infer{3}{4}) Consider any pure strategy profile~$\vec{\sigma}$ of
$(\calG,v_0)$ with payoff~$\vec{x}$ such that\linebreak
$\Prob^{\vec{\sigma}}_{v_0}(\Reach(W_i))=0$ for each \pli with
$x_i=0$. We show that $\vec{\sigma}$~is favourable:
$\Prob^{\vec{\sigma}}_{v_0}(\Win_i\mid xv\cdot V^\omega)\geq
\val_i^\calG(v)$ for each \pli and each history~$xv$ of $(\calG,v_0)$ that is
consistent with~$\vec{\sigma}$. There are two cases: If $x_i=1$, then
$\Prob^{\vec{\sigma}}_{v_0}(\Win_i\mid xv\cdot V^\omega)=1$ for all
histories~$xv$ consistent with~$\vec{\sigma}$, and the inequality holds.
Otherwise, $x_i=0$ and $\Prob^{\vec{\sigma}}_{v_0}(\Reach(W_i))=0$.
Hence, $\val_i^\calG(v)=0$ for all histories~$xv$ consistent
with~$\vec{\sigma}$, and the inequality holds as~well.
Now, by \cref{lemma:nash-threat}, we can extend~$\vec{\sigma}$ to a
pure Nash equilibrium with payoff~$\vec{x}$.

(\infer{4}{1}) Trivial.

Under the additional assumption that all objectives are
\omg-regular, the implications (\infer{2}{5}) and
(\infer{5}{6}) are proven analogously (using 
\cref{lemma:nash-threat-regular} instead of \cref{lemma:nash-threat}); the
implication (\infer{6}{1}) is trivial.
\end{proof}

As an immediate consequence of \cref{prop:qual-nash}, we can conclude that
pure finite-state strategies are as powerful as arbitrary randomised strategies
as far as the existence of Nash equilibria with binary payoffs in finite SMGs
with \omg-regular objectives is concerned.

\begin{corollary}\label{cor:nash-finite-state}
Let $(\calG,v_0)$ be a finite SMG with \omg-regular objectives, and
let $\vec{x}\in\{0,1\}^\Pi$. There exists a Nash equilibrium of $(\calG,v_0)$
with payoff~$\vec{x}$ \iff there exists a pure finite-state Nash equilibrium of
$(\calG,v_0)$ with payoff~$\vec{x}$.
\end{corollary}

\begin{proof}
The claim follows from \cref{prop:qual-nash} and the fact that every SMG
with \omg-regular objectives can be reduced to one with parity objectives
(using finite memory).
\end{proof}

\subsection{Computational Complexity}

We can now give an algorithm that decides \EQualNE for SMGs with Muller
objectives. The algorithm relies on \cref{prop:qual-nash}, which allows us to
reduce \EQualNE to an MDP problem.

Formally, given a Muller SMG
$\calG=(\Pi,V,(V_i)_{i\in\Pi},\Delta,\chi,(\calF_i)_{i\in\Pi})$ and a binary
payoff $\vec{x}=(x_i)_{i\in\Pi}$, we~define the Markov decision
process~$\calG(\vec{x})$ as follows: Let
$Z\subseteq V$ be the set of all vertices~$v$ such that
$\val_i^\calG(v)=0$ for each \pli with $x_i=0$; the~set of vertices of
$\calG(\vec{x})$ is precisely the set~$Z$, with the set of vertices
controlled by \pl0
being $Z_0\coloneq\bigcup_{i\in\Pi} (V_i\cap Z)$; if $Z=\emptyset$, we~define
$\calG(\vec{x})$ to be a trivial MDP with the empty set as its objective.
The~transition relation
of~$\calG(\vec{x})$ is the restriction of~$\Delta$ to transitions between
$Z$-states. Note~that the transition relation of~$\calG(\vec{x})$ is
well-defined since $Z$~is a subarena of $\calG$. Finally, the single
objective in~$\calG(\vec{x})$ is $\Reach(T)$ where $T\subseteq Z$ is the
union of all end components $U\subseteq Z$ with payoff~$\vec{x}$.

\begin{lemma}\label{lemma:nash-muller}
Let $(\calG,v_0)$ be a finite Muller SMG, and let $\vec{x}
\in\{0,1\}^\Pi$. Then $(\calG,v_0)$ has a Nash equilibrium with
payoff~$\vec{x}$ \iff $\val^{\calG(\vec{x})}(v_0)=1$.
\end{lemma}

\begin{proof}
$(\Rightarrow)$ Assume that $(\calG,v_0)$ has a Nash equilibrium with
payoff~$\vec{x}$. By \cref{prop:qual-nash}, there exists a
strategy profile~$\vec{\sigma}$ of $(\calG,v_0)$ with payoff~$\vec{x}$ such
that $\Prob_{v_0}^{\vec{\sigma}}(\Reach(V\setminus Z))=0$.
We claim that $\Prob^{\vec{\sigma}}_{v_0}(\Reach(T))=1$.
Otherwise, by \cref{lemma:end-components}, there would exist an end component
$U\subseteq Z$ such that $\Prob^{\vec{\sigma}}_{v_0}(\{\pi\in V^\omega:
\Inf(\pi)=U\})>0$, and $U$~is either not winning for some \pli with
$x_i=1$ or it is winning for some \pli with $x_i=0$. But 
then $\vec{\sigma}$~cannot have payoff~$\vec{x}$, a contradiction.
Now, since $\Prob_{v_0}^{\vec{\sigma}}(\Reach(V\setminus Z))=0$, the
strategy profile~$\vec{\sigma}$
induces a strategy~$\sigma$ in~$\calG(\vec{x})$ such that
$\Prob^\sigma_{v_0}(X)=\Prob^{\vec{\sigma}}_{v_0}(X)$ for every Borel set
$X\subseteq Z^\omega$. In particular, $\Prob^\sigma_{v_0}(\Reach(T))=1$ and
hence $\val^{\calG(\vec{x})}(v_0)=1$.

$(\Leftarrow)$ Assume that $\val^{\calG(\vec{x})}(v_0)=1$ (in particular,
$v_0\in Z$), and let $\sigma$ be an optimal strategy in $(\calG(\vec{x}),v_0)$.
From~$\sigma$, using \cref{lemma:end-component-strategy}, we can devise a
strategy~$\sigma'$ such that $\Prob^{\sigma'}_{v_0}
(\{\pi\in V^\omega:\text{$\Inf(\pi)$ has payoff~$\vec{x}$}\})=1$.
Finally, $\sigma'$~can be extended to
a strategy profile~$\vec{\sigma}$ of $(\calG,v_0)$ with payoff~$\vec{x}$ such
that $\Prob_{v_0}^{\vec{\sigma}}({\Reach(V\setminus Z)})=0$. By 
\cref{prop:qual-nash}, this implies that $(\calG,v_0)$ has a Nash equilibrium
with payoff~$\vec{x}$.
\end{proof}

Since the values of an MDP with a reachability objective can be computed in
polynomial time, the difficult part lies in computing the MDP
$\calG(\vec{x})$ from $\calG$ and $\vec{x}$ (\ie its domain~$Z$ and
the target set~$T$). For Muller SMGs, polynomial space suffices to
achieve this. In fact, \EQualNE is \PSpace-complete for these
games.

\begin{theorem}\label{thm:qualne-muller-pspace}
\EQualNE is \PSpace-complete for Muller SMGs.
\end{theorem}

\begin{proof}
Hardness follows from \cref{thm:muller-pspace}. To prove membership in
\PSpace, we describe a polynomial-space algorithm for deciding \EQualNE on
Muller SMGs: On input~$\calG,v_0,\vec{x}$, the
algorithm starts by computing for each \pli with $x_i=0$ the set of
vertices~$v$ such that $\val_i^\calG(v)=0$, which can be done in polynomial
space by \cref{thm:muller-pspace}. The intersection of these sets is the
domain~$Z$ of the Markov decision process~$\calG(\vec{x})$. If $v_0$ is not
contained in this intersection, the algorithm immediately rejects. Otherwise,
the algorithm determines the union~$T$ of all end components with
payoff~$\vec{x}$ contained in~$Z$ by enumerating all subsets of~$Z$, one
at a time, and checking which ones are end components with payoff~$\vec{x}$.
Finally, the algorithm computes (in~polynomial time) the value
$\val^{\calG(\vec{x})}(v_0)$ of the MDP~$\calG(\vec{x})$ from~$v_0$ and
accepts if this value is~$1$. In all other cases, the algorithm rejects.
The correctness of the algorithm follows immediately from
\cref{lemma:nash-muller}.
\end{proof}

For games with Streett objectives, \EQualNE becomes \NP-complete; we start by
proving the upper bound.

\begin{theorem}\label{thm:qualne-streett-np}
\EQualNE is in \NP for Streett SMGs.
\end{theorem}

\begin{proof}
We describe a nondeterministic polynomial-time algorithm for solving \EQualNE:
On input~$\calG,v_0,\vec{x}$, the algorithm starts by guessing a subarena~$Z'
\subseteq V$ and for each \pli with $x_i=0$ a positional strategy~$\tau_i$ of
the coalition~$\Pi\setminus\{i\}$ in the coalition game~$\calG_i$.
In~the next step, the algorithm checks (in polynomial time)
whether $\val^{\tau_i}(v)=1$ for each vertex $v\in Z'$ and
each \pli with $x_i=0$. If not, the algorithm rejects immediately.
Otherwise, the algorithm proceeds by guessing (at most) $n\coloneq |V|$ subsets
$U_1,\dots,U_n\subseteq Z'$ and checks whether they are end components
with payoff~$\vec{x}$ (which can be done in polynomial time). If yes, the
algorithm sets $T'\coloneq\bigcup_{j=1}^n
U_j$ and computes (in polynomial time) the value~$\val^{\calG(\vec{x})}(v_0)$
of the MDP~$\calG(\vec{x})$ from~$v_0$ with $Z'$ substituted for~$Z$ and $T'$
substituted for~$T$. If this value equals~$1$, the algorithm accepts; otherwise,
it rejects.

It remains to be shown that the algorithm is correct: On the one hand,
if~$(\calG,v_0)$ has a Nash equilibrium with payoff~$\vec{x}$, then the run
of the algorithm where it guesses $Z'=Z$, globally optimal positional
strategies~$\tau_i$ (which exist by \cref{thm:positional-optimal}) and end
components~$U_i$ such that $T'=T$ will be accepting
since then, by \cref{lemma:nash-muller}, $\val^{\calG(\vec{x})}(v_0)=1$.
On the other hand, in any accepting run of the algorithm we have
$Z'\subseteq Z$ and $T'\subseteq T$, and the computed value cannot be
higher than $\val^{\calG(\vec{x})}(v_0)$; hence,
$\val^{\calG(\vec{x})}(v_0)=1$, and \cref{lemma:nash-muller} guarantees the
existence of a Nash equilibrium with payoff~$\vec{x}$.
\end{proof}

The matching lower bound does even hold for deterministic two-player Streett
games and was established in \cite{Ummels08}. 

\begin{theorem}\label{thm:qualne-streett-np-hard}
\EQualNE is \NP-hard for deterministic two-player Streett games.
\end{theorem}

\begin{proof}
The proof is accomplished by a variant of the proof for \NP-hardness of the
qualitative decision problem for deterministic two-player zero-sum
Rabin-Streett games \citep{EmersonJ99} and by a reduction from \SAT. Given a
Boolean formula~$\phi=C_1\wedge\dots\wedge C_m$
in conjunctive normal form, where \wlg $m\geq 1$ and each clause is nonempty,
we~construct a deterministic two-player Streett
game~$\calG$ as follows: For each clause~$C$, the game~$\calG$ has
a vertex~$C$, which is controlled by \pl0, and for each literal~$L$
occurring in~$\phi$, there is a vertex~$L$, which is controlled by \pl1. There
are edges from a clause to each literal that occurs in this clause, and from a
literal to each clause occurring in~$\phi$.
The structure of the game is depicted in \cref{fig:qualne-streett-np-hard}.
\begin{figure}
\begin{tikzpicture}[x=2cm,y=3cm,->,bend angle=10]
\node (c1) at (0.5,0) [play,label={above:$0$}] {$C_1$};
\node (d1) at (2,0) {$\cdots$};
\node (c2) at (3.5,0) [play,label={above:$0$}] {$C_m$};
\node (l1) at (0,-1) [play,label={below:$1$}] {$X_1$};
\node (l2) at (1,-1) [play,label={below:$1$}] {\tiny$\neg X_1$};
\node (d4) at (2,-1) {$\cdots$};
\node (l3) at (3,-1) [play,label={below:$1$}] {$X_n$};
\node (l4) at (4,-1) [play,label={below:$1$}] {\tiny$\neg{X_n}$};

\draw (l1) to [bend right] (c1);
\draw (l1) to (c2);
\draw (l2) to [bend right] (c1);
\draw (l2) to [bend right] (c2);
\draw (l3) to (c1);
\draw (l3) to [bend right] (c2);
\draw (l4) to [bend right] (c1);
\draw (l4) to [bend right] (c2);
\draw[dashed] (c1) to [bend right] (l1);
\draw[dashed] (c1) to [bend right] (l2);
\draw[dashed] (c1) to [bend right] (l3);
\draw[dashed] (c1) to (l4);
\draw[dashed] (c2) to [bend right] (l1);
\draw[dashed] (c2) to (l2);
\draw[dashed] (c2) to [bend right] (l3);
\draw[dashed] (c2) to [bend right] (l4);
\end{tikzpicture}
\caption{\label{fig:qualne-streett-np-hard} Reducing \SAT to \EQualNE
for games with Streett objectives}
\end{figure}
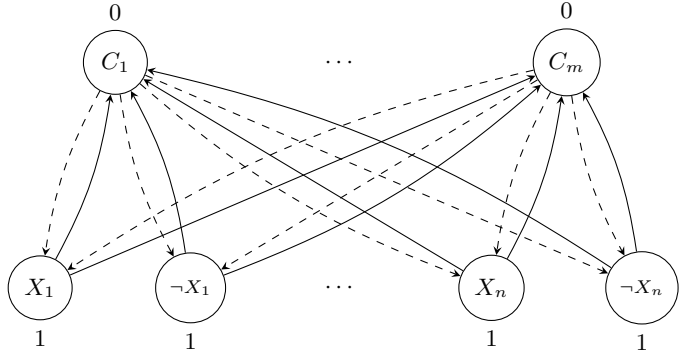
\Pl0's objective is given by the empty Streett objective, \ie she
wins every play of the game, whereas \pl1's objective consists of all Streett
pairs of the form $(\{X\},\{\neg X\})$ or $(\{\neg X\},\{X\})$, \ie she wins
if, for each variable~$X$, either $X$~and~$\neg X$ are both visited infinitely
often or neither of them is.

Clearly, $\calG$ can be constructed from~$\phi$ in polynomial time. We claim
that $\phi$~is satisfiable \iff
$(\calG,C_1)$ has a Nash equilibrium with payoff~$(1,0)$.

($\Rightarrow$) Assume that $\phi$~is satisfiable, and consider the
following positional strategy~$\sigma_0$ of \pl0: whenever the play reaches a
clause, then $\sigma_0$~plays to a literal that is mapped to true by the
satisfying assignment. This strategy ensures that for each variable~$X$ at most
one of the literals $X$ or~$\neg X$ is
visited infinitely often. Hence, $(\sigma_0,\sigma_1)$ is a Nash
equilibrium of $(\calG,C_1)$ with payoff~$(1,0)$ for every
strategy~$\sigma_1$ of \pl1.

($\Leftarrow$) Let $(\sigma_0,\sigma_1)$ be a Nash equilibrium of $(\calG,C_1)$
with payoff $(1,0)$, and assume that $\phi$~is not satisfiable.
Consider the two-player zero-sum Rabin-Streett game~$\tilde{\calG}$,
which is derived from~$\calG$ by setting \pl0's objective to the
complement of \pl1's objective.
We claim that \pl1 has a winning strategy in $(\tilde{\calG},C_1)$,
which she could use to improve her payoff in $(\calG,C_1)$, a contradiction to
$(\sigma_0,\sigma_1)$ being a Nash equilibrium. By determinacy, we only need
to show that \pl0 does not have a winning strategy.
Let $\tau$ be an optimal positional strategy of \pl0 in $(\tilde{\calG},C_1)$
(which exists by \cref{thm:positional-optimal}). Since $\phi$ is unsatisfiable,
there must exist a variable~$X$ and clauses
$C$ and~$C'$ such that $\tau(C)=X$ and $\tau(C')=\neg X$. But~\pl1
can counter this strategy by playing from~$X$ to~$C'$ and from any other
literal to $C$. Hence, $\tau$~is not winning.
\end{proof}

For games with Rabin objectives, the situation is more delicate. One might
think that, because of the duality of Rabin and Streett objectives,
\EQualNE is in \coNP for SMGs with Rabin objectives.\footnote{In fact,
\citet{UmmelsW09a} claimed that the problem is in \coNP.}
However, as we will see later, this~is rather unlikely, and we can only
show that the problem lies in the class~\PNPlog of problems solvable by a
deterministic polynomial-time algorithm that may perform a logarithmic
number of queries to an NP~oracle.
In fact, the same upper bound holds for games with a Streett
\emph{or} a Rabin objective for each player.

\begin{theorem}\label{thm:qualne-rabin-ptonp}
\EQualNE is in \PNPlog for Streett-Rabin SMGs.
\end{theorem}

\begin{proof}
Let us describe a polynomial-time algorithm performing a logarithmic number
of queries to an \NP oracle for the problem. On input $\calG,v_0,\vec{x}$,
the~algorithm starts by determining for each vertex~$v$ and each
\emph{Rabin player}~$i$ with $x_i=0$ whether $\val_i^\calG(v)=0$.
Naively implemented, this
requires a super-logarithmic number of queries to the oracle. To reduce
the number of queries, we use a neat trick, due to \citet{Hemachandra89}.
Let us denote by $R$~and~$S$ the set of players $i\in\Pi$ with $x_i=0$ who
have a Rabin, respectively a Streett objective. Instead of looping through
all pairs of a vertex and a player, we~start by determining the number~$r$ of
all pairs~$(v,i)$ such that $i\in R$ and $\val_i^\calG(v)=0$.
It is not difficult to see that this number can be computed
using binary search by performing only a logarithmic number of queries
to an \NP oracle, which we can use for deciding whether $\val_i^\calG(v)>0$
(\cref{cor:rabin-np}). Then we perform one more query; we ask whether
for each \pl{i\in R\cup S} there exists a set $Z_i\subseteq V$ as
well as sets $U_1,\dots,U_{\abs{V}}\subseteq V$ and positional strategies
$(\sigma_i)_{i\in R}$ and $(\tau_i)_{i\in S}$, where $\sigma_i$ is a
strategy of \pli and $\tau_i$ is a strategy of the
coalition~$\Pi\setminus\{i\}$ in the coalition game~$\calG_i$, with the
following properties:
\begin{enumerate}[(1)]
 \item $Z\coloneq\bigcap_{i\in R\cup S}Z_i$ is a subarena of~$\calG$
with $v_0\in Z$, and $\sum_{i\in R} |Z_i|=r$;
 \item $\val^{\sigma_i}(v)>0$ for each \pl{i\in R} and each
$v\in V\setminus Z_i$;
 \item $\val^{\tau_i}(v)=1$ for each \pl{i\in S} and each $v\in Z_i$;
 \item each~$U_j$ is an end component of $\calG\restrict Z$ with
payoff~$\vec{x}$;
 \item the value from~$v_0$ of the MDP that is obtained from~$\calG$ by
restricting to vertices inside~$Z$ and imposing the objective
$\Reach(\bigcup\{U_1,\dots,U_{\abs{V}}\})$ equals~$1$.
\end{enumerate}
This query can be decided by an \NP oracle by guessing suitable
sets and strategies and verifying (1)--(5) in polynomial time. If the answer to
the query is \emph{yes}, the algorithm accepts; otherwise it rejects.

Obviously, the algorithm runs in polynomial time. To see that the algorithm is
correct, first note that for each \pl{i\in R} the set~$Z_i$ does not only
include all $v\in V$ such that $\val_i^\calG(v)=0$, but also excludes all
other vertices. Otherwise, there would exist
a vertex $v\in Z_i$ with $\val_i^\calG(v)>0$.
But then the number of pairs $(v,i)$ with $i\in R$ and $\val_i^\calG(v)=0$
would be strictly less than~$r$, a~contradiction. Now, the correctness of
the algorithm follows with the same reasoning as in the proof of
\cref{thm:qualne-streett-np}.
\end{proof}

\begin{remark}
For a bounded number of players, \EQualNE is in \coNP for SMGs with Rabin
objectives.
\end{remark}

Regarding lower bounds for \EQualNE in SMGs with Rabin objectives, we start
by proving that the problem is \coNP-hard, even for deterministic two-player
games.

\begin{theorem}\label{thm:qualne-rabin-conp-hard}
\EQualNE is \coNP-hard for deterministic two-player Rabin games.
\end{theorem}

\begin{proof}
The proof is similar to the proof of \cref{thm:qualne-streett-np-hard} and
is accomplished by a reduction from the unsatisfiability problem for Boolean
formulae in conjunctive normal form.
Given a Boolean formula~$\phi=C_1\wedge\dots\wedge C_m$ in
conjunctive normal form, where \wlg $m\geq 1$ and each clause is nonempty,
we~construct a deterministic two-player Rabin game~$\calG$ as follows. The
arena of~$\calG$ is the same as in the proof of
\cref{thm:qualne-streett-np-hard}, depicted in
\cref{fig:qualne-streett-np-hard}. However, this~time
\pl1 wins every play of the game (her objective consists of the
single Rabin pair $(V,\emptyset)$), and \pl0's objective consists of all
\emph{Rabin} pairs of the form
$(\{X\},\{\neg X\})$ or $(\{\neg X\},\{X\})$.

Clearly, $\calG$~can be constructed from~$\phi$ in polynomial time. We claim
that the $\phi$ is unsatisfiable \iff $(\calG,C_1)$ has a Nash equilibrium with
payoff~$(0,1)$.

($\Rightarrow$) Assume that $\phi$~is unsatisfiable, and consider the
two-player zero-sum Rabin-Streett game~$\tilde{\calG}$,
which is derived from~$\calG$ by setting \pl1's objective to the
complement of \pl0's objective.
Let $\sigma_1$ be a globally optimal strategy for \pl1 in
this game. We claim that $\sigma_1$~is winning in $(\calG_0,C_1)$.
Consequently, $(\sigma_0,\sigma_1)$ is a Nash equilibrium of $(\calG,C_1)$
with payoff $(0,1)$ for every strategy~$\sigma_0$ of \pl0.
Otherwise, \pl0 would have a positional
winning strategy in $(\tilde{\calG},C_1)$. But a positional strategy~$\tau$ of
\pl0 picks for each clause a literal contained in this clause. Since $\phi$~is
unsatisfiable, there must exist a variable~$X$ and clauses $C$~and~$C'$ such
that $\tau(C)=X$ and $\tau(C')=\neg X$. \Pl1 could counter this strategy by
playing from~$X$ to~$C'$ and from any other literal to~$C$, a~contradiction.

($\Leftarrow$) Let $(\sigma_0,\sigma_1)$ be a Nash equilibrium of
$(\calG,C_1)$ with payoff $(0,1)$, and assume that $\phi$~is satisfiable.
Consider the following positional strategy~$\tau$ of \pl0:
whenever the play reaches a clause, then $\tau$~plays to
a literal that is mapped to true by the satisfying assignment.
This strategy ensures that for each variable~$X$ at most
one of the literals $X$ or~$\neg X$ is visited infinitely often. Since
the construction of~$\calG$ ensures that, under any strategy profile,
at least one literal is visited infinitely often, $\tau$~ensures a
winning play for \pl0. Hence, \pl0 can improve her payoff by playing~$\tau$
instead of~$\sigma_0$, a~contradiction to the fact that $(\sigma_0,\sigma_1)$
is a Nash equilibrium.
\end{proof}

The next result shows that \EQualNE is not only \coNP-hard for Rabin
games, but also \NP-hard.
In fact, it is even \NP-hard to decide whether in a deterministic Rabin game
there exists a play that fulfils the objective of each player.

\begin{proposition}\label{prop:rabin-np-hard}
The problem of deciding, given a deterministic Rabin game, whether there
exists a play that is won by each player is \NP-hard.
\end{proposition}

\begin{proof}
We reduce from \SAT: given a Boolean formula~$\phi=C_1\wedge\dots\wedge C_m$
in conjunctive normal form over propositional variables
$X_1,\dots,X_n$, where \wlg $m\geq 1$ and each clause is nonempty,
we show how to construct in polynomial time a deterministic $(n+1)$-player
Rabin game~$\calG$ such that $\phi$~is satisfiable \iff there exists a play
of~$\calG$ that is won by each player. The game has
vertices $C_1,\dots,C_m$ and, for each clause~$C$ and each literal
$L$ that occurs in~$C$, a vertex $(C,L)$. All vertices are controlled by \pl0.
There are edges from a clause~$C_j$ to each vertex $(C_j,L)$ such that
$L$~occurs in~$C_j$ and from there to $C_{(j\bmod m)+1}$. The arena of~$\calG$
is schematically depicted in \cref{fig:np-hardness-game-1}.
\begin{figure}
\begin{tikzpicture}[x=1.4cm,y=1.3cm,->]
\useasboundingbox (-0.6,2.2) rectangle (6.1,-2.2);
\tikzstyle{play}=[circle,draw,minimum size=0.7cm]
\tikzstyle{dummy}=[circle,minimum size=0.7cm]

\node (c1) at (0,0) [play] {$C_1$};
\node (l11) at (1,1) [play] {$L_{11}$};
\node (l12) at (1,0) {\vdots};
\node (l13) at (1,-1) [play] {$L_{1k}$};
\node (c2) at (2,0) [play] {$C_2$};
\node (l21) at (3,-1) [dummy] {};
\node (l22) at (3.33,0) [dummy] {\dots};
\node (l23) at (3,1) [dummy] {};
\node (l24) at (3.5,-1) [dummy] {};
\node (l25) at (3.5,1) [dummy] {};
\node (c3) at (4.5,0) [play] {$C_m$};
\node (l31) at (5.5,1) [play] {$L_{m1}$};
\node (l32) at (5.5,0) [dummy] {\vdots};
\node (l33) at (5.5,-1) [play] {$L_{mk}$};

\draw (-0.6,0) to (c1);
\draw (c1) to (l11); \draw (c1) to (l13);
\draw (l11) to (c2); \draw (l13) to (c2);
\draw (c2) to (l21); \draw (c2) to (l23);
\draw (l24) to (c3); \draw (l25) to (c3);
\draw (c3) to (l31); \draw (c3) to (l33);
\draw (l31) .. controls +(135:2cm) and +(75:4cm) .. (c1);
\draw (l33) .. controls +(-135:2cm) and +(-75:4cm) .. (c1);
\end{tikzpicture}
\caption{\label{fig:np-hardness-game-1}Reducing SAT to deciding the existence
of a play winning for all players in a deterministic Rabin game}
\end{figure}
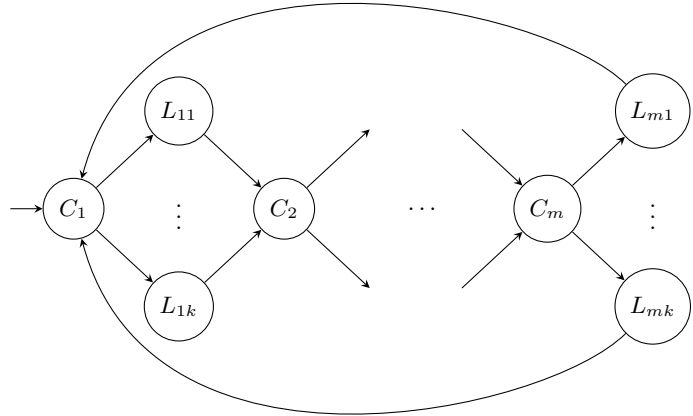
The Rabin objectives are defined as follows:
\begin{iteMize}{$-$}
 \item \pl0 wins every play of~$\calG$;
 \item \pl{i\neq 0} wins if each vertex of the form $(C,X_i)$ is
visited only finitely often \emph{or} each vertex of the form $(C,\neg X_i)$
is visited only finitely often.
\end{iteMize}

Clearly, $\calG$~can be constructed from~$\phi$ in polynomial time. To
establish the reduction, we need to show that $\phi$~is satisfiable \iff
there exists a play of~$\calG$ that is won by each player.

($\Rightarrow$) Assume that $\alpha\colon\{X_1,\dots,X_n\}\to\{\true,\false\}$
is a satisfying assignment of~$\phi$. Clearly, the positional strategy of \pl0
where from each clause~$C$ she plays to a fixed vertex $(C,L)$ such that
$L$~is mapped to true by~$\alpha$ induces a play that is won by each player.

$(\Leftarrow$) Assume that there exists a play~$\pi$ of~$\calG$ that is
won by each player. Obviously, it is not possible that both a vertex
$(C,X_i)$ and a vertex $(C',\neg X_i)$ are visited infinitely often in~$\pi$
since this would violate \pli's objective. Consider the variable
assignment that maps~$X$ to true if some vertex $(C,X)$ is visited infinitely
often in~$\pi$. This assignment satisfies the formula because, by the
construction of~$\calG$, for each clause~$C$ there exists a literal~$L$
in~$C$ such that the vertex $(C,L)$ is visited infinitely often
in~$\pi$.
\end{proof}

It follows from \cref{thm:qualne-rabin-conp-hard,prop:rabin-np-hard} that,
unless $\NP=\coNP$, \EQualNE is
not contained in $\NP\cup\coNP$, even for deterministic Rabin games.
With a little more effort, one can show that \EQualNE is
\DP-hard for deterministic Rabin games (see \cite{Ummels10}).
Finally, for stochastic Rabin games, we can show that \EQualNE is
\PNPlog-complete.

\begin{theorem}\label{thm:qualne-rabin-pnplog-hard}
\EQualNE is \PNPlog-hard for Rabin SMGs.
\end{theorem}

\begin{proof}
\Citet{Wagner90} and, independently, \citet{BussH91} showed that \PNPlog is
the closure of \NP \wrt \emph{polynomial-time Boolean formula reducibility}.
The canonical complete
problem for this class is to decide, given a Boolean combination~$\alpha$ of
statements of the form ``$\phi$~is satisfiable'', where $\phi$~ranges over all
Boolean formulae, whether $\alpha$~evaluates to true. We~claim that for every
such statement~$\alpha$ we can construct in polynomial time a Rabin SMG
$(\calG,v_0)$ such that $\alpha$~evaluates to true \iff $(\calG,v_0)$ has a
Nash equilibrium with payoff
$(0,1,\dots,1)$. The game~$\calG$ is constructed by induction on the
complexity of~$\alpha$; \wlg, we assume that negations are only applied to
atoms. If $\alpha$~is of the form ``$\phi$~is satisfiable'' or ``$\phi$~is
not satisfiable'', then the existence of a suitable game~$\calG$ follows
from \cref{prop:rabin-np-hard} or \cref{thm:qualne-rabin-conp-hard},
respectively.

Now, let $\alpha=\alpha_1\wedge\alpha_2$, and assume that we already have
constructed suitable games $(\calG_1,v_1)$ and $(\calG_2,v_2)$,
played by the same players $0,1,\dots,n$. The game~$\calG$ is the
disjoint union of $\calG_1$ and~$\calG_2$ combined with one new stochastic
vertex~$v_0$.
From~$v_0$, the game moves with probability~$\frac{1}{2}$ each to
$v_1$ or~$v_2$. Obviously, $(\calG,v_0)$ has a Nash equilibrium with
payoff $(0,1,\dots,1)$ \iff both $(\calG_1,v_1)$ and $(\calG_2,v_2)$ have such
an equilibrium.

Finally, let $\alpha=\alpha_1\vee\alpha_2$, and assume that we already have
constructed suitable games $(\calG_1,v_1)$ and $(\calG_2,v_2)$, again
played by the same players $0,1,\dots,n$. As~in the previous
case, the game~$\calG$
is the disjoint union of $\calG_1$ and~$\calG_2$ combined with one new
vertex~$v_0$, which has transitions to both $v_1$ and~$v_2$. However, this
time $v_0$~is controlled by \pl1. Obviously, $(\calG,v_0)$ has a Nash
equilibrium with payoff~$(0,1,\dots,1)$ \iff at least one of the games
$(\calG_1,v_1)$ and $(\calG_2,v_2)$ has such an equilibrium.
\end{proof}

Our next aim is to prove that \EQualNE is in $\UP\cap\coUP$ for
parity SMGs. We will make use of \cref{algo:parity-end-comp},
which computes for a game~$\calG$ with priority functions
$(\Omega_i)_{i\in\Pi}$ and ${\vec{x}\in\{0,1\}^\Pi}$ the union
of all end components with payoff~$\vec{x}$.
\begin{algorithm}[t]
\setlength{\baselineskip}{3ex}
\begin{tabbing}
\hspace*{1em}\=\hspace{1em}\=\hspace{1em}\=\hspace{1em}\=
\hspace{1em}\=\hspace{1em}\= \kill

\emph{Input:} parity SMG $\calG=(\Pi,V,(V_i)_{i\in\Pi},\Delta,\chi,
(\Omega_i)_{i\in\Pi})$, $\vec{x}=(x_i)_{i\in\Pi}\in\{0,1\}^\Pi$ \\
\emph{Output:} $\bigcup\{U\subseteq V:\text{$U$ is an end component
of~$\calG$ with payoff~$\vec{x}$}\}$ \\[\medskipamount]

\keyw{output} $\FindEC(V)$ \\[\medskipamount]

\+\keyw{procedure} $\FindEC(X)$ \\
$Z\coloneq\emptyset$ \\
compute all end components of~$\calG$ maximal in~$X$ \\
\+\keyw{for each} such end component~$U$ \keyw{do} \\
$P\coloneq\{i\in\Pi:\min\Omega_i(\chi(U))\equiv x_i\bmod 2\}$ \\
\+\keyw{if} $P=\emptyset$ \keyw{then} \\
(\textasteriskcentered\ $U$ is an end component with payoff~$\vec{x}$
\textasteriskcentered) \\
$Z\coloneq Z\cup U$ \-\\
\+\keyw{else} \\
(\textasteriskcentered\ $U$ has the wrong payoff \textasteriskcentered) \\
$Y\coloneq\bigcap_{i\in P}\{v\in U:\Omega_i(\chi(v))>\min\Omega_i(\chi(U))\}$
  \\
$Z\coloneq Z\cup\FindEC(Y)$ \-\\
\keyw{end if} \-\\
\keyw{end for} \\
\keyw{return} $Z$ \-\\
\keyw{end procedure}
\end{tabbing}
\vspace*{-2ex}
\caption{\label{algo:parity-end-comp}Finding end components in parity SMGs}
\end{algorithm}
The algorithm is a straightforward adaptation of the algorithm for computing
the union of all winning end components in a Streett MDP
\citep{ChatterjeeAH05}.
At the heart of the algorithm lies the procedure $\FindEC$ that returns on
input $X\subseteq V$ the union of all end components with payoff~$\vec{x}$
that are contained in~$X$.
The~procedure starts by computing all end components maximal in~$X$. If
such an end component~$U$ has payoff~$\vec{x}$, all vertices in~$U$ can
be added to the result of the procedure. Otherwise, there exists a
player~$i$ such that either $x_i=0$ and the least priority for \pli in~$U$
is odd or $x_i=1$ and the least priority for \pli in~$U$ is even. Each end
component with payoff~$\vec{x}$ inside~$U$ must exclude all vertices with this
least priority. Hence, we call the procedure recursively on the subset of~$U$
that results from removing these vertices.

Note that on input~$X$, the total number of recursive calls to the
procedure $\FindEC$ is bounded by~$\abs{X}$. Since,
additionally, the set of all end components maximal in a set~$X$ can be
computed in polynomial time, this proves that
\cref{algo:parity-end-comp} runs in polynomial time.

\begin{theorem}\label{thm:qualne-parity-up}
\EQualNE is in $\UP\cap\coUP$ for parity SMGs.
\end{theorem}

\begin{proof}
A \UP algorithm that decides
\EQualNE for parity SMGs works as follows:
On input $\calG,v_0,\vec{x}$, the~algorithm starts by guessing, for
each \pli with $x_i=0$, the set~$Z_i$ of vertices~$v$ with
$\val^{\calG}_i(v)=0$. Then, for each $v\in V$, the guess whether
$v\in Z_i$ or $v\nin Z_i$ is verified by running the \UP algorithm
for the respective problem. If some guess was not correct, the
algorithm rejects immediately. Otherwise, it constructs the
subarena $Z\coloneq\bigcap_{i\in\Pi:x_i=0} Z_i$ and uses
\cref{algo:parity-end-comp} to determine the union~$T$ of all end components
with payoff~$\vec{x}$. If $v_0\nin Z$, the algorithm rejects immediately.
Otherwise, it computes in polynomial time the
value~$\val^{\calG(\vec{x})}(v_0)$ of the MDP~$\calG(\vec{x})$ from~$v_0$.
If this value equals~$1$, the
algorithm accepts; otherwise, it rejects. Analogously, an algorithm for
the complement of \EQualNE accepts \iff $v_0\nin Z$ or
$\val^{\calG(\vec{x})}(v_0)<1$.

Obviously, both algorithms run in polynomial time. Moreover, on each input
there exists at most one accepting run because the algorithms only accept
if each of the sets~$Z_i$ has been guessed correctly. Finally, their
correctness follows from \cref{lemma:nash-muller}.
\end{proof}

Recall from \cref{sect:algorithmics} that it is an open question whether
the qualitative decision problem for parity \TwoSGs admits a polynomial-time
algorithm. Such an algorithm would allow us compute the domain of
the MDP~$\calG(\vec{x})$ efficiently, which would imply that \EQualNE
is in \PTime for parity SMGs. In fact, given a class~$\calC$ of parity
\TwoSGs for which the qualitative decision problem is in \PTime, we~can
easily derive a class of parity SMGs for which \EQualNE is in \PTime,
namely the class~$\calC^*$ of all parity SMGs such that for each \pli the
coalition game~$\calG_i$ is in~$\calC$.

\begin{theorem}\label{thm:qualne-parity-ptime}
Let $\calC$ be a class of finite parity \TwoSGs
such that the qualitative decision problem is decidable in \PTime for games
in~$\calC$. Then \EQualNE is in \PTime for games in~$\calC^*$.
\end{theorem}

\begin{proof}
Consider the algorithm given in the proof of \cref{thm:qualne-parity-up}.
For each \pli, the set~$Z_i$ can be computed in polynomial time if
$\calG_i\in\calC$, and there is no need to guess this set. The
resulting deterministic algorithm still runs in polynomial time.
\end{proof}

By \cref{thm:parity-bounded-ptime}, for each $d\in\bbN$, the qualitative
decision problem for parity \TwoSGs with at most $d$~priorities belongs to
\PTime.
Hence, it follows from \cref{thm:qualne-parity-ptime} that \EQualNE is
decidable in polynomial time for parity~SMGs with at most $d$~priorities.
In~particular, \EQualNE is in \PTime for (co-)\dbr B\"uchi~SMGs.
 
\begin{corollary}
For each $d\in\bbN$, \EQualNE is decidable in polynomial time for parity SMGs
with at most $d$~priorities.
\end{corollary}

\section{Conclusion}

We have analysed the complexity of deciding whether a stochastic
multiplayer game with $\omega$-regular objectives has a Nash equilibrium
whose payoff falls into a certain interval.
Our results demonstrate that this problem is more complicated for
multiplayer games than for two-player zero-sum games.
In~particular, the problem of deciding the existence
of a Nash equilibrium where \pl0 wins almost surely is
undecidable for simple stochastic multiplayer games, whereas the same problem is
decidable in polynomial time for two-player zero-sum simple stochastic games.
On the positive side, we have shown that the strictly qualitative fragment
of \NE has a complexity that is comparable to the complexity of the
qualitative decision problem for two-player zero-sum games.

Several directions for future research come to mind: First, one can study 
other restrictions of \NE that might be decidable. For example, it is
plausible that the restriction of \NE to games with two players is decidable.
Second, it would be interesting to extend our results to other game models
such as \emph{concurrent games} \cite{Shapley53,AlfaroHK07}
or games with quantitative payoff functions.


\end{document}